\documentclass[11pt,a4paper,twoside,openright]{book}
\usepackage{amsmath}
\usepackage{rotating}
\usepackage{changepage}
\usepackage[utf8x]{inputenc}
\usepackage{graphicx}
\usepackage[T1]{fontenc} 
\usepackage[english]{babel}
\usepackage{pdflscape}
\usepackage{float}
\usepackage{lmodern}
\usepackage{epigraph}
\newenvironment{changemargin}[2]{%
  \begin{list}{}{%
    \setlength{\topsep}{0pt}%
    \setlength{\leftmargin}{#1}%
    \setlength{\rightmargin}{#2}%
    \setlength{\listparindent}{\parindent}%
    \setlength{\itemindent}{\parindent}%
    \setlength{\parsep}{\parskip}%
  }%
  \item[]}{\end{list}}
\title{The dynamics on the three-dimensional boundary of the 4D Topological BF model}
\author{Andrea Amoretti}
\date{21 July 2012}
\begin{document}
\begin{titlepage}
\maketitle
\end{titlepage}
\newpage\null\thispagestyle{empty}\newpage 
\thispagestyle{empty}
\renewcommand{\chaptermark}[1]{%
\markboth{#1}{}} 
\renewcommand{\sectionmark}[1]{\markright{\thesection.\ #1}}
\pagenumbering{Roman}
\tableofcontents

\chapter*{Introduction}
\addcontentsline{toc}{chapter}{Introduction}
\chaptermark{Introduction} 
Topological Quantum Field Theories (TQFT) are characterized by the nature of their observables, (correlation functions), which do not depend on local features of the space on which these theories are defined, but only on global quantities. In particular, this means that the observables are independent of any metric which may be used to define the classical theory.

The origin of these theories has a purely mathematical nature and it was only later that their considerable physical relevance was realized. In fact, at the end of the 70's it was found that it is possible to analyze some features of Topology by using techniques of field theory. The birth of the TQFT's can be traced back to the works of Schwarz and Witten.
It was Schwarz who showed in 1978 \cite{schwarz1978} that Ray-Singer torsion, a particular topological invariant, could be represented as the partition function of a certain quantum field theory.
Quite distinct from this observation was the work of Witten in 1982 \cite{Witten:1982im}, in which a framework was given for understanding Morse theory in terms of supersymmetric quantum mechanics.\\
Ever since then, techniques of field theory have been systematically used in order to tackle mathematical problems, especially concerning the three and four dimensional manifolds \cite{donaldson}.\\

Specifically, this thesis deals with two topological models belonging to the topological theories of the Schwartz type: the Chern-Simons (CS) model \cite{schwarz1978, Schwarz:1979ae}  and the BF theory \cite{Birmingham, Horowitz}.\\
The CS model has considerable physical interest as well as mathematical. Indeed, this three-dimensional theory is closely related to the bi-dimensional conformal theories \cite{Moore} and to the three-dimensional quantum gravity \cite{Witten:1988hc}.\\
Unlike the CS action, which can be defined only in three dimensions, the BF action exists in any space-time dimension and is similar, but in some cases also coincides with, the $d$-dimensional Einstein action. This action has appeared in various forms in different contexts, being responsible for novel statistics for stringy objects \cite{Aneziris:1990gm} or the theory of Josephson junction arrays \cite{Diamantini:1995yb} as well as for the classification of 2-knots \cite{Cattaneo:1995tw}.\\

More precisely, this thesis is concerned with the introduction of a boundary in these topological models. In the three following chapters, the non-abelian CS model, the non-abelian BF theory in three space-time dimensions and its abelian four-dimensional version will be analyzed. All these models shall be defined on a flat space with a planar boundary.\\
The effect of a boundary is extremely interesting in TQFT's. Indeed, topological theories are known to have local observables only when the base manifold has a boundary \cite{Moore, Witten:1988hf}.\\
The introduction of a boundary in the CS theory has been investigated in \cite{Moore, Witten:1988hf}, with the result that the local observables arising from the presence of the boundary are two-dimensional conserved chiral currents generating the Ka\v{c}-Moody algebra \cite{kac} of the Wess-Zumino-Witten model \cite{Witten:1983ar}.\\
The existence of a Ka\v{c}-Moody algebra with central extension in the CS model with a boundary has led to some interesting applications due to the connection of the CS theory with several physical systems. In fact, following the equivalence between CS theory and three-dimensional gravity theory with a cosmological constant \cite{Witten:1988hc, Deser:1983nh}, it was stressed that the algebra plays a crucial role in understanding the statistical origin of the
entropy of a black hole \cite{Banados:1992wn}. It is thus possible to use the algebra to compute
the BTZ black hole (negative cosmological constant) entropy \cite{Banados:1998ta} and the
Kerr-de Sitter space (positive cosmological constant) entropy \cite{Park:1998qk}.\\

Regarding the non-abelian BF theory in three dimensions, similarly, the existence of chiral currents living on the boundary
and satisfying a Ka\v{c}-Moody algebra with central extension \cite{maggiore} was shown.\\
Furthermore, also in the abelian four-dimensional BF theory the existence of an algebra of boundary observables \cite{Balachandran:1992qg, balachandran} was proved.\\
Both the three dimensional and the four-dimensional BF model with a boundary were used to find the boundary actions for theories of gravity with
first order formulations \cite{momen}.\\

However, the branch of physics in which the study of these theories with a boundary have aroused the most interest in recent decades is condensed matter, due to the discovery of a new state of matter in the early 1980's, which can not be described in terms of a symmetry breaking but in terms of the topological order.\\

The starting point was, in 1981, the discovery of the quantum Hall effect (QHE) \cite{Tsui:1999zz}, a phase peculiar of three space-time dimensions. In the quantum Hall state, an external magnetic field perpendicular to an electron
gas in two space dimensions causes the electrons to circulate in quantized orbits. The “bulk” of the electron gas is an insulator, but along its edge, electrons circulate in a direction that depends
on the orientation of the magnetic field. The circulating edge states of the QHE are different from ordinary states of matter because they persist even
in the presence of impurities.\\
In contrast to the QHE, where the magnetic
field breaks time-reversal (T) symmetry, a new class of T invariants systems, i.e., topological
insulators (TI), has been predicted \cite{Kane:2005zz, Bernevig:2006zz} and experimentally observed \cite{konig:2007} in three space-time dimensions, leading
to the quantum spin Hall effect (QSHE). At the boundary of these systems, one has helical
states, namely electrons with opposite spin propagating in opposite directions. The presence
of these edge currents with opposite chiralities leads to extremely peculiar current-voltage
relationships in multi-terminal measurements. The experimental observations support these theoretical predictions \cite{konig:2007}. Also non trivial
realization of TI's in four space-time dimensions have been conjectured and realized \cite{Hasan:2010xy}.\\

The low energy sector of these kind of materials is well described in terms of the CS model and the BF theory. Moreover, it is important to study these models in the presence of a boundary in order to analyze the dynamics of the edge states of these systems.  \\
In fact, it is well known that the edge dynamics of the QHE is successfully described by the abelian CS theory with a boundary both in the
integer and in the fractional regimes \cite{Zhang:1992eu, Wen:1995qn}. In the case of three dimensional TI's an abelian doubled
CS \cite{Bernevig:2006zz}, which is equivalent to a BF effective theory, has been introduced \cite{cho, Santos:2011bf, magnoli}. Moreover, the authors of \cite{cho} argued that the abelian four-dimensional BF theory could describe some features of TI's in three space dimensions.\\

As is evident, the BF model seems to have a great potential in describing T invariants topological states of matter and from this point came the inception of this thesis, whose main purpose is to systematically describe the dynamics on the boundary of the four-dimensional BF model in terms of principles of field theory.\\

To fulfill this task, it is necessary to face a further problem, i.e., how the boundary can be introduced in the TQFT's.\\
Indeed, the inclusion of a boundary in field theory is a highly complex
task if one wishes to preserve locality and power counting, the most basic
ingredients of Quantum Field Theory.\\
In 1981 K. Symanzik \cite{syma} addressed this question: his key idea was to add to
the bulk action a local boundary term which modifies the propagators of the
fields in such a way that nothing propagates from one side of the boundary
to the other. He called this property “separability” and showed that it
requires the realizations of a well identified class of boundary conditions that
can be implemented by a local bilinear interaction.\\
These ideas inspired the authors of \cite{ maggiore,magnoli,emery, collina,maxw} and \cite{ferraro}, who used a
closely related approach to compute the chiral current algebra residing on the
boundary of the three-dimensional CS model (\cite{emery, collina, ferraro}), of the three-dimensional BF theory (\cite{maggiore,magnoli}), and of the Maxwell-Chern-Simons Theory \cite{maxw}.\\
Indeed, the authors of \cite{collina} added to the action local boundary terms compatible with power counting, using a covariant gauge fixing. In \cite{maggiore,magnoli, emery,maxw} and \cite{ferraro}, on
the other hand, a non-covariant axial gauge was preferred.\\
The main reason for the latter choice was that the main advantage of a
covariant gauge, i.e. Lorentz invariance, already fails due to the presence of
the boundary. On the other hand, the axial choice does not completely fix
the gauge \cite{bassetto}, and a residual gauge invariance exists, implying the existence
of one Ward identity in the CS theory and two Ward identities in the BF model, which play a crucial role, since, when restricted to the
boundary, they generate the chiral current algebras previously mentioned.\\
In \cite{maggiore, emery,collina} and \cite{ferraro} the explicit calculation of the propagators of the theory
seems to be necessary. In \cite{maxw} a different method was applied, which was also later used in \cite{magnoli}. This technique avoids the direct calculation of the Green functions which, in some cases, may be difficult and not very useful.\\
The basic idea of this method is to integrate the equation of motion with a boundary term (derived from the boundary potential previously mentioned),
in the proximity of the boundary and to consider the expressions obtained separately for the right side and for the left side of the boundary, in agreement with 
the Symanzik’s idea of separability. In this way, it is possible to
determine, rather than impose, the boundary conditions on the propagators, which can be expressed as boundary conditions on the fields \cite{syma}.\\
However, this method has some complications when two or more fields are involved, as in the BF theory, as it provides some boundary conditions which are not acceptable.\\
One of the original finding of this thesis is to establish a set of criteria for discarding the non-acceptable boundary conditions furnished by this method, as illustrated in Chapter 2.\\ 
The techniques developed in Chapter 2 have been extremely useful in order to study the four-dimensional BF model, and to characterize the dynamics on the boundary of this theory.\\
The characterization of the boundary dynamic of the BF theory in four space-time dimensions is the most original finding of this thesis and is illustrated in Chapter 3.\\
\\
This thesis is organized into three chapters.\\
In \textbf{Chapter 1} the introduction of the boundary in the non-abelian CS model is illustrated. The purpose of this chapter is to describe the techniques utilized in \cite{emery, ferraro}, which are largely used in the following chapters.\\
In \textbf{Chapter 2} the three-dimensional non-abelian BF theory with a boundary is analyzed. The techniques used in \cite{maggiore}, in which the direct calculation of the propagators is necessary, and the technique used in \cite{magnoli}, which avoids the computation of the Green functions, are compared.\\
The original finding of this chapter is the determination of certain criteria in order to discard the non acceptable boundary conditions furnished by the method used in \cite{magnoli}.\\
In \textbf{Chapter 3} the techniques developed in the previous chapters are applied to the abelian BF theory in four space-time dimensions with a boundary. The most original finding of this chapter, (and of the entire thesis), is the characterization of the boundary dynamics in terms of canonical commutation relations generated by the algebra of local boundary observables, which exists due to the residual gauge invariance of the bulk theory. 

\chapter{The Chern-Simons model}
\pagenumbering{arabic}
\label{CScap}
In this chapter we want to illustrate how it is possible to introduce a boundary in a field theory in the simple case of the non-abelian Chern-Simons model, in which only one gauge field is involved. We shall review the method and the findings obtained in \cite{emery, ferraro}. The purpose of this chapter is to illustrate some techniques which are largely used in Chapter 2 and 3.\\

In Section \ref{sec1} the classical Chern-Simons theory and its features are described.\\
In Section \ref{sec2} the Symanzik's method for the introduction of the boundary is illustrated. The boundary conditions are derived by using the technique illustrated in \cite{maxw}, which avoids the direct calculation of the propagators of the theory. For completeness, the Green functions of the model are calculated using the method illustrated in \cite{emery}. At the end of the Section, the Ka\v{c}-Moody algebra of local boundary observables is derived according to what was done in \cite{emery}.
\section{The classical theory}
\label{sec1}
We denote with $M$ the three-dimensional flat space-time, with $A^a_{\mu}$ a generic gauge field and with $f^{abc}$ the structure constants of the non-abelian gauge group $G$, which we suppose to be simple and compact.\\
The Chern-Simons action \cite{Moore,swartz,witten}
\begin{equation}
\label{csaction}
S_{CS}=-\frac{k}{4 \pi} \int_M d^3 x \epsilon^{\mu \nu \rho} \{ A^a_\mu \partial_\nu A^a_\rho+\frac{1}{3} f^{abc} A^a_\mu A^b_\nu A^c_\rho \}
\end{equation}
is defined as the most general local and covariant action which respects the power-counting, which is invariant under the gauge transformation
\begin{equation}
\label{gauge}
\begin{array}{l}
\displaystyle A_\mu^a \rightarrow A_\mu^a -(D_\mu \theta)^a\\
\displaystyle (D_\mu X)^a \equiv \partial _\mu X^a+ f^{abc}A_\mu^a X^c
\end{array}
\end{equation}
and which has a dimensionless coupling constant $k$.\\
In this chapter we will discuss the theory in presence of a boundary $x_2=0$ and, for this reason, it is convenient to use the light-cone coordinate system
\begin{equation}
\label{conoluce1}
\begin{cases}
u=x_2\\
z=\frac{1}{\sqrt{2}}(x_0+x_1)\\
\overline{z}=\frac{1}{\sqrt{2}}(x_0-x_1),
\end{cases}
\end{equation}
which induces similar definitions in the space of the fields:
\begin{equation}
 \label{conoluce2}
\begin{cases}
A_u=A_2^a\\
A^a=\frac{1}{\sqrt{2}}(A_0^a+A_1^a)\\
\overline{A}^a=\frac{1}{\sqrt{2}}(A_0^a-A_1^a).
\end{cases}
\end{equation}
With these conventions, it is possible to rewrite the action \eqref{csaction} as follows:
\begin{equation}
S_{CS}=-\frac{k}{2\pi} \int_M du d^2 z \{ A^a \overline{\partial} A^a_u+ A^a_u \partial \overline{A}^a+\overline{A}^a \partial_u A^a+f^{abc}A^a\overline{A}^b A_u^c \}
\end{equation}
As it is known, it is necessary to choose a gauge by adding an appropriate gauge-fixing term to the action \eqref{csaction} in order to make the theory consistent. For the purpose of this chapter it is convenient to choose an axial gauge $A_u^a=0$, which corresponds to the gauge-fixing term:
\begin{equation}
\label{gf}
S_{GF}=\int_M du d^2 z \{ b^a A_u^a+\overline{c}^a(\partial_u c^a+f^{abc}A_u^b c^c)\},
\end{equation}
where $c^a$ and $\overline{c}^a$ are respectively the ghost and anti-ghost fields and $b^a$ are the Lagrange multipliers.\\
Having done that, the action $S_{TOT}=S_{CS}+S_{GF}$ is invariant under the BRS transformations:
\begin{equation}
\begin{array}{l}
\displaystyle s A_\mu^a = -(D_\mu c)^a\\
\displaystyle s c^a=\frac{1}{2} f^{abc}c^b c^c \\
\displaystyle s \overline{c}^a= b^a\\
\displaystyle s b^a=0.
\end{array}
\end{equation}
Moreover, as usual, the gauge-fixing term is a BRS-variation:\\
 $S_{GF}=s \int dy d^2z \overline{c}^aA^a_u$.\\

As regards the discrete symmetries, once defined the action of the parity transformation on the coordinates as follows:
\begin{equation}
\label{coordipar}
P: \{ z,\overline{z},u \} \leftrightarrow \{ \overline{z},z,-u \} 
\end{equation}
it is possible to define only one transformation on the space of the fields which leaves $S_{TOT}$ unchanged:
\begin{equation}
\label{parity}
 P:\begin{cases}
 z \leftrightarrow \overline{z}\\ u \leftrightarrow -u \\ A^a \leftrightarrow \overline{A}^a \\ A_u^a \leftrightarrow -A_u^a \\ b^a \leftrightarrow -b^a \\ \overline{c}^a \rightarrow -c^a \\ c^a \rightarrow \overline{c}^a 
 \end{cases}  
\end{equation}
In what follows we will call this transformation \textit{parity}.\\
In regard to the time-reversal transformation, once defined the action of this symmetry on the coordinates as
\begin{equation}
\label{coorditime}
 T: \{z,\overline{z},u \} \leftrightarrow \{-\overline{z},-z, u \},
\end{equation}
it is impossible to find a transformation in the space of the fields which leaves $S_{TOT}$ unchanged. Therefore, the Chern-Simons action is not invariant under the time-reversal transformation.\\
Furthermore, each field of the theory carries a charge called Faddeev-Popov charge, (the values for each field are shown in Table \ref{dim}), and it is easy to see that each term in $S_{TOT}$ has a Faddeev-Popov charge equal to zero.

The gauge-fixing term $S_{GF}$ is not covariant but the action $S_{TOT}$ is invariant under the two dimensional Lorentz transformations on the plane $\{z,\overline{z}\}$. In particular, this is equivalent to saying that the helicity is conserved \cite{emery}. The mass dimensions, the values of the Faddeev-Popov charge and of the helicity of the fields of the theory are listed in Table \ref{dim}.
\begin{table}[H]
\centering
\begin{tabular}{|c|c|c|c|c|c|c|c|c|c|}
\hline
 &$z$&$\overline{z}$& $u$&$A^a$&$\overline{A^a}$&$A^a_u$& $c^a$&$\overline{c}^a$&$b^a$\\ \hline
dim &-1&-1&-1&1&1&1&1&1&2\\ \hline
helicity&-1&1&0&1&-1&0&0&0&0\\ \hline
$\Phi \Pi$&0&0&0&0&0&0&1&-1&0\\ \hline
\end{tabular}
\label{dim}
\caption{mass dimension, $\Phi$-$\Pi$ charge and helicity}
\end{table}

At the classical level, the theory is scale invariant since all the parameters introduced are dimensionless.\\
Finally, we note that $S_{TOT}$ has a residual gauge invariance on the plane $\{z,\overline{z}\}$:
\begin{equation}
\begin{array}{l}
\displaystyle \delta A_\mu^a=-\partial_{\mu} \omega^a-f^{abc}A_\mu^b \omega^c, \\
\displaystyle \delta \varphi^a= -f^{abc} \varphi^b \omega^c, \qquad \forall \varphi \ne A_\mu^a
\end{array}
\end{equation}
where $\omega$ is a function of $z$ and $\overline{z}$. Then, the gauge is not completely fixed by \eqref{gauge} and, as we will see, the Ward identity, which expresses the residual gauge invariance of the theory, will play a key role in determining the algebra on the boundary $u=0$.\\
At the classical level, the generating functional of the connected Green functions $Z_c(J_{\varphi})$ is obtained from the classical action by a Legendre transformation:
\begin{equation}
Z_c(J_\varphi)=S_{TOT}(\varphi)+\int_M du d^2z \sum_\varphi J^a_\varphi \varphi^a,
\end{equation}
where $J^a_{\varphi}$ are the sources for the fields, denoted collectively by $\varphi^a$. Consequently, the equations of motion for the fields and for the Lagrange multipliers are:
\begin{equation}
\label{campi}
 \begin{array}{l}
\displaystyle \frac{k}{2\pi}(\partial_u \overline{A}^a-\overline{\partial}A_u^a-f^{abc}\overline{A}^bA_u^c)+J_A^a=0  \\
\displaystyle \frac{k}{2\pi}(\partial A_u^a-\partial_uA^a+f^{abc}A^bA_u^c)+J_{\overline{A}}^a=0\\
\displaystyle \frac{k}{2\pi}(\overline{\partial}A^a-\partial \overline{A}^a-f^{abc}A^b \overline{A}^c+\frac{2\pi}{k}(b^a-f^{abc}\overline{c}^bc^c))+J_{A_u}^a=0\\
\displaystyle A_u^a+J_b^a=0,
 \end{array}
\end{equation}
while for the ghost fields we obtain:
\begin{equation}
\label{ghost}
 \begin{array}{l}
  \displaystyle \partial_u \overline{c}^a+f^{abc}A^b_u \overline{c}^c-J_c^a=0\\
  \displaystyle \partial_u c^a+f^{abc}A^b_uc^c-J_{\overline{c}}^a=0.
 \end{array}
\end{equation}
Due to the linearity of the equations of motion for the ghost \eqref{ghost}, the Slavnov identity, which express the invariance under BRS transformations, takes the form of a local Ward identity:
\begin{equation}
\label{wardlocale}
 \partial J_A^a+\overline{\partial} J_{\overline{A}}^a+\partial_u J_{A_u}+\partial_u \frac{\delta Z_c}{\delta J^a_b}-\sum_{\varphi}f^{abc}J^b_{\varphi}\frac{\delta Z_c}{\delta J^a_{\varphi}}=0.
\end{equation}
\section{The boundary}
\label{sec2}
In this section we will introduce a boundary in the theory \cite{collina,ferraro,emery} and, for this purpose, we will choose the plane $u=0$. In order to do this, we breaks the equation of motions \eqref{campi} and \eqref{ghost} with a boundary term which respects some basic requirements \cite{maxw}.\\
\textbf{Locality:} the boundary contribution must be local. This means that all possible breaking terms have the form
\begin{equation}
\delta^{(n)}(u)X(z,\overline{z},u),
\end{equation}
where $\delta^{(n)}(u)$ is the $n$-order derivative of the Dirac delta function with respect to its argument, and $X(z,\overline{z},u)$ is a local functional.\\
\textbf{Separability:} this condition, called also \textit{decoupling condition} \cite{syma}, refers to a Symanzik's original idea according to which the $n$-point Green functions which involve two fields computed in points belonging to opposite sides of space must vanish. In particular, the propagators of the theory must satisfy:
\begin{equation}
\begin{array}{l}
\displaystyle uu'<0 \; \Rightarrow \; \; \Delta_{AB}(x,x')=\mathcal{h}T(\varphi_{A}(x)\varphi_{B}(x'))\mathcal{i}=0\\
\displaystyle x\equiv (z,\overline{z},u).
\end{array}
\end{equation}
This property is satisfied by propagators which split according to
\begin{equation}
\label{prop}
\Delta_{AB}(x,x')= \theta_+ \Delta_{AB+}(x,x')+\theta_- \Delta_{AB-}(x,x'),
\end{equation}
where $\Delta_{AB+}(x,x')$ and $\Delta_{AB-}(x,x')$ are respectively the propagators for the right and the left side of the space-time and $\theta_\pm \equiv \theta(\pm u)\theta(\pm u')$, where $\theta(x)$ is the step function, defined as usual.\\
The condition \eqref{prop} induces the decoupling of the generating functional of the connected Green functions $Z_c(J_{\varphi})$ according to
\begin{equation}
\label{funzionale}
Z_c(J_{\varphi})=Z_c(J_{\varphi})_++Z_c(J_{\varphi})_-,
\end{equation}
where $Z_c(J_{\varphi})_+$ and $Z_c(J_{\varphi})_-$ are the generators of the connected Green functions for the right side and for the left side of the space-time, respectively.\\
\textbf{Linearity:} finally, we require that the boundary contributions to the equations of motions are linear in the fields because, in general, the symmetries of the classical action, if only linearly broken at the classical level, nonetheless retains the exact symmetries of the quantum action \cite{weinberg}.\\

It is convenient to define the insertions of the fields on the boundary
\begin{equation}
\begin{split}
\label{cbordo}
 &\varphi_\pm= \lim_{u \rightarrow 0^\pm} \frac{\delta Z_c}{\delta J_{\varphi}}(Z)\\
 & Z \equiv \{ z, \overline{z} \}
\end{split}
\end{equation}
in order to simplify the next formulas. We note that, under parity, the fields \eqref{cbordo} transform as:
\begin{equation}
\begin{split}
 &A_\pm^a \leftrightarrow \overline{A}_\mp^a \\ 
 &A_{u\pm} \leftrightarrow -A_{u\mp}^a \\ 
 &\overline{c}^a_\pm \rightarrow -c^a_\mp \\ 
 &c^a_\pm \rightarrow \overline{c}^a_\mp
\end{split}
\end{equation}
Moreover, we require that the boundary term in the equations of motion preserves parity invariance, power counting and conservation of Faddeev-Popov charge, in addition to the conditions illustrated above. With these assumptions, the most general equation of motion for the ghost fields with a boundary term are:
\begin{equation}
\label{grotte}
 \begin{array}{l}
  \displaystyle \partial_u \overline{c}^a+f^{abc}A^b_u \overline{c}^c-J_c^a=\delta(u)(\mu_+ \overline{c}^a_++\mu_- \overline{c}^a_-)\\
  \displaystyle \partial_u c^a+f^{abc}A^b_uc^c-J_{\overline{c}}^a=-\delta(u)(\mu_- c^a_++\mu_+ c^a_-),\\
 \end{array}
\end{equation}
where $\mu_+$ and $\mu_-$ are constant parameters which we will determine later.\\
We set a first constraint on the parameters $\mu_+$ and $\mu_-$ by requiring that the equations of motion \eqref{grotte} are compatible with each other. This condition is nothing more than the commutation property which equations of motion in general obey, and it is equivalent to require that the boundary term is derived from an action. In other words, the compatibility condition implies that:
\begin{equation}
\label{compgo}
 \frac{\delta^2 Z_c}{\delta c^a_{\pm} \delta \overline{c}^a_{\pm}}=-\frac{\delta^2 Z_c}{\delta \overline{c}^a_{\pm} \delta c^a_{\pm} },
\end{equation}
where the minus sign on the right hand side of the equation \eqref{compgo} is due to the fact that we have differentiated with respect to Grassmann variables.\\
The compatibility condition \eqref{compgo} sets an algebraic constraint on the parameters $\mu_+$ and $\mu_-$:
\begin{equation}
 \mu_+=\mu_- \equiv \mu.
\end{equation}
In the same way, the equations of motion for the gauge fields which respect the condition just illustrated are:
\begin{equation}
\label{campirotte}
 \begin{array}{l}
\displaystyle \frac{k}{2\pi}(\partial_u \overline{A}^a-\overline{\partial}A_u^a-f^{abc}\overline{A}^bA_u^c)+J_A^a=-\delta(u)(\lambda_- \overline{A}_+^a+\lambda_+ \overline{A}_-^a)  \\
\displaystyle \frac{k}{2\pi}(\partial A_u^a-\partial_uA^a+f^{abc}A^bA_u^c)+J_{\overline{A}}^a=-\delta(u)(\lambda_+ A_+^a+\lambda_- A_-^a)\\
\displaystyle \frac{k}{2\pi}(\overline{\partial}A^a-\partial \overline{A}^a-f^{abc}A^b \overline{A}^c+\frac{2\pi}{k}(+b^a-f^{abc}\overline{c}^bc^c))+J_{A_u}^a=0\\
\displaystyle A_u^a+J_b^a=0,
 \end{array}
\end{equation}
where we have introduced a minus sign on the right hand side for convenience in the following developments and the quantities $\lambda_{\pm}$ are constant parameters. The right hand side of the third equation, as that of the fourth one, has not been modified due to the gauge choice $A_u^a=0$.\\
In this case, the compatibility condition implies that:
\begin{equation}
 \frac{\delta Z_{c\pm}}{\delta A^a_{\pm} \delta \overline{A}_\pm^a}=\frac{\delta Z_{c\pm}}{\delta \overline{A}_\pm^a \delta A^a_{\pm} },
\end{equation}
which yields:
\begin{equation}
 \lambda_+=\lambda_-\equiv \lambda.
\end{equation}
Then, the most general equations of motion for the ghost and gauge fields with a boundary term are:
 \begin{align}
\label{rottetot1}
&\frac{k}{2\pi}(\partial_u \overline{A}^a-\overline{\partial}A_u^a-f^{abc}\overline{A}^bA_u^c)+J_A^a=-\lambda \delta(u)( \overline{A}_+^a+ \overline{A}_-^a)  \\
\label{rottetot2}
&\frac{k}{2\pi}(\partial A_u^a-\partial_uA^a+f^{abc}A^bA_u^c)+J_{\overline{A}}^a=-\lambda \delta(u)( A_+^a+ A_-^a)\\
\label{rottetot3}
&\frac{k}{2\pi}(\overline{\partial}A^a-\partial \overline{A}^a-f^{abc}A^b \overline{A}^c+\frac{2\pi}{k}(+b^a-f^{abc}\overline{c}^bc^c))+J_{A_u}^a=0\\
& A_u^a+J_b^a=0\\
\label{rottetot4}
& \partial_u \overline{c}^a+f^{abc}A^b_u \overline{c}^c-J_c^a=\mu\delta(u)( \overline{c}^a_++ \overline{c}^a_-)\\
\label{rottetot5}
& \partial_u c^a+f^{abc}A^b_uc^c-J_{\overline{c}}^a=-\mu\delta(u)( c^a_++ c^a_-),
& \end{align}
where $\mu$ and $\lambda$ are constant parameters to be determined by the boundary conditions.
\subsection{The boundary conditions}
In this section we will find the boundary conditions for the gauge and ghost fields following the method used in \cite{maxw}. This method is economical compared to other methods \cite{collina,sassarini,emery} because it makes the direct calculation of the propagators  unnecessary. As we will see in the next chapter, it presents some complications when the boundary conditions involve two different fields (BF theory), but this is not the case.\\
In order to obtain the boundary conditions, we will integrate the equations of motion \eqref{rottetot1}-\eqref{rottetot5} with respect to the coordinate $u$ in the infinitesimal interval $[-\epsilon,\epsilon]$ and we will evaluate the expressions obtained in the weak limit $\epsilon \rightarrow0$. In doing so, we will find an algebraic system which involves the parameters $\mu$ and $\lambda$ and the insertions $\varphi^a_{\pm}$. In order to solve the system we will use the separability condition \cite{syma}, requiring that the system can be solved separately on the right side and on the left side of the boundary.
\subsubsection{The ghost sector}
If we apply the method just described to the equation \eqref{rottetot4} we obtain:
\begin{equation}
  \overline{c}^a(Z,\epsilon)-\overline{c}^a(Z,-\epsilon)-\int_{-\epsilon}^{\epsilon}duf^{abc}J_b^b\overline{c}^a =\mu(\overline{c}^a_++\overline{c}^a_-),
\end{equation}
which, in the weak limit, becomes:
\begin{equation}
\label{c2}
\overline{c}^a_+-\overline{c}^a_-=\mu(\overline{c}^a_++\overline{c}^a_-).
\end{equation}
Equally, we find 
\begin{equation}
\label{c1}
 c^a_+-c^a_-=-\mu(c^a_++c^a_-).
\end{equation}
for the equation \eqref{rottetot5}. Next, we impose the separability condition on the equations \eqref{c2} and \eqref{c1}, obtaining the following systems of equation:
\begin{equation}
\label{sistem}
 \begin{cases} (1-\mu)\overline{c}^a_+=0 \\ (1+\mu) c^a_+=0, \end{cases} \qquad \begin{cases} (1+\mu)\overline{c}^a_-=0 \\ (1-\mu)c^a_-=0. \end{cases}
\end{equation}
Notice that the second system in \eqref{sistem} can be obtained from the first one by a parity transformation.\\
The independent solutions for the systems \eqref{sistem} are:
\begin{equation}
\label{solutiong}
 \begin{split}
&I: \qquad \mu=1 \qquad c^a_+=\overline{c}^a_-=0\\
I&I: \qquad \mu=-1 \qquad c^a_-=\overline{c}^a_+=0.
 \end{split}
\end{equation}
If we substitute the solutions \eqref{solutiong} into the equations \eqref{rottetot4} and \eqref{rottetot5} we obtain:
\begin{equation}
 \begin{array}{l}
  \displaystyle \partial_u \overline{c}^a+f^{abc}A^b_u \overline{c}^c-J_c^a=\delta(u) \overline{c}^a_+(z,\overline{z})\\
\displaystyle \partial_u c^a+f^{abc}A^b_uc^c-J_{\overline{c}}^a=-\delta(u) c^a_-(z,\overline{z})
 \end{array}
\end{equation}
for the solution $I$ and
\begin{equation}
 \begin{array}{l}
  \displaystyle \partial_u \overline{c}^a+f^{abc}A^b_u \overline{c}^c-J_c^a=-\delta(u) \overline{c}^a_-(z,\overline{z})\\
\displaystyle \partial_u c^a+f^{abc}A^b_uc^c-J_{\overline{c}}^a=\delta(u) c^a_+(z,\overline{z}).
 \end{array}
\end{equation}
for the solution $II$.
\subsubsection{The gauge sector}
We now apply the same method to the equations of motion for the gauge fields \eqref{rottetot1} and \eqref{rottetot2}, finding the following relations:
\begin{equation}
\label{sistemg}
 \begin{array}{l}
\displaystyle  \frac{k}{2\pi}(\overline{A}^a_+-\overline{A}^a_-)=-\lambda(\overline{A}^a_++\overline{A}^a_-)\\
\displaystyle \frac{k}{2\pi} (A^a_--A^a_+)=-\lambda(A^a_++A^a_-).
\end{array}
\end{equation}
If we impose the separability condition, we find the following algebraic systems:
\begin{equation}
\label{sistemgg}
\begin{cases} (\lambda+\frac{k}{2\pi})\overline{A}^a_+=0 \\  (\lambda-\frac{k}{2\pi})A^a_+=0, \end{cases} \qquad \begin{cases} (\lambda+\frac{k}{2\pi})A^a_-=0 \\ (\lambda-\frac{k}{2\pi}) \overline{A}^a_-=0. \end{cases}
\end{equation}
The independent solutions of the two systems \eqref{sistemgg} are:
\begin{equation}
\label{solutiongg}
 \begin{split}
  &I: \qquad \lambda=\frac{k}{2\pi} \qquad A_-^a=\overline{A}^a_+=0\\
 I&I: \qquad \lambda=-\frac{k}{2\pi} \qquad \overline{A}^a_-=A^a_+=0.
 \end{split}
\end{equation}
Subsequently, the equations of motion for the gauge fields are:
\begin{equation}
\label{campirotte1}
 \begin{array}{l}
\displaystyle \frac{k}{2\pi}(\partial_u \overline{A}^a-\overline{\partial}A_u^a-f^{abc}\overline{A}^bA_u^c)+J_A^a=-\frac{k}{2\pi}\delta(u) \overline{A}_-^a  \\
\displaystyle \frac{k}{2\pi}(\partial A_u^a-\partial_uA^a+f^{abc}A^bA_u^c)+J_{\overline{A}}^a=-\frac{k}{2\pi}\delta(u) A_+^a\\
\displaystyle \frac{k}{2\pi}(\overline{\partial}A^a-\partial \overline{A}^a-f^{abc}A^b \overline{A}^c+\frac{2\pi}{k}(+b^a-f^{abc}\overline{c}^bc^c))+J_{A_u}^a=0\\
\displaystyle A_u^a+J_b^a=0,
 \end{array}
\end{equation}
for the solution $I$ and
\begin{equation}
\label{campirotte2}
 \begin{array}{l}
\displaystyle \frac{k}{2\pi}(\partial_u \overline{A}^a-\overline{\partial}A_u^a-f^{abc}\overline{A}^bA_u^c)+J_A^a=\frac{k}{2\pi}\delta(u) \overline{A}_+^a  \\
\displaystyle \frac{k}{2\pi}(\partial A_u^a-\partial_uA^a+f^{abc}A^bA_u^c)+J_{\overline{A}}^a=\frac{k}{2\pi}\delta(u) A_-^a\\
\displaystyle \frac{k}{2\pi}(\overline{\partial}A^a-\partial \overline{A}^a-f^{abc}A^b \overline{A}^c+\frac{2\pi}{k}(+b^a-f^{abc}\overline{c}^bc^c))+J_{A_u}^a=0\\
\displaystyle A_u^a+J_b^a=0,
 \end{array}
\end{equation}
for the solution $II$.
\subsection{The propagators}
For completeness, in this section we will derive the propagators of the theory
\begin{equation}
\label{propagator}
 \Delta_{\varphi_1,\varphi_2}(x,x')= \mathcal{h}T(\varphi_1(x),\varphi_2(x'))\mathcal{i},
\end{equation}
taking into account the decomposition \eqref{prop}.
\subsubsection{The ghost sector}
The free equations of motion for the ghost fields without a boundary term are:
\begin{equation}
 \begin{array}{l}
  \displaystyle \partial_u \overline{c}^a-J_c^a=0\\
  \displaystyle \partial_u c^a-J_{\overline{c}}^a=0.
 \end{array}
\end{equation}
Let us differentiate the first equation with respect to $J^b_c(x')$ and the second equation with respect to $J^b_{\overline{c}}(x')$ obtaining the following relations for the ghost propagators:
\begin{equation}
\label{eqprop}
 \begin{array}{l}
  \displaystyle \partial_u \Delta_{\overline{c}c}^{ab}(x',x)=\delta^{ab}\delta^{(3)}(x-x')\\
  \displaystyle \partial_u \Delta_{c\overline{c}}^{ab}(x',x)=\delta^{ab}\delta^{(3)}(x-x').\\
 \end{array}
\end{equation}
As we have already observed,  the most general form of the propagator \eqref{propagator} which respects the separability condition is:
\begin{equation}
\label{propdef}
 \Delta_{AB}^{ab}(x,x')= \theta_+ \Delta_{AB+}^{ab}(x,x')+\theta_- \Delta_{AB-}^{ab}(x,x').
\end{equation}
$\Delta_{AB+}^{ab}(x,x')$ and $\Delta_{AB-}^{ab}(x,x')$ are respectively the propagators for the right side and for the left side of the space-time with respect to the boundary. They have to be solutions for the equations \eqref{eqprop}. Having done that, we find:
\begin{equation}
\label{propmatr}
 \Delta_{AB+}^{ab}= \delta^{ab}\begin{pmatrix} 0 & -T_{\rho}(x,x') \\ T_{\rho}(x',x) & 0 \end{pmatrix},
\end{equation}
where we have used the matrix notation and the indices $A$ and $B$ denote the pair of fields $(c^a, \overline{c}^a)$. $\Delta_{AB-}$ is obtained from \eqref{propmatr} via  a parity transformation and $T_{\rho}(x,x')$ indicates the tempered distribution
\begin{equation}
\label{Trho}
 T_{\rho}(x,x')=(\theta(u-u')+\rho)\delta^{(2)}(x-x'),
\end{equation}
where $\rho$ is a constant parameter to be determined by the boundary conditions. As observed in \cite{emery}, $\rho$ could be an arbitrary function of the variables $z$ and $\overline{z}$. The choice $\rho=const$ was made in order to satisfy the conditions of conservation of helicity, power-counting and regularity.\\
The next step is to consider the free equations of motion with a boundary term and to substitute the propagators just obtained into these equations in order to determine the possible values of $\rho$ taking into account the boundary conditions which we have found in the previous section.\\
Regarding the boundary condition $I$ in \eqref{solutiong}, the free ghost equations with a boundary term are:
\begin{equation}
 \begin{array}{l}
  \displaystyle \partial_u \overline{c}^a-J_c^a=\delta(u) \overline{c}^a_+(z,\overline{z})\\
\displaystyle \partial_u c^a-J_{\overline{c}}^a=-\delta(u) c^a_-(z,\overline{z}).
 \end{array}
\end{equation}
Consequently, the equation for $\Delta_{\overline{c}c}(x,x')$ is:
\begin{equation}
\label{equprop}
 \partial_u \Delta_{\overline{c}c}(x,x')-\delta^{(3)}(x-x')=-\delta(u)\Delta_{\overline{c}c}(0^+,x').
\end{equation}
The substitution of the solution for the propagators \eqref{propagator} into the equation \eqref{equprop} leads to:
\begin{equation}
\begin{array}{l}
 \displaystyle -\delta^{(2)}(Z-Z')[\theta(u')\delta(u)(\theta(u-u')+\rho)+\theta(u')\theta(u)\delta(u-u')+\\
\displaystyle -\theta(-u')\delta(u)(\theta(u-u')+\rho)+\theta(-u)\theta(-u')\delta(u-u')]+\\
+\delta^{(3)}(x-x')=\displaystyle -\delta(u)\Delta_{\overline{c}c}(0^+,X).
\end{array}
\end{equation}
If we require that $X'$ belongs to the half-space $u'<0$, we find:
\begin{equation}
 \delta^{(2)}(Z-Z')\delta(u)(1+\rho)=0,
\end{equation}
which yields:
\begin{equation}
 \rho=-1,
\end{equation}
In regard to the boundary condition $II$, the free equations of motion are:
\begin{equation}
 \begin{array}{l}
  \displaystyle \partial_u \overline{c}^a-J_c^a=-\delta(u) \overline{c}^a_-(z,\overline{z})\\
\displaystyle \partial_u c^a-J_{\overline{c}}^a=\delta(u) c^a_+(z,\overline{z}).
 \end{array}
\end{equation}
Via arguments similar to those just made, we find:
\begin{equation}
 \rho=0.
\end{equation}
Summarizing, the two boundary conditions for the ghost sector are:
\begin{equation}
 \begin{split}
  &I: \qquad \rho=-1 \qquad \mu=1\\
  I&I: \qquad \rho=0 \qquad \mu=-1.
 \end{split}
\end{equation}
\subsubsection{The gauge sector}
The free equations of motion for the fields $A^a$, $\overline{A}^a$, $A_u^a$ and for the Lagrange multipliers $b^a$ without a boundary term are:
\begin{equation}
\label{campilib}
 \begin{array}{l}
  \displaystyle \frac{k}{2\pi}(\partial_u \overline{A}^a+\overline{\partial}J_b^a)+J_A^a=0  \\
\displaystyle -\frac{k}{2\pi}(\partial J_b^a+\partial_uA^a)+J_{\overline{A}}^a=0\\
\displaystyle \frac{k}{2\pi}(\overline{\partial}A^a-\partial \overline{A}^a)+J_{A_u}^a+b^a=0\\
\displaystyle A_u^a+J_b^a=0.
 \end{array}
\end{equation}
If we differentiate the previous equations with respect to the sources of the fields, we find the following system of equations for the propagators:
\begin{equation}
\label{equazioniprop}
 \begin{split}
&\Delta^{ba}_{AA_u}(x',x)=\Delta^{ba}_{A_uA}(x',x)=0\\
&\Delta^{ba}_{\overline{A}A_u}(x',x)=\Delta^{ba}_{A_u \overline{A}}(x',x) =0  \\
&\Delta^{ba}_{A_uA_u}(x',x)=0 \\
&\Delta^{ba}_{bA_u}(x',x)=-\delta^{ab}\delta^{(3)}(x-x')=\Delta^{ba}_{A_ub}(x',x)\\
&\partial_u \Delta^{ba}_{b \overline{A}}(x',x)=-\delta^{ab} \overline{\partial}\delta^{(3)}(x'-x) \\
&\partial_u \Delta^{ba}_{b A}(x',x)=-\delta^{ab} \partial\delta^{(3)}(x'-x) \\
&\partial_u \Delta^{ba}_{A\overline{A}}(x',x)=-\frac{2\pi}{k}\delta^{ab}\delta^{(3)}(x-x')\\
&\partial_u \Delta^{ba}_{\overline{A}A}(x',x)=-\frac{2\pi}{k}\delta^{ab}\delta^{(3)}(x-x')\\
&\partial_u \Delta^{ba}_{\overline{A}\overline{A}}(x',x)=\partial_u \Delta^{ba}_{AA}(x',x)=0\\
&\frac{k}{2\pi}(\overline{\partial}\Delta^{ba}_{AA}(x',x)-\partial\Delta^{ba}_{A\overline{A}}(x',x))+\Delta^{ba}_{Ab}(x',x)=0\\
&\Delta^{ba}_{bb}(x',x)=-\frac{k}{2\pi}(\overline{\partial}\Delta^{ba}_{bA}(x',x)-\partial\Delta^{ba}_{b\overline{A}}(x',x)).
 \end{split}
\end{equation}
The general solution of the system \eqref{equazioniprop} can be written in the form \eqref{propdef}, where:
\begin{equation}
\label{propagatoregen}
\begin{split}
&\Delta^{ab+}_{AB}= \\
& \delta^{ab} \begin{pmatrix} \frac{2\pi}{k} \frac{\alpha}{2\pi i(z-z')^2} & \frac{2\pi}{k} T_{\gamma}(x',x) & 0 & \partial T_{\gamma-\alpha}(x',x)\\ -\frac{2\pi}{k} T_{\gamma}(x,x') & \frac{2\pi}{k}  \frac{\beta}{2\pi i (\overline{z}-\overline{z}')^2}&0 & -\overline{\partial}T_{\gamma-\beta}(x,x')\\ 0 & 0 & 0 & -\delta^{(3)}(x-x')\\ -\partial T_{\gamma-\alpha}(x,x') & \overline{\partial}T_{\gamma-\beta}(x',x) & -\delta^{(3)}(x-x') & \frac{k}{2\pi} (-1-2\gamma+\alpha+\beta) \partial \overline{\partial} \delta^2 \end{pmatrix}
\end{split}
\end{equation}
The indices $A$ and $B$  denote the ordered set of fields $(A,\overline{A},A_u,b)$ and $\Delta_{AB}^{ab-}$ is obtained from \eqref{propagatoregen} by a parity transformation. The quantities $\alpha$, $\beta$ and $\gamma$ are constant parameters to be determined by the boundary conditions and $T_{\rho}$ is the tempered distribution defined in \eqref{Trho}.\\
In order to solve this system \eqref{equazioniprop} we have used the following representation of the delta function:
\begin{equation}
 \overline{\partial}\frac{1}{z-z'-i\epsilon (\overline{z}-\overline{z}')} \equiv \overline{\partial}\frac{1}{z-z'}=2\pi i \delta^{(2)}(Z-Z'),
\end{equation}
where we have defined:
\begin{equation}
\label{unosuzeta}
 \frac{1}{(z-z')^2} \equiv - \partial \frac{1}{(z-z')}.
\end{equation}
Via arguments similar to those just made, it is possible to determine the values of the parameters $\alpha$, $\beta$ and $\gamma$ for the two boundary conditions \eqref{solutiongg}. Therefore, we obtain:
\begin{equation}
\begin{split}
 &\; I: \; \; \;  \gamma  =0, \qquad \alpha \; \; \text{arbitrary}, \qquad \beta=0, \qquad \; \; \; \; \lambda=\frac{k}{2\pi};\\
&II: \; \; \; \gamma  =-1, \qquad \alpha=0, \qquad \beta \; \; \text{arbitrary}, \qquad \lambda=-\frac{k}{2\pi}.
\end{split}
\end{equation}
\subsection{The boundary algebra}
In this section we will explain how the Ward identity \eqref{wardlocale} implies the existence  of a Ka\v{c}-Moody algebra \cite{kac} of conserved chiral currents on the boundary. In order to do this we will use the technique illustrated in \cite{maxw}.\\
Firstly, we will analyze how the presence of a boundary term in the equations of motion modifies the Ward identity \eqref{wardlocale}.\\
If we consider the solution $I$ in \eqref{solutiongg} and the corresponding equations of motion \eqref{campirotte1}, using all the precautions mentioned in \cite{emery} (the discussion of which is beyond the purposes of this thesis), the boundary breaking of the Ward identity \eqref{wardlocale} can be formally obtained from the equations of motion:
\begin{equation}
\label{ward1}
 \partial J_A^a+\overline{\partial}J_{\overline{A}}^a+\partial_u J_{A_u}^a+\partial_u b^a-\sum_{\varphi} f^{abc} J_{\varphi}^b \varphi^c=-\delta(u)\frac{k}{2\pi}(\overline{\partial}A^a_++\partial \overline{A}^a_-). 
\end{equation}
As regards the solution $II$, we achieve in the same way:
\begin{equation}
\label{ward2}
  \partial J_A^a+\overline{\partial}J_{\overline{A}}^a+\partial_u J_{A_u}^a+\partial_u b^a-\sum_{\varphi} f^{abc} J_{\varphi}^b \varphi^c=\delta(u)\frac{k}{2\pi}(\overline{\partial}A^a_-+\partial \overline{A}^a_+).
\end{equation}
The equations \eqref{ward1} and \eqref{ward2} are the Ward identities which express the residual gauge invariance on the boundary $u=0$.\\
In what follows we will derive the algebra for the solution $I$ and we will only list the results for the solution $II$ since this solution can be obtained from the first one via a parity transformation.\\
By now, we postulate that
\begin{equation}
\label{multiplier}
\int_{-\mathcal{1}}^{+\mathcal{1}} du \partial_u b^a(x)=0,
\end{equation}
since this condition is not guaranteed by the axial gauge. Consequently, we obtain from the equation \eqref{ward1} the following integrated Ward identity:
\begin{equation}
\label{wardglobale}
 \int du (\partial J_A^a+\overline{\partial} J_{\overline{A}}^a-\sum_{\varphi}f^{abc}J^b_{\varphi} \varphi^c )=-\frac{k}{2\pi}(\overline{\partial}A^a_++\partial \overline{A}^a_-).
\end{equation}
We note that our choice \eqref{multiplier} on the behavior of the Lagrange multipliers is legitimate if, and only if, the right hand side of the identity \eqref{wardglobale} does not vanish. In fact, in this case the boundary term in \eqref{wardglobale} plays the role of the gauge-fixing: if it vanished, we would be forced to reintroduce the term $\int_{-\mathcal{1}}^{\mathcal{1}} du \partial_u b^a$ in order to define the propagators of the theory. We will discuss this argument in detail later, when we analyze the three-dimensional BF model. Indeed, the Chern-Simons model has only one field on the boundary and, for this reason, there are no boundary conditions which lead to an ill-defined Ward identity as is the case for the BF model where there are two different fields on the boundary. In this case the problem is relevant and needs some precautions which we will discuss in the next chapter.\\
Then, if we rewrite the  identity \eqref{wardglobale} in the functional way we find:
\begin{equation}
 \label{wardglobalefunz}
 \int du (\partial J_A^a+\overline{\partial} J_{\overline{A}}^a-\sum_{\varphi}f^{abc}J^b_{\varphi} \frac{\delta Z_c}{\delta J^c_{\varphi}} )=-\frac{k}{2\pi} \left( \overline{\partial}\frac{\delta Z_{c+}}{\delta J_{A}^a(x)} \Big|_{u=0^+}+\partial\frac{\delta Z_{c-}}{\delta J_{\overline{A}}^a(x)}\Big|_{u=0^-} \right) 
\end{equation}
Having done that, we differentiate the previous expression with respect to $J^b_{A}(x')$, with $x'$ lying on the right side of the boundary $u=0$, and we evaluate the expression obtained at the vanishing sources, achieving the following expression:
\begin{equation}
\label{lab1}
\begin{split}
 &-\frac{2\pi}{k}\int du (\delta^{ab} \partial \delta^{(3)} (x-x')-f^{abc} \delta^{(3)}(x-x')A^c) =\\
&\left( \overline{\partial}\frac{\delta^2 Z_{c+}}{\delta J_A^b(x')\delta J_{A}^a(x)} \Big|_{u=u'=0^+}\right)_{J_{\varphi}=0}+\left(\partial\frac{\delta^2 Z_{c-}}{\delta J_A^b(x') \delta J_{\overline{A}}^a(x)}\Big|_{u=0^-,u'=0^+} \right)_{J_{\varphi}=0}
\end{split}
\end{equation}
The second term on the right hand side of the equation \eqref{lab1} vanishes due to the separability condition and the previous identity can be rewritten as follows:
\begin{equation}
\label{lab2}
-\frac{2\pi}{k}(\delta^{ab} \partial \delta^{(2)} (Z-Z')-f^{abc} \delta^{(2)}(Z-Z')A^c_+)=\overline{\partial} \mathcal{h}T(A_+^b(Z')A_+^a(Z))\mathcal{i}
\end{equation}
Keeping in mind the definition of the T-ordered product, we have to specify which variable plays the role of the time in the light-cone coordinates. This problem was discussed in literature by Dirac \cite{dirac}. In his article Dirac highlighted that both $z$ and $\overline{z}$ can be taken as the time coordinate. In our case a good choice is to take $\overline{z}$ as the time coordinate. Following this, we obtain:
\begin{equation}
\begin{split}
\label{commut}
&\overline{\partial} \mathcal{h}T(A_+^b(Z')A_+^a(Z))\mathcal{i}=\\
&\overline{\partial}\mathcal{h}\theta(\overline{z}'-\overline{z})A_+^b(Z')A_+^a(Z)+\theta(\overline{z}-\overline{z}')A_+^a(Z)A_+^b(Z')\mathcal{i}=\\
&\mathcal{h}-\delta(\overline{z}-\overline{z}')A_+^b(Z')A_+^a(Z)+\theta(\overline{z}'-\overline{z})A_+^b(Z') \overline{\partial}A_+^a(Z)\mathcal{i}+\\
&+\mathcal{h}\delta(\overline{z}-\overline{z}')A_+^a(Z)A_+^b(Z')+\theta(\overline{z}-\overline{z}') \overline{\partial}A_+^a(Z)A_+^b(Z')\mathcal{i}=\\
&\mathcal{h}\theta(\overline{z}'-\overline{z})A^b_+(Z')\overline{\partial}A^a_+(Z)+\theta(\overline{z}-\overline{z}')\overline{\partial}A^a_+(Z)A^b_+(Z')\mathcal{i}\\
&+\delta(\overline{z}-\overline{z}')\mathcal{h}[A_+^a(Z),A_+^b(Z')]\mathcal{i},
\end{split}
\end{equation}
where we have used the relation $\overline{\partial}\theta(\overline{z})=\delta(\overline{z})$.\\
Moreover, keeping into account the separability condition, if we evaluate the identity \eqref{wardglobale} at the vanishing sources we find the \textit{chirality conditions} \cite{collina, emery}:
\begin{align}
\label{chira1}
&\overline{\partial}A^a_+=0 \; \; \Rightarrow \; \; A^a_+=A^a_+(z)\\
&\partial \overline{A}^a_-=0 \; \; \Rightarrow \; \; \overline{A}^a_-=\overline{A}^a_-(\overline{z}).
\end{align}
If we substitute the previous conditions into the equation \eqref{lab2} we find:
\begin{equation}
-\frac{2\pi}{k}\delta(\overline{z}-\overline{z}')\mathcal{h}(\delta^{ab} \partial \delta (z-\overline{z}')-f^{abc} \delta(z-z')A^c_+)\mathcal{i}=\delta(\overline{z}-\overline{z}')\mathcal{h}[A_+^a(z),A_+^b(z')]\mathcal{i},
\end{equation}
which yields:
\begin{equation}
\label{zumin}
 [A_+^a(z),A_+^b(z')]=-\frac{2\pi}{k}(\delta^{ab} \partial \delta (z-\overline{z}')-f^{abc} \delta(z-z')A^c_+(z)).
\end{equation}
Then, if we express the previous relation in term of the fields
\begin{equation}
 K_+^a(z) \equiv \frac{k}{2\pi}A_+^a(z),
\end{equation}
we find:
\begin{equation}
\label{kacmoody}
 [K_+^a(z),K_+^b(z')]=f^{abc} \delta(z-z')K^c_+-\frac{k}{2\pi}\delta^{ab}\partial \delta(z-z'),
\end{equation}
which is a Ka\v{c}-Moody algebra \cite{kac} generated by the chiral currents $A^a_+$ \cite{collina,emery}. As previously mentioned, the symmetry under parity implies the existence of an algebra on the opposite side of the boundary:
\begin{equation}
[\overline{K}_-^a(\overline{z}),\overline{K}_-^b(\overline{z}')]=f^{abc} \delta(\overline{z}-\overline{z}')\overline{K}^c_--\frac{k}{2\pi}\delta^{ab}\overline{\partial} \delta(\overline{z}-\overline{z}').
\end{equation}
It is important to note that, in calculating the algebra \eqref{kacmoody}, the choice of $\overline{z}$ as the time coordinate was crucial to calculate the quantity $\overline{\partial}\mathcal{h}T(A^b_+(z')A^a_+(z))\mathcal{i}$ in \eqref{commut}. We will analyze theories which involve more than one field and generate more complex algebra later in this thesis. In these cases, depending on the commutator which you want to calculate, it will be necessary to choose $z$ or $\overline{z}$ as time coordinate. Since the choice between $z$ and $\overline{z}$ is irrelevant from the dynamic point of view, it does not affect the existence of the algebra on the boundary. However, once we have decided the side of the boundary to which we refer, the choice is fixed by the technique used in \eqref{commut}. Ergo, the choice of the time variable must be understood as a declaration to be made before the expansion of the T-ordered product rather than a relevant physical choice. In particular, depending on the declaration we 
have 
made, we obtain informations about the equal-$z$ commutators but not about the equal-$\overline{z}$ ones. We shall develop this argument later, when the problem will become relevant.\\
In regard to the solution $II$, it generates an equivalent algebra except for the exchange $ \{ z, K^a_+, -\frac{k}{2 \pi} \} \leftrightarrow \{ \overline{z}, \overline{K}^a_+, \frac{k}{2 \pi} \} $.\\

Finally, we make a comment on the chiral currents. The identity \eqref{chira1}, combined with the boundary condition $\overline{A}_+^a=0$, implies:
\begin{equation}
\label{conservazione}
 \overline{\partial}A^a_++\partial\overline{A}_+^a=0,
\end{equation}
which is the conservation relation of the field $A_{\mu}$ in the light-cone coordinate system. In fact, if we rewrite the equation \eqref{conservazione} in the Cartesian coordinate system we find:
\begin{equation}
\partial_0 A_{0+}^a-\partial_1 A_{1+}^a=0,
\end{equation}
which is the continuity equation for the densities $A_{0+}^a$ and $A_{1+}^a$ \cite{maxw}.

\chapter{The three-dimensional BF model}
In Chapter 1 we have illustrated how to introduce a boundary in the Chern-Simons theory using the Symanzik's method. We have derived the boundary conditions for the fields of the theory using the technique illustrated in \cite{maxw}, which avoids the direct computation of the propagator. In the simple case of the Chern-Simons theory, in which only one gauge field is involved, there are no complications in using this technique and the boundary conditions which we have found are the same obtained in \cite{emery} by directly calculating the Green functions.\\

In this chapter we shall apply the Symanzik's method to introduce a boundary in the three-dimensional non-abelian BF model. In this theory, two gauge fields are involved and we will see that, for this reason, the method which avoids the calculations of the propagators furnishes more boundary conditions than the method used in \cite{maggiore}, which uses the direct computation of the propagators. We shall prove that the additional boundary conditions are non-acceptable and we shall establish a set of criteria in order to directly discard these boundary conditions without the need to compare the results with that provided by the propagators.\\

In Section 2.1 the classical three-dimensional BF theory and its features are described.\\
In Section \ref{secb2} the Symanzik's method for the introduction of the boundary is illustrated. In order to find the boundary conditions, the method used in \cite{magnoli} and the method used in \cite{maggiore} are compared and a set of criteria in order to discard the non-acceptable boundary condition furnished by the first technique are established. At the end of the section, the Ka\v{c}-Moody algebra of local boundary observable is derived according to what was done in \cite{maggiore}.\\
In Section \ref{secb3} the Chern-Simons theory and the three-dimensional BF model are compared and it is pointed out that the BF model in three space-time dimensions with a cosmological constant is equivalent to two Chern-Simons theories with coupling constants of the opposite sign.
\label{capBF}
\label{secb1}
\section{The classical theory}
We denote with $M$ the three-dimensional flat space-time, with $A^a_{\mu}$ and $B^a_{\mu}$ two generic gauge fields and with $f^{abc}$ the structure constants of the gauge group $G$, which we suppose to be simple and compact. The most general action on $M$ which is invariant under the symmetries
\begin{equation}
\label{trasf1}
 \begin{split}
  &\delta^{(1)} A_{\mu}^a=-(D_{\mu} \theta)^a\\
  &\delta^{(1)} B_{\mu}^a=-f^{abc}B_{\mu}^b \theta^c
 \end{split}
\end{equation}
and
\begin{equation}
\label{trasf2}
 \begin{split}
  &\delta^{(2)} A_{\mu}^a=-\lambda f^{abc}B_{\mu}^b \alpha^c\\
  &\delta^{(2)} B_{\mu}^a=-(D_{\mu} \alpha)^a,
 \end{split}
\end{equation}
($\theta^a$ and $\alpha^a$ are local parameters), is the three-dimensional BF action:
\begin{equation}
 \label{BF}
S_{BF}=\frac{1}{2}\int_M d^3x \epsilon^{\mu \nu \rho} \{ F_{\mu \nu}^a B_{\rho}^a+ \frac{\lambda}{3}f^{abc}B_{\mu}^aB_{\nu}^bB_{\rho}^c \},
\end{equation}
where $F^a_{\mu \nu}$ is the two-form
\begin{equation}
  \partial_{\mu} A_{\nu}^a - \partial_{\nu} A^a_{\mu}+f^{abc}A_{\mu}^bA_{\nu}^c
\end{equation}
and $\lambda$ is a positive constant known as the \textit{cosmological constant} in previous literature.\\

We use the light-cone coordinate system in order to render the introduction of a boundary in the theory  more transparent as we did in the previous chapter. Consequently, we redefine the fields of the theory as follows:
\begin{equation}
 \begin{cases} A^a_u=A_2^a \\ A^a=\frac{1}{\sqrt{2}}(A^a_0+A^a_1) \\ \overline{A}^a=\frac{1}{\sqrt{2}}(A^a_0-A^a_1), \end{cases} \qquad \begin{cases} B^a_u=B_2^a \\ B^a=\frac{1}{\sqrt{2}}(B^a_0+B^a_1) \\ \overline{B}^a=\frac{1}{\sqrt{2}}(B^a_0-B^a_1). \end{cases}
\end{equation}
With these conventions, the action \eqref{BF} can be rewritten as:
\begin{equation}
\label{gfb}
\begin{split}
 S_{BF}=&\int_M du d^2z \{B^a(\overline{\partial}A^a_u-\partial_u \overline{A}^a+f^{abc}\overline{A}^bA_u^c)+\overline{B}^a(\partial_u A^a-\partial A_u^a+f^{abc}A_u^bA^c)\\
&+B_u^a(\partial \overline{A}^a-\overline{\partial}A^a+f^{abc}A^b\overline{A}^c)+\lambda f^{abc} B^a \overline{B}^b B_u^c \}.
\end{split}
\end{equation}

It is necessary to choose a gauge in order to make the theory consistent and, since in this chapter we want to study the BF theory with a boundary $u=0$, it is convenient to choose the axial gauge \cite{brandhuber}
\begin{equation}
 \begin{split}
  &A_u^a=0\\
  & B_u^a=0.
 \end{split}
\end{equation}
We can introduce the conditions above by adding the following gauge fixing term to the action:
\begin{equation}
\label{gaugefixing}
\begin{split}
 S_{GF}=& \int_M du d^2z \{b^aA_u^a+\overline{c}^a(\partial_u c^a+f^{abc}A_u^bc^c+\lambda f^{abc}B_u^b\phi^c)\\
        &+d^aB^a_u+\overline{\phi}^a(\partial_u\phi^a+f^{abc}A_u^b\phi^c+f^{abc}B_u^bc^c) \},
\end{split}
\end{equation}
where $c^a$, $\phi^a$, $\overline{c}^a$ and $\overline{\phi}^a$ are the ghost and anti-ghost fields and $b^a$ and $d^a$ are the Lagrange multipliers.\\
Then, the action $S_{TOT}=S_{GF}+S_{BF}$ is invariant under the BRS transformations \cite{maggiore2}:
\begin{equation}
\label{BRS11}
 \begin{split}
  &s A_{\mu}^a=-(D_{\mu}c)^a-\lambda f^{abc} B_{\mu}^b \phi^c\\
  &s B_{\mu}^a=-(D_{\mu}\phi)^a -f^{abc} B_{\mu}^b c^c\\
&sc^a=\frac{1}{2}f^{abc}(c^bc^c+\lambda \phi^b \phi^c)\\
  &s \phi^a= f^{abc} \phi^b c^c\\
&s \overline{c}^a=b^a\\
  &s \overline{\phi}^a=d^a\\
&sb^a=0\\
  &sd^a=0,
 \end{split}
\end{equation}
and the gauge-fixing term $S_{GF}$ is, as usual, a BRS variation:\\
$S_{GF}=s \int \overline({c}^a A_u^a + \overline{\phi}^a B_u^a)$.\\

As regards the discrete symmetries, once defined the action of the parity transformation on the coordinates as in \eqref{coordipar}, it is possible to define only one transformation in the space of the fields which leaves $S_{TOT}$ unchanged:
\begin{equation}
\label{parityb}
 \begin{split}
  &A^a \leftrightarrow \overline{A}^a \qquad \qquad \; \; \;  B^a  \leftrightarrow \overline{B}^a\\
  &A_u^a \leftrightarrow -A_u^a \qquad \qquad  B_u^a  \leftrightarrow -B_u^a\\
  &c^a \rightarrow \overline{\phi}^a \qquad \qquad \; \; \; \; \; \phi^a \rightarrow  \overline{c}^a\\
  &\overline{c}^a \rightarrow -\phi^a \qquad \qquad \; \;  \overline{\phi}^a \rightarrow -c^a\\
  &b^a \rightarrow -b^a \qquad \qquad \; \; \;  d^a \rightarrow -d^a.
 \end{split}
\end{equation}
Thus far, we will call this symmetry \textit{parity}.\\
Differently from what happens in the Chern-Simons model, it is possible to define one transformation in the space of the fields which leaves $S_{TOT}$ unchanged under the Time-Reversal transformation \eqref{coorditime}:
\begin{equation}
\label{timereversal}
 \begin{split}
  &A^a \leftrightarrow -\overline{A}^a \qquad \; \; \; \; \;  B^a  \leftrightarrow \overline{B}^a\\
  &A_u^a \leftrightarrow A_u^a \qquad \qquad  B_u^a  \leftrightarrow -B_u^a\\
  &c^a \rightarrow c^a \qquad \qquad \; \;  \phi^a \rightarrow  -\phi^a\\
  &\overline{c}^a \rightarrow \overline{c}^a \qquad \qquad \; \;  \overline{\phi}^a \rightarrow -\overline{\phi}^a\\
  &b^a \leftrightarrow b^a \qquad \qquad \; \;  d^a \leftrightarrow -d^a.
 \end{split}
\end{equation}
In this chapter we will label this transformation \textit{Time-Reversal}.\\
Moreover, each field of the theory carries a Faddeev-Popov charge, (Table \ref{dimb}), and it is straightforward to see that each term in $S_{TOT}$ has a Faddeev-Popov charge equal to zero.\\
The addition of the term $S_{GF}$ to $S_{BF}$ breaks the covariance of the action, but preserves the conservation of helicity on the plane $\{ z, \overline{z} \} $.\\
The mass dimensions, the values of the Faddeev-Popov charge and of the helicity of the fields of the theory are listed in Table \ref{dimb}.
\begin{table}[H]
\centering
\begin{tabular}{|c|c|c|c|c|c|c|c|c|c|c|c|c|c|c|c|}
\hline
 &$z$&$\overline{z}$& $u$&$A^a$&$\overline{A^a}$&$A^a_u$& $c^a$&$\overline{c}^a$&$b^a$&$B^a$&$\overline{B}^a$&$B_u^a$&$\phi^a$&$\overline{\phi}^a$&$d^a$\\ \hline
dim &-1&-1&-1&1&1&1&1&1&2&1&1&1&1&1&2\\ \hline
helicity&-1&1&0&1&-1&0&0&0&0&1&-1&0&0&0&0\\ \hline
$\Phi \Pi$&0&0&0&0&0&0&1&-1&0&0&0&0&1&-1&0\\ \hline
\end{tabular}
\label{dimb}
\caption{dimension, $\Phi$-$\Pi$ charge and helicity}
\end{table}
It is possible to derive the equations of motion from the generating functional of the connected Green functions $Z_c$ in a similar way to what has been calculated for the Chern-Simons model, finding the following equations:
\begin{equation}
\label{motoc}
 \begin{split}
  & \overline{\partial}B_u^a-\partial_u \overline{B}^a+f^{abc} \overline{A}^b B_u^c-f^{abc}A_u^b \overline{B}^c+J_A^a=0\\
&\partial_u B^a-\partial B_u^a+f^{abc}A_u^bB^c-f^{abc}A^bB_u^c+J^a_{\overline{A}}=0\\
&\partial \overline{B}^a-\overline{\partial}B^a+f^{abc}A^b \overline{B}^c-f^{abc} \overline{A}^b B^c + b^a-f^{abc} \overline{c}^b c^c-f^{abc} \overline{\phi}^b \phi^c+J^a_{A_u}=0\\
&A_u^a+J_b^a=0\\
&\overline{\partial}A_u^a-\partial_u \overline{A}^a+f^{abc} \overline{A}^b A_u^c+ \lambda f^{abc} \overline{B}^b B_u^c+J^a_B=0\\
&\partial_u A^a-\partial A_u^a +f^{abc}A_u^bA^c-\lambda f^{abc}B^bB^c_u+J_{\overline{B}}^a=0\\
&\partial \overline{A}^a-\overline{\partial} A^a+f^{abc}A^b \overline{A}^c+\lambda f^{abc}B^b \overline{B}^c+d^a-f^{abc} \overline{\phi}^bc^c - \lambda f^{abc} \overline{c}^b \phi^c +J^a_{B_u}=0\\
&B_u^a+J^a_d=0,
\end{split}
\end{equation}
while the equations of motion for the ghost fields are:
\begin{equation}
\label{motog}
 \begin{split}
  &\partial_u \overline{c}^a+f^{abc} A_u^b \overline{c}^c+f^{abc} B_u^b \overline{\phi}^c-J_c^a=0\\
  &\partial_u c^a+f^{abc}A_u^bc^c+ \lambda f^{abc} B_u^b \phi^c-J^a_{\overline{c}}=0\\
&\partial_u \overline{\phi}^a+f^{abc}A_u^b \overline{\phi}^c+\lambda f^{abc} B_u^b \overline{c}^c-J_{\phi}^a=0 \\
&\partial_u \phi^a+f^{abc}A_u^b \phi^c+f^{abc} B_u^b c^c-J^a_{\overline{\phi}}=0.
 \end{split}
\end{equation}

As it is known \cite{bassetto}, the gauge-fixing term $S_{GF}$ does not completely fix the gauge and the residual gauge invariance on the plane $u=0$ is functionally described by the two local Ward identities, (one for each symmetry $\delta^{(1)}$ and $\delta^{(2)}$):
\begin{align}
\label{wardbfff}
 &H^a(x)Z_c(J_{\varphi})=-\partial_{\mu}J^{a \mu}_A-\partial_u \frac{\delta Z_c}{\delta J^a_b}+\sum_{\varphi}f^{abc}J^b_{\varphi} \frac{\delta Z_c}{\delta J^a_{\varphi}}=0 \\
 \label{wardbfff1}
 &N^a(x)Z_c(J_{\varphi})=-\partial_{\mu}J^{a \mu}_B-\partial_{u}\frac{\delta Z_c}{\delta J_d^a} \notag \\
 &+f^{abc} \Big[ J^{b \mu}_B \frac{\delta}{\delta J^{c \mu}_A}+J^b_{\phi} \frac{\delta}{\delta J^a_{\phi}}+J^b_b \frac{\delta}{\delta J^c_d}+ J^b_{\overline{c}} \frac{\delta}{\delta J^c_{\overline{\phi}}}+ \notag \\
 &+\lambda \Big( J^{b \mu}_A \frac{\delta}{\delta J^{c \mu}_B}+J_c^{b \mu} \frac{\delta}{\delta J_{\phi}^c}+J^b_{\overline{\phi}} \frac{\delta}{\delta J_{\overline{c}}^c}+J^b_d \frac{\delta}{\delta J^c_b} \Big) \Big] Z_c=0,
\end{align}
which play a key role in determining the algebra on the boundary $u=0$, as in the Chern-Simons model. 
\section{The boundary}
\label{secb2}
In this section we will introduce the most general local boundary term which respects the conditions of separability, invariance under parity transformation, power-counting, conservation of helicity and of the Faddeev-Popov charge as we did for the Chern-Simons model in the previous chapter. In what follows, we shall maintain all the conventions adopted in Chapter \ref{CScap}.\\
In order to do this, we write the most general equations of motion for the ghost fields with a boundary term:
\begin{equation}
\label{ghostbfrotte}
 \begin{split}
  &\partial_u \overline{c}^a+f^{abc} A_u^b \overline{c}^c+f^{abc} B_u^b \overline{\phi}^c-J_c^a=\delta(u) [\mu_+ \overline{c}^a_++\mu_- \overline{c}^a_-+k_+ \overline{\phi}^a_++k_- \overline{\phi}^a_-]\\
&\partial_u \overline{\phi}^a+f^{abc}A_u^b \overline{\phi}^c+\lambda f^{abc} B_u^b \overline{c}^c-J_{\phi}^a=\delta(u) [\alpha_+ \overline{c}^a_++\alpha_- \overline{c}^a_-+\beta_+ \overline{\phi}^a_++\beta_- \overline{\phi}^a_-] \\
&\partial_u \phi^a+f^{abc}A_u^b \phi^c+f^{abc} B_u^b c^c-J^a_{\overline{\phi}}=-\delta(u) [\mu_+ \phi_-^a+\mu_- \phi_+^a+k_+ c_-^a+ k_- c_+^a]\\
  &\partial_u c^a+f^{abc}A_u^bc^c+ \lambda f^{abc} B_u^b \phi^c-J^a_{\overline{c}}=-\delta(u) [\alpha_+ \phi^a_-+ \alpha_- \phi^a_++ \beta_+ c^a_-+\beta_-c^a_+],
 \end{split}
\end{equation}
where $\mu_{\pm}$, $\alpha_{\pm}$, $k_{\pm}$ and $\beta_{\pm}$ are constant parameters which we will fix by imposing the conditions listed above.\\
If we impose the separability condition, we find the following algebraic constraints on the parameters:
\begin{equation}
 \begin{split}
  &k_+=k_- \equiv k\\
  &\beta_-=\mu_+\\
  &\beta_+=\mu_-\\
  &\alpha_+=\alpha_- \equiv \nu.\\
 \end{split}
\end{equation}
The issue can be further simplified by noting that the action $S_{TOT}$ is invariant under a discrete symmetry which involves only the ghost sector:
\begin{equation}
\label{simmetry}
 \begin{split}
  &\overline{c}^a \leftrightarrow \phi^a\\
  &c^a \leftrightarrow \overline{\phi}^a.
 \end{split}
\end{equation}
If we require that the introduction of the boundary preserves this symmetry, we obtain two more conditions on the parameters:
\begin{equation}
 \begin{split}
  &\mu_+=-\mu_- \equiv \mu\\
  &k=\nu=0.
 \end{split}
\end{equation}
This completed, the most general equations of motion for the ghost fields with a  boundary term are:
\begin{equation}
\label{ghostbfrottef}
 \begin{split}
   &\partial_u \overline{c}^a+f^{abc} A_u^b \overline{c}^c+f^{abc} B_u^b \overline{\phi}^c-J_c^a=\mu \delta(u) ( \overline{c}^a_+- \overline{c}^a_-)\\
&\partial_u \overline{\phi}^a+f^{abc}A_u^b \overline{\phi}^c+\lambda f^{abc} B_u^b \overline{c}^c-J_{\phi}^a= \mu \delta(u) (\overline{\phi}^a_- -\overline{\phi}^a_+) \\
&\partial_u \phi^a+f^{abc}A_u^b \phi^c+f^{abc} B_u^b c^c-J^a_{\overline{\phi}}=-\mu \delta(u) ( \phi_-^a- \phi_+^a)\\
  &\partial_u c^a+f^{abc}A_u^bc^c+ \lambda f^{abc} B_u^b \phi^c-J^a_{\overline{c}}=-\mu \delta(u) (c^a_+- c^a_-),
 \end{split}
\end{equation}
where $\mu$ is a constant parameter to be determined with the boundary conditions.\\
In the same way, and keeping in mind our gauge choice $A^a_u=B^a_u=0$, the most general equations of motion for the gauge fields satisfying the conditions outlined above are:
\begin{equation}
\label{campirottebf}
 \begin{split}
  & \overline{\partial}B_u^a-\partial_u \overline{B}^a+f^{abc} \overline{A}^b B_u^c-f^{abc}A_u^b \overline{B}^c+J_A^a=\delta(u) [ \alpha_1 (\overline{A}^a_++\overline{A}^a_-)+\alpha_2 \overline{B}^a_++\alpha_3 \overline{B}^a_- ]\\
&\partial_u B^a-\partial B_u^a+f^{abc}A_u^bB^c-f^{abc}A^bB_u^c+J^a_{\overline{A}}=\delta(u) [\alpha_1(A^a_++A^a_-)+\alpha_3B^a_++\alpha_2B^a_-]\\
&\partial \overline{B}^a-\overline{\partial}B^a+f^{abc}A^b \overline{B}^c-f^{abc} \overline{A}^b B^c + b^a-f^{abc} \overline{c}^b c^c-f^{abc} \overline{\phi}^b \phi^c+J^a_{A_u}=0\\
&A_u^a+J_b^a=0\\
&\overline{\partial}A_u^a-\partial_u \overline{A}^a+f^{abc} \overline{A}^b A_u^c+ \lambda f^{abc} \overline{B}^b B_u^c+J^a_B=\delta(u) [\alpha_3 \overline{A}^a_++ \alpha_2 \overline{A}^a_-+ \alpha_4 (\overline{B}^a_++\overline{B}^a_-)]\\
&\partial_u A^a-\partial A_u^a +f^{abc}A_u^bA^c-\lambda f^{abc}B^bB^c_u+J_{\overline{B}}^a=\delta(u)[\alpha_2A^a_++\alpha_3A^a_-+\alpha_4(B^a_++B^a_-)]\\
&\partial \overline{A}^a-\overline{\partial} A^a+f^{abc}A^b \overline{A}^c+\lambda f^{abc}B^b \overline{B}^c+d^a-f^{abc} \overline{\phi}^bc^c - \lambda f^{abc} \overline{c}^b \phi^c +J^a_{B_u}=0\\
&B_u^a+J^a_d=0,
 \end{split}
\end{equation}
where $\alpha_1$, $\alpha_2$, $\alpha_3$ and $\alpha_4$ are constant parameters to be determined by the boundary conditions.\\
Furthermore, if we want that the boundary term preserves the Time-Reversal symmetry we must impose the additional condition $\alpha_2=-\alpha_3$. Nonetheless, we do not require that the boundary preserves this symmetry although it is a symmetry of the bulk, since one of the purposes of this thesis is to study the effects of the Time-Reversal breaking on the boundary. 
\subsection{The boundary conditions: two methods compared}
In this section we will derive the boundary condition with two different methods, and comparing them. Initially, we will use the technique used in \cite{maggiore} for the BF model and in \cite{emery} for the Chern-Simons theory, which utilizes the direct calculation of the propagators of the theory. Then, we will obtain the same boundary conditions by using the method described in \cite{magnoli, maxw} which we have used in the previous chapter for the Chern-Simons model. As we will see, this technique is equivalent to the first, (with certain precautions), and it avoids the direct calculation of the propagators. For this reason, it is practical when the calculation of the propagators is difficult as in the Maxwell-Chern-Simons theory \cite{maxw}.
\subsubsection{The method of the propagators: the gauge sector}
Let us illustrate the first method for the simple case of the ghost fields.\\
The free equations of motion are:
\begin{equation}
\label{libereghost}
 \begin{split}
  &\partial_u \overline{c}^a(x)-J_c^a(x)=0 \qquad  \partial_u c^a(x)-J_{\overline{c}}^a(x)=0\\
  &\partial_u \overline{\phi}^a(x)-J_{\phi}^a(x)=0 \qquad  \partial_u \phi^a(x)-J_{\overline{\phi}}^a(x)=0.
 \end{split}
\end{equation}
Keeping in mind the definition \eqref{propagator}, the equations \eqref{libereghost} lead to the following relations involving the propagators:
\begin{equation}
\label{sistemghost}
\begin{split}
 &\partial_{u'} \Delta_{c \overline{c}}^{ab}(x,x')=\delta^{ab} \delta^{(3)}(x-x') \qquad  \partial_{u'} \Delta_{cc}^{ab}(x,x')=0\\
 &\partial_{u'} \Delta_{c \overline{\phi}}^{ab}(x,x')=0 \qquad \qquad \qquad \; \; \; \; \; \;  \partial_{u'} \Delta_{c \phi}^{ab}(x,x')=0\\
 &\partial_{u'} \Delta_{\overline{c} c}^{ab}(x,x')= \delta^{ab} \delta^{(3)}(x-x') \qquad \partial_{u'} \Delta_{\overline{c} \phi}^{ab}(x,x')=0\\
 &\partial_{u'} \Delta_{\phi \overline{\phi}}^{ab}(x,x')= \delta^{ab} \delta^{(3)}(x-x') \qquad  \partial_{u'}\Delta_{\phi \overline{c}}^{ab}(x,x')=0\\
 &\partial_{u'} \Delta_{\overline{\phi} \phi}^{ab}(x,x')= \delta^{ab} \delta^{(3)}(x-x') \qquad  \partial_{u'}\Delta_{ \overline{\phi} c}^{ab}(x,x')=0.
\end{split}
\end{equation}
The most general solution of the system \eqref{sistemghost} which satisfies the conditions of separability, compatibility, power-counting, conservation of helicity and which respects the symmetry \eqref{simmetry} can be written in the form $\Delta_{AB}^{ab}(x,x')= \theta_+ \Delta_{AB+}^{ab}(x,x')+\theta_- \Delta_{AB-}^{ab}(x,x')$, with:
\begin{equation}
\label{proppos}
 \Delta_{AB+}^{ab}= \delta^{ab} \begin{pmatrix} 0 & -T_{\rho}(x,x') & 0 & 0 \\ T_{\rho}(x',x) & 0 & 0 &0 \\ 0 & 0 & 0 & -T_{-(1+\rho)}(x,x')\\ 0 & 0 & T_{-(1+\rho)}(x',x) & 0 \end{pmatrix},  
\end{equation}
where $T_{\rho}(x,x')$ is the tempered distribution $(\theta(u-u')+\rho)\delta^{(2)}(x-x')$; it depends on the parameter $\rho$ which we will determine later. $\Delta^{ab}_{AB-}$ is obtained from the \eqref{proppos} by a parity transformation. As previously done,  we have used the matrix notation and the indices $A$ and $B$ denote the ordered set of fields $(c^a, \overline{c}^a, \phi^a, \overline{\phi}^a)$.\\
Subsequently, we require that the solution which we have found is compatible with the equations of motion with a boundary term \eqref{ghostbfrotte}. The calculations are quite similar to those made in the previous chapter to find the values of the parameters of the propagators. In this case the derivation presents some additional technical difficulties due to the fact that the parameters which appear in the equations of motion are still undetermined, thus, they will be part of the solution.\\
In this way, we obtain the following algebraic system for the parameters $\mu$ and $\rho$:
\begin{equation}
 \begin{cases} (1+ \rho)(1-\mu)=0\\ \rho(1+\mu)=0, \end{cases}
\end{equation}
which has two independent solutions:
\begin{equation}
 \begin{split}
  &I: \qquad \rho=0, \qquad \mu=1\\
  I&I: \qquad \rho=-1 \qquad \mu=-1.
 \end{split}
\end{equation} 
An analysis of how the propagators just obtained impose the boundary conditions for the ghost fields must be carried out. We will investigate in detail the solution $I$ and we will just list the results for the solution $II$.\\
If we substitute the parameters $\rho=0$ and $\mu=1$ into the equation \eqref{proppos} we find:
\begin{equation}
\label{propsol1}
 \Delta_{AB+}^{ab}=\delta^{ab} \begin{pmatrix} 0 & -\theta(u-u')\delta^2 & 0 & 0 \\ \theta(u'-u) \delta^2 & 0 & 0 & 0 \\ 0 & 0 & 0 & \theta(u'-u) \delta^2 \\ 0 & 0 & -\theta(u-u') \delta^2 & 0 \end{pmatrix},
\end{equation}
where $\delta^{(2)}$ denotes the distribution $\delta^{(2)}(Z-Z')$.\\
It is straightforward to see that the solution \eqref{propsol1} implies the following conditions for an arbitrary ghost field $\xi^a(x)$:
\begin{equation}
 \begin{split}
  &\lim_{u \rightarrow 0^+} \mathcal{h} c^a(x) \xi^b(x') \mathcal{i}=0\\
  &\lim_{u \rightarrow 0^+} \mathcal{h} \overline{\phi}^a(x) \xi^b(x') \mathcal{i}=0.
 \end{split}
\end{equation}
This finding can be interpreted as a Dirichlet boundary condition for the fields \cite{emery}: $c^a$ and $\overline{\phi}^a$ tend to zero when approaching the boundary from the right side. In other words:
\begin{equation}
 c^a_+=\overline{\phi}^a_+=0.
\end{equation}
In the same way, we derive the boundary conditions for the left side of the boundary from $\Delta_{AB-}^{ab}$:
\begin{equation}
 c^a_-=\overline{\phi}^a_-=0.
\end{equation}
Consequently,  the equations of motion for the solution $I$ are:
\begin{equation}
 \begin{split}
   &\partial_u \overline{c}^a+f^{abc} A_u^b \overline{c}^c+f^{abc} B_u^b \overline{\phi}^c-J_c^a= \delta(u) ( \overline{c}^a_+- \overline{c}^a_-)\\
&\partial_u \overline{\phi}^a+f^{abc}A_u^b \overline{\phi}^c+\lambda f^{abc} B_u^b \overline{c}^c-J_{\phi}^a= 0 \\
&\partial_u \phi^a+f^{abc}A_u^b \phi^c+f^{abc} B_u^b c^c-J^a_{\overline{\phi}}= \delta(u) ( \phi_+^a- \phi_-^a)\\
  &\partial_u c^a+f^{abc}A_u^bc^c+ \lambda f^{abc} B_u^b \phi^c-J^a_{\overline{c}}=0.
 \end{split}
\end{equation}
Furthermore, the boundary conditions for the solution $II$ are:
\begin{equation}
 \phi^a_+=\phi^a_-=\overline{c}^a_+=\overline{c}^a_-=0,
\end{equation}
which yield the following equations of motion:
\begin{equation}
 \begin{split}
   &\partial_u \overline{c}^a+f^{abc} A_u^b \overline{c}^c+f^{abc} B_u^b \overline{\phi}^c-J_c^a=0\\
&\partial_u \overline{\phi}^a+f^{abc}A_u^b \overline{\phi}^c+\lambda f^{abc} B_u^b \overline{c}^c-J_{\phi}^a=  \delta(u) (\overline{\phi}^a_+ -\overline{\phi}^a_-) \\
&\partial_u \phi^a+f^{abc}A_u^b \phi^c+f^{abc} B_u^b c^c-J^a_{\overline{\phi}}=0\\
  &\partial_u c^a+f^{abc}A_u^bc^c+ \lambda f^{abc} B_u^b \phi^c-J^a_{\overline{c}}= \delta(u) (c^a_+- c^a_-).
 \end{split}
\end{equation}
\subsubsection{The method of the propagators: the gauge sector}
With regard to the gauge fields, the method of the propagators does not present additional conceptual difficulties. The complications are purely technical, due to the large number of fields and parameters to be considered. In this section we will only list the results obtained in \cite{maggiore} focusing on the number of independent solutions which this method provides, (for more details on the calculations of the propagators see Appendix \ref{appA}).\\
Characteristically, the most general propagators for the gauge fields can be written in the form \eqref{propdef} and $\Delta^{ab}_{AB+}$ is obtained from $\Delta^{ab}_{AB-}$ by a parity transformation, (Appendix \ref{appA}). $\Delta^{ab}_{AB+}$ and $\Delta^{ab}_{AB-}$ depend on ten parameters $a_i, \; \; i=1,...,10$ which we shall fix by imposing the compatibility of the propagators with the equations of motion \eqref{campirottebf} and with the Ward identities \eqref{wardbfff} and \eqref{wardbfff1}.\\
If we require that the propagators are compatible with the equations of motion, we obtain the following algebraic system of sixteen non-linear equations involving the ten parameters $a_i$ and the four parameters $\alpha_i$ of the equations of motion \eqref{campirottebf}:
\begin{equation}
\label{condizionieq}
 \begin{split}
  &(a_3-a_4)(\alpha_2+1)+a_2\alpha_1=0\\
  &(1+a_7)(1-\alpha_3)-a_2\alpha_1=0\\
  &(a_6-a_7)(1+\alpha_2)+a_5\alpha_1=0\\
  &a_4(1-\alpha_3)-a_1\alpha_1=0\\
  &a_9(1+\alpha_2)+a_7\alpha_1=0\\
  &a_9(1-\alpha_3)-\alpha_1(1+a_3-a_4)=0\\
  &a_{10}(1+\alpha_2)+\alpha_1(a_6-a_7)=0\\
  &a_8(1-\alpha_3)-a_4\alpha_1=0\\
  &a_1(1-\alpha_2)-a_4\alpha_4=0\\
  &a_5(1+\alpha_3)+\alpha_4(a_6-a_7)=0\\
  &a_2(1-\alpha_2)-\alpha_4(1+a_7)=0\\
  &a_2(1+\alpha_3)+\alpha_4(a_3-a_4)=0\\
  &a_4(1-\alpha_2)-\alpha_4a_8=0\\
  &(a_6-a_7)(1+\alpha_3)+\alpha_4a_{10}=0\\
  &(1+a_3-a_4)(1-\alpha_2)-\alpha_4a_9=0\\
  &a_7(1+\alpha_3)+\alpha_4a_9=0.
 \end{split}
\end{equation}
The introduction of a boundary in the theory leads to a breaking of the Ward identities which we need to exist only at the classical level. We can reach the goal by setting the breaking to be linear in the fields \cite{weinberg}. As highlighted in \cite{maggiore}, the request can be satisfied by fixing:
\begin{equation}
\label{line}
\begin{split}
 & \alpha_3=\alpha_2\\
 & \alpha_4=\lambda \alpha_1.
\end{split}
\end{equation}

The previous constraints are peculiar to the non-abelian theory. In the abelian theory the problem of linearity does not arise and it is not necessary to introduce additional constraints on the parameters. So, the system is less constrained, although there are as many acceptable solution as in the non-abelian theory.\\
this completed, using all the precautions illustrated in \cite{emery}, the boundary breaking of the Ward identities \eqref{wardbfff} and \eqref{wardbfff1} can be formally derived from the equations of motion \eqref{campirottebf}, obtaining the following expressions:
\begin{equation}
\label{wardlocalebf}
 \begin{split}
  H^a(x)Z_c(J_{\varphi})=&-\alpha_1 \delta(u) (\partial \overline{A}^a_++\partial \overline{A}^a_-+\overline{\partial}A^a_++\overline{\partial}A^a_-) \\ 
 &-\alpha_2 \delta(u) (\partial \overline{B}^a_++\partial \overline{B}^a_-+\overline{\partial}B^a_++\overline{\partial}B^a_-)\\
N^a(x)Z_c(J_{\varphi})=&-\lambda \alpha_1 \delta(u) (\partial \overline{B}^a_++\partial \overline{B}^a_-+\overline{\partial}B^a_++\overline{\partial}B^a_-)\\
 &-\alpha_2 \delta(u) (\partial \overline{A}^a_++\partial \overline{A}^a_-+\overline{\partial}A^a_++\overline{\partial}A^a_-)
 \end{split}
\end{equation}
This point forward, we postulate that:
\begin{equation}
\label{lagrange}
\begin{split}
 &\int_{-\mathcal{1}}^{\mathcal{1}} du \partial_u b^a(x)=0\\
 &\int_{-\mathcal{1}}^{\mathcal{1}} du \partial_u d^a(x)=0
\end{split}
\end{equation}
Consequently, the Ward identities \eqref{wardlocalebf} can be rewritten in the integrated form as follows:
 \begin{align}
\label{wardbf1}
  &\int_{-\mathcal{1}}^{\mathcal{1}} du H^a(x)Z_c(J_{\varphi})=&-\alpha_1(\partial \overline{A}^a_++\partial \overline{A}^a_-+\overline{\partial}A^a_++\overline{\partial}A^a_-) \notag \\
& &-\alpha_2(\partial \overline{B}^a_++\partial \overline{B}^a_-+\overline{\partial}B^a_++\overline{\partial}B^a_-)\\
\label{wardbf2}
&\int_{-\mathcal{1}}^{\mathcal{1}} du N^a(x)Z_c(J_{\varphi})=&-\lambda \alpha_1 (\partial \overline{B}^a_++\partial \overline{B}^a_-+\overline{\partial}B^a_++\overline{\partial}B^a_-) \notag \\
& &-\alpha_2(\partial \overline{A}^a_++\partial \overline{A}^a_-+\overline{\partial}A^a_++\overline{\partial}A^a_-),
 \end{align}
where the terms $\partial_u b^a$ and $\partial_u d^a$ on the left hand side vanish due to the conditions \eqref{lagrange}.\\
If we require that the propagators are compatible with the Ward identities \eqref{wardbf1} and \eqref{wardbf2}, we obtain eight more algebraic equations involving the parameters $a_i$ and $\alpha_i$:
\begin{equation}
\label{condizioniward}
 \begin{split}
  &\alpha_1a_2+\alpha_2(a_3-a_4)-\alpha_1a_1-\alpha_2a_4=1\\
  &\alpha_1a_2+(1+a_7)\alpha_2-a_5\alpha_1-(a_6-a_7)\alpha_2=1\\
  &\alpha_1a_7+\alpha_2a_9-\alpha_1a_4-a_8\alpha_2=0\\
  &\alpha_1(1+a_3-a_4)+\alpha_2a_9-(a_6-a_7)\alpha_1-a_{10}\alpha_2=0\\
  &\alpha_2a_2+\lambda \alpha_1(a_3-a_4)-a_1\alpha_2-\lambda \alpha_1 a_4=0\\
  &\alpha_2a_2+\lambda \alpha_1 (1+a_7)-a_5\alpha_2-\lambda(a_6-a_7)\alpha_1=0\\
  &\alpha_2a_7+\lambda \alpha_1 a_9-\alpha_2a_4-\lambda \alpha_1a_8=1\\
  &\alpha_2(1+a_3-a_4)+\lambda \alpha_1 a_9 -(a_6-a_7)\alpha_2-\lambda \alpha_1 a_{10}=1.
 \end{split}
\end{equation}
There are four independent solutions for the equations \eqref{condizionieq} and \eqref{condizioniward} which we list in the following tables:
\begin{table}[H]
 \centering
  \begin{tabular}{|c|c|c|c|c|}
   \hline
 &$\alpha_1$&$\alpha_2$&$\alpha_3$&$\alpha_4$\\ \hline
$I$&0&1&1&0\\ \hline
$II$&0&-1&-1&0\\ \hline
 $III$&$\frac{1}{\sqrt{\lambda}}$&0&0&$\sqrt{\lambda}$ \\ \hline
 $IV$&$-\frac{1}{\sqrt{\lambda}}$&0&0&$-\sqrt{\lambda}$ \\ \hline
  \end{tabular}
\label{tabellasoluzioni1}
\caption{solutions of the equations \eqref{condizionieq} and \eqref{condizioniward} for the parameters $\alpha_i$.}
\end{table}
\begin{table}[H]
\centering
 \begin{tabular}{|c|c|c|c|c|c|c|c|c|c|c|}
\hline
  & $a_1$& $a_2$& $a_3$& $a_4$ & $a_5$& $a_6$& $a_7$ & $a_8$&$a_9$&$a_{10}$\\ \hline
 $I$&0&0&-1&-1&0&0&0&0&0&0\\ \hline
$II$&0&0&-1&0&0&0&-1&0&0&0\\ \hline
$III$&$\frac{1}{2}\sqrt{\lambda}$&$-\frac{1}{2}\sqrt{\lambda}$&-1&$-\frac{1}{2}$&$\frac{1}{2}\sqrt{\lambda}$&0&$-\frac{1}{2}$&$\frac{1}{2\sqrt{\lambda}}$&$-\frac{1}{2\sqrt{\lambda}}$&$\frac{1}{2\sqrt{\lambda}}$\\ \hline
$IV$&$-\frac{1}{2}\sqrt{\lambda}$&$\frac{1}{2}\sqrt{\lambda}$&-1&$-\frac{1}{2}$&$-\frac{1}{2}\sqrt{\lambda}$&0&$-\frac{1}{2}$&$-\frac{1}{2\sqrt{\lambda}}$&$\frac{1}{2\sqrt{\lambda}}$&$-\frac{1}{2\sqrt{\lambda}}$\\ \hline
 \end{tabular}
\label{tabellasoluzioni2}
\caption{solutions of the equations \eqref{condizionieq} and \eqref{condizioniward} for the parameters $a_i$.}
\end{table}
It is now possible to derive the boundary conditions for the gauge fields from the propagators just obtained, as was done for the ghost fields, giving the boundary conditions for the gauge fields:
\begin{equation}
\label{condiziosine}
 \begin{split}  
&I: \; \; \overline{A}^a_+=A^a_-=\overline{B}^a_+=B^a_-=0\\
I&I: \; \; A_+^a=\overline{A}_-^a=B_+^a=\overline{B}_-^a=0\\
II&I: \; \; A^a_+-\sqrt{\lambda}B^a_+=\overline{A}^a_++\sqrt{\lambda}\overline{B}^a_+=\overline{A}^a_--\sqrt{\lambda}\overline{B}^a_-=A^a_-+\sqrt{\lambda}B^a_-=0\\
IV& \;: \; \; A^a_++\sqrt{\lambda}B^a_+=\overline{A}^a_+-\sqrt{\lambda}\overline{B}^a_+=\overline{A}^a_-+\sqrt{\lambda}\overline{B}^a_-=A^a_--\sqrt{\lambda}B^a_-=0
 \end{split}
\end{equation}
The four previous conditions suggest an interesting observation in the role of T in the three-dimensional BF theory with a boundary. As we know, T is a symmetry of the bulk and the solutions \eqref{condiziosine} identify two different behaviors of the boundary with respect to the Time-Reversal: the solutions $I$ and $II$ are transformed into each other if we apply T ``vertically'', but they are not T-invariant separately. On the other hand, T acts on the solutions $III$ and $IV$ ``horizontally'' since they are themselves T-invariant. Consequently, if we need the boundary to preserve the T-invariance in accordance with the bulk, we are forced to choose the solution $III$ and $IV$. Conversely, if there are certain physical reasons which require that the boundary breaks the T-invariance, we must choose the solutions $I$ and $II$. The choice is influenced by the presence of the cosmological constant $\lambda$. Indeed, the solutions $III$ and $IV$ exist if, and only if, $\lambda > 0$ and then, it is possible to 
introduce a boundary which preserves the T symmetry only if this condition is met. This argument becomes more relevant if we consider that, as we will see later, the action of the three-dimensional BF model with a cosmological constant $\lambda>0$ can be rewritten as the sum of two Chern-Simons action with coupling constants of the opposite sign, (remember that the Chern-Simons action is not invariant under the Time-Reversal transformation).\\
For clarity, we summarize the afore-mentioned in the following table.
\begin{table}[H]
\begin{tabular}{|l|c|c|}
\hline
\textbf{T preserved} & $\lambda \neq 0$ & solutions $III$ and $IV$ \\
\textbf{T broken} & $\lambda=0$,$\lambda\neq0$ & solutions $I$ and $II$ \\ \hline
\end{tabular}
\end{table}
\subsubsection{The algebraic method: the ghost sector}
In this section we will derive the same boundary conditions by using the method which we illustrated in the previous chapter for the Chern-Simons theory \cite{magnoli,maxw}.\\
We integrate the equations of motion \eqref{ghostbfrottef} with respect to $u$ in the infinitesimal interval $[-\epsilon,\epsilon]$ and we evaluate the expression obtained in the weak limit $\epsilon \rightarrow 0$. Consequently, if we impose the separability condition, we arrive at the following systems of equations which involve the insertions of the ghost fields on the boundary and the parameter $\mu$:
\begin{equation}
 \begin{cases} (1-\mu)\overline{c}^a_+=0\\ (1+\mu)\overline{\phi}^a_+=0\\ (1-\mu)\phi^a_+=0 \\ (1+\mu) c^a_+=0 \end{cases}, \qquad  \begin{cases} (1-\mu)\overline{c}^a_-=0\\ (1+\mu)\overline{\phi}^a_-=0\\ (1-\mu)\phi^a_-=0 \\ (1+\mu) c^a_-=0 \end{cases}.
\end{equation}
The previous systems have two independent solutions:
\begin{equation}
\label{soluzionima}
 \begin{split}
  &I: \qquad \mu=1, \qquad \overline{\phi}^a_+=\overline{\phi}^a_-=c^a_+=c^a_-=0\\
  I&I: \qquad \mu=-1 \qquad \overline{c}^a_+=\overline{c}^a_-=\phi^a_+=\phi^a_-=0.
 \end{split}
\end{equation}
We can derive the conditions on the parameter $\rho$ of the propagators from the \eqref{soluzionima}, similar to what was done for the propagators of the Chern-Simons theory in Chapter \ref{CScap}.\\
It is important to note that there are no differences between the results obtained with the method of the propagators and those obtained with the algebraic method in this case, because the symmetry \eqref{simmetry} decouples the equations of motion for the fields $c^a$ and $\phi^a$. As we will see, this is not the case for the gauge sector.
\subsubsection{The algebraic method: the gauge sector}
If we apply the algebraic method to the equations of motion for the gauge fields \eqref{campirottebf}, we find the following algebraic systems:
\begin{equation}
\label{sistemacampi}
 \begin{cases} \alpha_1 \overline{A}^a_++(1+\alpha_2) \overline{B}^a_+=0\\ (1+\alpha_2)\overline{A}^a_++\lambda \alpha_1 \overline{B}^a_+=0\\ \alpha_1 A^a_++(\alpha_2-1)B^a_+=0\\ (\alpha_2-1)A^a_++\lambda \alpha_1 B^a_+=0,  \end{cases} \qquad \begin{cases} \alpha_1 \overline{A}^a_-+(\alpha_2-1) \overline{B}^a_-=0\\ (\alpha_2-1) \overline{A}^a_-+ \lambda \alpha_1 \overline{B}^a_-=0\\ \alpha_1 A^a_-+(1+\alpha_2)B^a_-=0 \\ (\alpha_2+1)A^a_-+\lambda \alpha_1 B^a_-=0. \end{cases}
\end{equation}
where we have kept account of the linearity conditions \eqref{line}. We list the independent solutions of the previous systems in the following tables.
\begin{table}[H]
 \centering
  \begin{tabular}{|c|c|c|}
\hline
 & $\alpha_1$& $\alpha_2$\\ \hline
$I$&0&1 \\ \hline
$II$&0 &-1 \\ \hline
$III$& $\frac{1}{\sqrt{\lambda}}$&0 \\ \hline
$IV$& $-\frac{1}{\sqrt{\lambda}}$&0 \\ \hline   
$V$&$\pm \frac{1+\alpha_2}{\sqrt{\lambda}}$& $\ne-1,0$ \\ \hline
$VI$&$\pm \frac{1-\alpha_2}{\sqrt{\lambda}}$& $\ne1,0$ \\ \hline
$VII$&$ \alpha_1\ne0 , \lambda =0 $&-1 \\ \hline
$VIII$&$ \alpha_1\ne0 , \lambda =0 $&1 \\ \hline
$IX$&0&0 \\ \hline
  \end{tabular}
\label{sold1}
\caption{values of the parameters for the solutions of \eqref{sistemacampi}.}
\end{table}

\begin{table}[H]
 \centering
  \begin{tabular}{|c|c|c|c|c|c|c|c|c|}
\hline
   &$A^a_+$&$\overline{A}^a_+$&$B^a_+$&$\overline{B}^a_+$&$A^a_-$&$\overline{A}^a_-$&$B^a_-$&$\overline{B}^a_-$\\ \hline
$I$& &0 & & 0&0& & 0& \\ \hline
$II$&0& & 0& & &0& & 0 \\ \hline
$III$& & &$\frac{1}{\sqrt{\lambda}}A^a_+$&$-\frac{1}{\sqrt{\lambda}}\overline{A}^a_+$& & & $-\frac{1}{\sqrt{\lambda}}A^a_-$&$\frac{1}{\sqrt{\lambda}}\overline{A}^a_-$ \\ \hline
$IV$& & &$-\frac{1}{\sqrt{\lambda}}A^a_+$&$\frac{1}{\sqrt{\lambda}}\overline{A}^a_+$& & & $\frac{1}{\sqrt{\lambda}}A^a_-$&$-\frac{1}{\sqrt{\lambda}}\overline{A}^a_-$ \\ \hline   
$V$&0& & 0&$\mp \frac{1}{\sqrt{\lambda}}\overline{A}^a_+ $& & 0&$\mp \frac{1}{\sqrt{\lambda}}A^a_-$&0 \\ \hline
$VI$&&0 & $\pm \frac{1}{\sqrt{\lambda}}A^a_+ $&0&0 & &0&$\pm \frac{1}{\sqrt{\lambda}}\overline{A}^a_- $ \\ \hline
$VII$&0&0&0&&0&0&&0 \\ \hline
$VIII$&0&0&&0&0&0&0& \\ \hline
$IX$&0&0&0&0&0&0&0&0 \\ \hline
\end{tabular}
\label{sold}
\caption{values of the insertion of the fields for the solutions of \eqref{sistemacampi}.}
 \end{table}
As is evident, the solutions $I-IV$ are equal to those obtained with the method of the propagators. The solution $IX$ is meaningless since it implies that all the fields disappear on the boundary. On the other hand, we did not find the solutions $V-VIII$ with the method of the propagators and, in principle, there are no physical reasons to discard these solutions.\\
The next step will be to analyze the integrated Ward identities \eqref{wardbf1} and \eqref{wardbf2} for each solution listed in the Tables 2.4 and 2.5, demonstrating that only the solutions common to both methods are acceptable. As we shall see, the reason for this statement lies in the (unpublished) interpretation of the boundary as the gauge-fixing: we must discard the additional solutions furnished by the algebraic method since they correspond to a theory which has non-invertible propagators (so the propagators does not exist).
\subsection{The Ward identities}
Initially, we analyze the solutions $V-VIII$ and prove that there is always a way to redefine fields and sources so that at least one of the boundary terms of the Ward identities \eqref{wardbf1} and \eqref{wardbf2} becomes zero. Consequently, we demonstrate that this condition is not acceptable.\\
Regarding to the solution $V$, the identities \eqref{wardbf1} and \eqref{wardbf2} become:
\begin{equation}
\begin{split}
 &\int_{-\mathcal{1}}^{\mathcal{1}} du H^a(x)Z_c(J_{\varphi})= \mp \frac{1}{\sqrt{\lambda}}(\partial \overline{A}^a_++\overline{\partial}A^a_-)\\
 &\int_{-\mathcal{1}}^{\mathcal{1}} du N^a(x)Z_c(J_{\varphi})= \mp \sqrt{\lambda}(\partial \overline{B}^a_++\overline{\partial}B^a_-).
\end{split}
\end{equation}
Let us redefine fields and sources in the following way:
\begin{equation}
\label{jpq}
\begin{cases}  P^a=A^a+\sqrt{\lambda}B^a\\  Q^a=A^a-\sqrt{\lambda}B^a \end{cases} \Rightarrow  \begin{cases}  J^a_P=\frac{1}{2}(J^a_A+\frac{J^a_B}{\sqrt{\lambda}})\\   J^a_Q=\frac{1}{2}(J^a_A-\frac{J^a_B}{\sqrt{\lambda}}) \end{cases}.
\end{equation}
With these conventions, the boundary conditions for the fields $P$ and $Q$ are
\begin{equation}
 \overline{P}^a_+=P^a_+=Q^a_+=\overline{P}^a_-=\overline{Q}^a_-=P^a_-=0
\end{equation}
if $\alpha_1=\frac{1+\alpha_2}{\sqrt{\lambda}}$ and 
\begin{equation}
 \overline{Q}^a_+=P^a_+=Q^a_+=\overline{P}^a_-=\overline{Q}^a_-=Q^a_-=0
\end{equation}
if $\alpha_1=-\frac{1+\alpha_2}{\sqrt{\lambda}}$.\\
Then, in the first case the Ward identities are
\begin{equation}
\label{nonsens1}
 \begin{split}
  &\int_{-\mathcal{1}}^{\mathcal{1}} du( \sqrt{\lambda} H^a(x)-N^a(x))Z_c(J_{\varphi})=-(\partial \overline{Q}^a_++\overline{\partial}Q^a_-)\\
  &\int_{-\mathcal{1}}^{\mathcal{1}} du( \sqrt{\lambda} H^a(x)+N^a(x))Z_c(J_{\varphi})=0,
 \end{split}
\end{equation}
while in the second case we have:
\begin{equation}
\label{nonsens2}
\begin{split}
  &\int_{-\mathcal{1}}^{\mathcal{1}} du( \sqrt{\lambda} H^a(x)-N^a(x))Z_c(J_{\varphi})=0\\
  &\int_{-\mathcal{1}}^{\mathcal{1}} du( \sqrt{\lambda} H^a(x)+N^a(x))Z_c(J_{\varphi})=\partial \overline{P}^a_++\overline{\partial} P^a_-.
 \end{split}
\end{equation}
Let us now consider one of the Ward identities in \eqref{nonsens1} or in \eqref{nonsens2} in which the right hand side becomes zero, e.g. the first identity in \eqref{nonsens2}. If we differentiate this expression with respect to $J^b_Q(x')$, with $x'$ lying on the right side of the boundary $u=0$, we find, evaluating the expression obtained at vanishing sources:
\begin{equation}
 \delta^{ab} \delta(\overline{z}-\overline{z}') \delta'(z-z')=0,
\end{equation}
which is obviously a meaningless equation. It is important to note that, if we had not assumed the condition \eqref{lagrange},  by repeating the same calculations, we would have obtained an identity involving the propagators of the theory:
\begin{equation}
 \int_{-\mathcal{1}}^{\mathcal{1}}du (\partial \delta^{(3)}(x-x')\delta^{ab}+\sqrt{\lambda}\partial_u \Delta^{ba}_{Qb}(x',x)-\partial_u \Delta^{ba}_{Qd}(x',x))=0,
\end{equation}
In other words, it is possible to assume the conditions \eqref{lagrange} on the Lagrange multipliers if, and only if, the boundary term of the Ward identities \eqref{wardbf1} and \eqref{wardbf2} does not vanish even after a redefinition of fields and sources. In this respect, the boundary term is required to define the propagators of the theory and acts as the gauge-fixing. For this reason, when the boundary conditions involve more than one field, (as is the case for the gauge fields), each solution of the system \eqref{sistemacampi} for which it is possible to redefine fields and sources so that the boundary term in the Ward identities becomes zero is inconsistent with our choice on the behavior of $b^a$ and $d^a$. Via arguments similar to those just made, it is straightforward to see that, if we want to maintain the conditions \eqref{lagrange}, we need to exclude the solutions $V-X$.\\
If we analyze the solutions $I-IV$, we note that, in these cases, there are always two independent fields, (unlike that which occurs for the solutions $V-X$ where there is only one independent field), and, for this reason, there is no way to redefine fields and sources so that the boundary term of the Ward identities \eqref{wardbf1} and \eqref{wardbf2} vanishes. Therefore, solutions $I-IV$ are the only acceptable.\\

To summarize, each boundary condition which leads to three constraints on the gauge fields is not acceptable because, for these kind of solutions, it is always possible to redefine fields and sources so that the boundary term of at least one of the corresponding Ward identities becomes zero. Thus, the propagators of the theory are ill-defined for these solutions as we have just demonstrated. The only acceptable solutions are those which lead to two constraints on the fields: in these cases, there are two arbitrary fields on the boundary and it is impossible to find a linear combination of fields which leads to an ill-defined Ward identity. The method of the propagators automatically selects the acceptable boundary conditions since, if there was no gauge-fixing in the theory, (as is the case for the solutions $V-VIII$), the propagators would not be invertible (so the propagators would not exist). On the other hand, in the algebraic method we have to study the compatibility conditions case by case, analyzing 
the number of 
constraints on the insertions of the fields. Nonetheless, as mentioned earlier, there may be considerable technical difficulties in the calculation of the propagators and the algebraic method allows us to study the physics on the boundary, thereby, evading the direct calculation of the Green functions.
\subsection{The boundary algebra}
In this section we will derive the boundary algebras associated with the solutions $I-V$ as we have done for the Chern-Simons theory in Chapter \ref{CScap}.\\
Let us consider solution $I$; the corresponding Ward identities are:
 \begin{align}
\label{w1}
  &\int_{-\mathcal{1}}^{\mathcal{1}} du (\partial J^a_A+ \overline{\partial}J^a_{\overline{A}}-f^{abc}(J^{ \mu b}_{A} A_{\mu}^c+J^{ \mu b}_{B} B_{\mu}^c)+...)=(\overline{\partial}B^a_++\partial \overline{B}^a_-)\\
\label{w2}
  &\int_{-\mathcal{1}}^{\mathcal{1}} du [\partial J^a_B+ \overline{\partial}J^a_{\overline{B}}-f^{abc}(J^{\mu b}_B A^c_{\mu}+\lambda J^{\mu b}_A B^c_{\mu})+...]=(\overline{\partial}A^a_++\partial \overline{A}^a_-),
 \end{align}
where the ellipses denote terms which are irrelevant for our purposes.\\
Keeping into account the separability condition, if we evaluate the identities \eqref{w1} and \eqref{w2} at the vanishing sources we find:
\begin{equation}
\label{chirality}
 \overline{\partial}B^a_+=\partial \overline{B}^a_-=\overline{\partial}A^a_+=\partial \overline{A}^a_-=0.
\end{equation}
We now repeat the same steps which we did in Chapter \ref{CScap}: we differentiate the identity \eqref{w1} with respect to $J^a_B(x')$, with $x'$ lying on the right side of the boundary $u=0$, evaluating the expression obtained at the vanishing sources:
\begin{equation}
\label{primaalge}
\overline{\partial} \mathcal{h}T(B^b_+(z')B^a_+(z))\mathcal{i}=-f^{abc}\delta^{(2)}(Z-Z')B^c_+.
\end{equation}
Keeping in mind the technique used in \eqref{commut} and the condition \eqref{chirality}, we choose $\overline{z}$ as the time variable in order to gather information about the equal-$\overline{z}$ commutators but not about the equal-$z$ ones. Consequently, we obtain:
\begin{equation}
 [B^a_+(z),B^b_+(z')]=\delta(z-z')f^{abc}B^c_+(z),
\end{equation}
where we have redefined the field $B^a_+$: $B^a_+ \rightarrow -B^a_+$. In what follows, we also redefine the field $A^a_+$ in the same way: $A^a_+ \rightarrow -A^a_+$.\\
If we differentiate the identity \eqref{w1} with respect to $J^b_A(x')$ and we repeat the same steps, we find:
\begin{equation}
 [A^a_+(z),B^b_+(z')]=\delta(z-z')f^{abc}A^c_+(z)+\delta^{ab}\partial \delta(z-z').
\end{equation}
It is straightforward to see that it is possible to obtain the same commutator from the identity \eqref{w2}.\\
If we differentiate the identity \eqref{w2} with respect to $J_B^b(x')$ we obtain:
\begin{equation}
 [A^a_+(z),A^b_+(z')]=\lambda f^{abc} \delta(z-z')B^c_+(z).
\end{equation}
If $\lambda=0$ the resulting algebra is:
\begin{equation}
 \begin{split}
  &[A^a_+(z),A^b_+(z')]=0\\
  &[B^a_+(z),B^b_+(z')]=\delta(z-z')f^{abc}B^c_+(z)\\
  &[A^a_+(z),B^b_+(z')]=\delta(z-z')f^{abc}A^c_+(z)+\delta^{ab}\partial \delta(z-z').
 \end{split}
\end{equation}
On the other hand, if $\lambda \ne 0$ we can redefine the fields as follows:
\begin{equation}
 \begin{split}
  &K^a_+=\frac{1}{2}(\frac{1}{\sqrt{\lambda}}A^a_++B^a_+)\\
  &T^a_+=\frac{1}{2}(-\frac{1}{\sqrt{\lambda}}A^a_++B^a_+).
 \end{split}
\end{equation}
Keeping in mind the equation \eqref{chirality}, the chirality conditions become:
\begin{equation}
 \overline{\partial}K^a_+=\overline{\partial}T^a_+=0,
\end{equation}
and the resulting algebra for $T$ and $K$ is:
\begin{equation}
\label{algebra1}
 \begin{split}
  &[K^a_+(z),K^b_+(z')]=f^{abc}\delta(z-z')K^c_+(z)+\frac{1}{2 \sqrt{\lambda}}\delta^{ab}\partial \delta(z-z')\\
  &[T^a_+(z),T^b_+(z')]=f^{abc}\delta(z-z')T^c_+(z)-\frac{1}{2 \sqrt{\lambda}}\delta^{ab}\partial \delta(z-z')\\
  &[K^a_+(z),T^b_+(z')]=0,
 \end{split}
\end{equation}
which is the direct sum of two Ka\v{z}-Moody algebras with central charge $\pm \frac{1}{2\sqrt{\lambda}}$ respectively.\\
The solution $II$ leads to the following Ward identities:
\begin{equation}
\label{w3}
 \begin{split}
  &\int_{-\mathcal{1}}^{\mathcal{1}} du H^a(x)Z_c(J_{\varphi})=\partial \overline{B}^a_++\overline{\partial}B^a_-\\
  &\int_{-\mathcal{1}}^{\mathcal{1}} du N^a(x)Z_c(J_{\varphi})=\partial \overline{A}^a_++\overline{\partial}A^a_-.
\end{split}
\end{equation}
As is evident from the comparison between \eqref{w1} and \eqref{w3}, the identities \eqref{w3} generate an algebra equivalent to that generated by the identities \eqref{w3}, except for the exchange $\{z,A,B, \delta^{ab} \}\leftrightarrow \{ \overline{z},\overline{A},\overline{B}, -\delta^{ab} \}$. We note only that, in this case, we must choose $z$ as the time variable in order to use the technique \eqref{commut} and then, we have information about the equal-$z$ commutators but not about the equal-$\overline{z}$ commutators.\\
Regarding the solution $III$, the identities \eqref{wardbf1} and \eqref{wardbf2} become:
\begin{align}
\label{w5}
  &\int_{-\mathcal{1}}^{\mathcal{1}} du H^a(x)Z_c(J_{\varphi})=-\frac{1}{\sqrt{\lambda}}(\partial \overline{A}^a_++\partial \overline{A}^a_-+\overline{\partial}A^a_++\overline{\partial}A^a_-)\\
\label{w6} 
 &\int_{-\mathcal{1}}^{\mathcal{1}} du N^a(x)Z_c(J_{\varphi})=-\sqrt{\lambda}(\partial \overline{B}^a_++\partial \overline{B}^a_-+\overline{\partial}B^a_++\overline{\partial}B^a_-).
 \end{align}
In this case, the fields $A$ and $B$ do not generate any algebra. However, the boundary conditions \eqref{condiziosine} suggest the correct variables to use. Now, let us redefine fields and sources as in \eqref{jpq}. With these conventions, the boundary conditions for $P$ and $Q$ are:
\begin{equation}
 Q^a_+=\overline{P}^a_+=\overline{Q}^a_-=P^a_-=0.
\end{equation}
Consequently, we can rewrite the identities \eqref{w5} and \eqref{w6} in the following way:
\begin{equation}
\label{w7}
 \begin{split}
  &-\int_{-\mathcal{1}}^{\mathcal{1}} du (\sqrt{\lambda}H^a(x)+N^a(x))Z_c(J_{\varphi})=\\
  &2\sqrt{\lambda}\int_{-\mathcal{1}}^{\mathcal{1}} du (\partial J^a_P+\overline{\partial}J^a_{\overline{P}}-f^{abc}(J^b_PP^c+J^b_{\overline{P}}\overline{P}^c)+...)=\overline{\partial}P^a_++\partial \overline{P}^a_-,
 \end{split}
\end{equation}
\begin{equation}
\label{w8}
 \begin{split}
  &-\int_{-\mathcal{1}}^{\mathcal{1}} du (\sqrt{\lambda}H^a(x)-N^a(x))Z_c(J_{\varphi})=\\
  &2\sqrt{\lambda}\int_{-\mathcal{1}}^{\mathcal{1}} du (\partial J^a_Q+\overline{\partial}J^a_{\overline{Q}}-f^{abc}(J^b_QQ^c+J^b_{\overline{Q}}\overline{Q}^c)+...)=\overline{\partial}Q^a_-+\partial \overline{Q}^a_+,
 \end{split}
\end{equation}
It is straightforward to derive the chirality conditions from the previous identities, finding that:
\begin{equation}
 \overline{\partial}P^a_+=\partial \overline{P}^a_-=\overline{\partial}Q^a_-=\partial \overline{Q}^a_+=0
\end{equation}
The corresponding algebra is derived from the identities \eqref{w7} and \eqref{w8} in the same way as we just did with the solutions $I$ and $II$. We note only that, if we want to use the technique \eqref{commut}, we must choose $\overline{z}$ as the time variable to calculate $[P_+^a(z),P_+^b(z')]$ while, in order to calculate $[\overline{Q}^a_+(\overline{z}),\overline{Q}^b_+(\overline{z}')]$ it is necessary to choose $z$. Then, the first commutator is a equal-$\overline{z}$ commutator while the second one is a equal-$z$ commutator. The previous assertions are consistent if, and only if, $[P^a_+(z),\overline{Q}^b_+(\overline{z}')]=0$ (as is always the case).\\
If we redefine the fields as follows:
\begin{equation}
 \begin{split}
  &Q^a \rightarrow -2\sqrt{\lambda}Q^a\\
  &P^a \rightarrow -2\sqrt{\lambda}P^a,
 \end{split}
\end{equation}
the algebra associated with the solution $III$ is:
\begin{equation}
\label{algebra2}
 \begin{split}
  &[P^a_+(z),P^b_+(z')]=f^{abc}\delta(z-z')P^c_+(z)+\frac{1}{2\sqrt{\lambda}}\delta^{ab}\partial \delta(z-z')\\
  &[\overline{Q}^a_+(\overline{z}),\overline{Q}^b_+(\overline{z}')]=f^{abc}\delta(\overline{z}-\overline{z}')\overline{Q}^c_+(\overline{z})+\frac{1}{2\sqrt{\lambda}}\delta^{ab}\overline{\partial}\delta(\overline{z}-\overline{z}')\\
  &[P^a_+(z),\overline{Q}^b_+(\overline{z}')]=0,
 \end{split}
\end{equation}
which is the direct sum of two Ka\v{c}-Moody algebras with a central charge $\frac{1}{2\sqrt{\lambda}}$.\\
The algebra associated with solution $IV$ is equivalent to that just obtained, except for the exchange $\{ z, P, \overline{Q},\delta^{ab} \} \leftrightarrow \{ \overline{z}, \overline{P}, Q,-\delta^{ab} \}$.\\

In regard to the Time-Reversal, we noted in the previous section that only the solutions $III$ and $IV$ preserve this symmetry. As a consequence, the boundary algebras associated with these solutions are the direct sum of two algebras, (one is chiral, the other anti-chiral), which are transposed into each other by the Time-Reversal transformation.\\
\section{Chern-Simons theory vs BF theory}
\label{secb3}
In this section we focus on an observation in \cite{sorella}, according to which the action of the three-dimensional BF model \eqref{BF} with cosmological constant $\lambda>0$ can be rewritten as the sum of two Chern-Simons actions \eqref{csaction} with coupling constants of the opposite sign.\\
Let us consider the particular linear combinations of fields:
\begin{equation}
 \begin{split}
  &R^{a \mu}=A^{a \mu}+ \sqrt{\lambda} B^{a \mu}\\
  &S^{a \mu}=A^{a \mu}- \sqrt{\lambda} B^{a \mu}.
 \end{split}
\end{equation}
If we rewrite the BF action \eqref{BF} in terms of the fields $R^{a \mu}$ and $S^{a \mu}$, we find:
\begin{equation}
\label{bfcs}
\begin{split}
 S_{BF}&= \frac{1}{2 \sqrt{\lambda}} \int d^3x \epsilon^{\mu \nu \rho} \left[ \partial_{\mu} R^a_{\nu}R^a_{\rho}+\frac{1}{3}R^a_{\mu}R^b_{\nu}R^c_{\rho}-\left( \partial_{\mu} S^a_{\nu}S^a_{\rho}+\frac{1}{3}S^a_{\mu}S^b_{\nu}S^c_{\rho} \right) \right] \\
 &= \frac{1}{2\sqrt{\lambda}} \Big( S_{CS}(R)-S_{CS}(S) \Big),
\end{split}
\end{equation}
which is the sum of two Chern-Simons actions with coupling constants $\pm \frac{1}{2 \sqrt{\lambda}}$ respectively, as anticipated.\\
Initially, we note that it is possible to rewrite the BF action in the form \eqref{bfcs} if, and only if, the coupling constant $\lambda$ is positive. In a sense, we can affirm that the true three-dimensional BF model is that without a coupling constant (with $\lambda=0$). The three-dimensional BF model a with coupling constant is equivalent to two Chern-Simons theories.\\
Another important observation concerns the duality between the coupling constant and the cosmological constant. As is evident from \eqref{bfcs}, the BF model with strong coupling constant is equivalent to two Chern-Simons theories with weak coupling constants and vice-versa. \\

The introduction of a boundary in the theory can be achieved by repeating the steps illustrated in the previous chapters. In this way, the most general equations of motion with a boundary term which preserves only the symmetry under parity are:
\begin{equation}
\label{csbfcs}
 \begin{split}
&\frac{1}{2\sqrt{\lambda}}(\partial R^a_u-\partial_u R^a+f^{abc}R^bR^c_u)-J_{\overline{R}}^a=\delta(u)(\alpha_1 R_+^a+\alpha_2 R_-^a+\alpha_3S^a_++\alpha_4S^a_-)\\
&\frac{1}{2 \sqrt{\lambda}}(\partial_u \overline{R}^a-\overline{\partial}R^a_u-f^{abc}\overline{R}^bR_u^c)-J^a_R=\delta(u)(\alpha_1 \overline{R}^a_-+\alpha_2 \overline{R}^a_++\alpha_3 \overline{S}^a_-+\alpha_4 \overline{S}^a_+)\\
&\frac{1}{2 \sqrt{\lambda}}(\overline{\partial}R^a-\partial \overline{R}^a-f^{abc}R^b \overline{R}^c)+b^a-f^{abc} \overline{c}^b c^c-J_{R_u}^a=0\\
&R_u^a-J^a_b=0\\
&\frac{1}{2\sqrt{\lambda}}(\partial S^a_u-\partial_u S^a+f^{abc}S^bS^c_u)+J_{\overline{S}}^a=\delta(u)(\alpha_5 S_+^a+\alpha_6 S_-^a+\alpha_7R^a_++\alpha_8R^a_-)\\
&\frac{1}{2 \sqrt{\lambda}}(\partial_u \overline{S}^a-\overline{\partial} S^a_u-f^{abc}\overline{S}^bS_u^c)+J^a_S=\delta(u)(\alpha_5 \overline{S}^a_-+\alpha_6 \overline{S}^a_++\alpha_7 \overline{R}^a_-+\alpha_8 \overline{R}^a_+)\\
&\frac{1}{2 \sqrt{\lambda}}(\overline{\partial}S^a-\partial \overline{S}^a-f^{abc}S^b \overline{S}^c)+d^a-f^{abc} \overline{\phi}^b \phi^c+J_{S_u}^a=0\\
&S^a_u+J^a_d=0,
 \end{split}
\end{equation}
where $c^a$, $\phi^a$, $\overline{c}^a$ and $\overline{\phi}^a$ are the ghost and anti-ghost fields and $b^a$ and $d^a$ are the Lagrange multipliers, while the quantities $\alpha_i$, $i=1,...,8$ are arbitrary parameters.\\
Moreover, if we require that the equations \eqref{csbfcs} are compatible with each other, we find the following algebraic constraints on the parameters $\alpha_i$:
\begin{equation}
\label{parcsbfcs}
 \begin{split}
  &\alpha_1=\alpha_2=k\\
  &\alpha_5=\alpha_6=\rho,\\
  \end{split}
\end{equation}
while the linearity conditions \eqref{line} imply that:
\begin{equation}
\label{linebfcsbf}
 \alpha_3=\alpha_4=\alpha_7=\alpha_8=0.
\end{equation}
In order to demonstrate this assertion, we consider the equations of motion for the BF model with a boundary term \eqref{campirottebf}. For example, let us consider the equations
\begin{align}
\label{prova1}
  & \overline{\partial}B_u^a-\partial_u \overline{B}^a+f^{abc} \overline{A}^b B_u^c-f^{abc}A_u^b \overline{B}^c+J_A^a=\delta(u) [ \alpha_1 (\overline{A}^a_++\overline{A}^a_-)+\alpha_2 \overline{B}^a_++\alpha_3 \overline{B}^a_- ]\\
\label{prova2}
  &\overline{\partial}A_u^a-\partial_u \overline{A}^a+f^{abc} \overline{A}^b A_u^c+ \lambda f^{abc} \overline{B}^b B_u^c+J^a_B=\delta(u) [\alpha_3 \overline{A}^a_++ \alpha_2 \overline{A}^a_-+ \alpha_4 (\overline{B}^a_++\overline{B}^a_-)]
\end{align}
If we multiply the equation \eqref{prova1} by $\sqrt{\lambda}$ and add the result to the equation \eqref{prova2}, we find:
\begin{equation}
\begin{split}
 \overline{\partial} R^a_u-\partial_u \overline{R}^a+f^{abc}\overline{R}^bR^c_u+J_{\overline{R}}^a=\delta(u) & \{ \sqrt{\lambda}[\alpha_1 (\overline{A}^a_++\overline{A}^a_-)+\alpha_2 \overline{B}^a_++\alpha_3 \overline{B}^a_- ]+\\
&+[\alpha_3 \overline{A}^a_++ \alpha_2 \overline{A}^a_-+ \alpha_4 (\overline{B}^a_++\overline{B}^a_-)] \}.
\end{split}
\end{equation}
Consequently, if we impose the linearity conditions \eqref{line}, i.e. $\alpha_2=\alpha_3$ and $\alpha_4=\lambda \alpha_1$, we obtain:
\begin{equation}
 \overline{\partial} R^a_u-\partial_u \overline{R}^a+f^{abc}\overline{R}^bR^c_u+J_{\overline{R}}^a=\delta(u) (\sqrt{\lambda} \alpha_1+\alpha_2) (\overline{R}^a_-+ \overline{R}^a_+),
\end{equation}
which, keeping into account the constraints \eqref{parcsbfcs} and \eqref{linebfcsbf}, coincides with the second equation in \eqref{csbfcs} apart from an non-essential re-definition of fields and sources. The previous assertion can be extended to all the other equations in \eqref{csbfcs}, finding that:
\begin{equation}
\label{compcsbf}
 \begin{split}
&\frac{1}{2\sqrt{\lambda}}(\partial R^a_u-\partial_u R^a+f^{abc}R^bR^c_u)-J_{\overline{R}}^a=\delta(u)k( R_+^a+ R_-^a)\\
&\frac{1}{2 \sqrt{\lambda}}(\partial_u \overline{R}^a-\overline{\partial}R^a_u-f^{abc}\overline{R}^bR_u^c)-J^a_R=\delta(u)k(\overline{R}^a_-+ \overline{R}^a_+)\\
&\frac{1}{2 \sqrt{\lambda}}(\overline{\partial}R^a-\partial \overline{R}^a-f^{abc}R^b \overline{R}^c)+b^a-f^{abc} \overline{c}^b c^c-J_{R_u}^a=0\\
&R_u^a-J^a_b=0\\
&\frac{1}{2\sqrt{\lambda}}(\partial S^a_u-\partial_u S^a+f^{abc}S^bS^c_u)+J_{\overline{S}}^a=\delta(u)\rho( S_+^a+ S_-^a)\\
&\frac{1}{2 \sqrt{\lambda}}(\partial_u \overline{S}^a-\overline{\partial} S^a_u-f^{abc}\overline{S}^bS_u^c)+J^a_S=\delta(u)\rho( \overline{S}^a_-+ \overline{S}^a_+)\\
&\frac{1}{2 \sqrt{\lambda}}(\overline{\partial}S^a-\partial \overline{S}^a-f^{abc}S^b \overline{S}^c)+d^a-f^{abc} \overline{\phi}^b \phi^c+J_{S_u}^a=0\\
&S^a_u+J^a_d=0.
 \end{split}
\end{equation} 
In this way, the linearity conditions, which have a rather technical origin in the extension of the Ward identities to the quantum level, find an alternative and intuitive interpretation: if we re-write the BF action with a boundary term as the sum of two Chern-Simons actions, the equations of motion decouple on the boundary.\\

The determination of the boundary conditions and the calculation of the boundary algebra proceed similarly to what has been done in the previous sections. We would only to point out that, in this case, the algebraic method does not furnish extra boundary conditions, since, as just proved, the equations of motion for $R$ and $S$ decouple and it is possible to apply the algebraic method to the two Chern-Simons theories for $R$ and $S$ separately, (we noted in Chapter \ref{CScap} that, if the boundary conditions involve only one field, the algebraic method does not furnish any extra solutions).\\
We briefly illustrate the results furnished by the application of the algebraic method to the equations of motion \eqref{compcsbf} for completeness.\\
If we consider the equations for the fields $R$ and $S$ separately, we obtain
\begin{table}[H]
\centering
 \begin{tabular}{|c|c|c|c|c|}
\hline
  $k$&$R_+^a$&$R_-^a$&$\overline{R}^a_+$&$\overline{R}^a_-$ \\ \hline
 $\frac{1}{2 \sqrt{\lambda}}$&$0$&&&$0$ \\ \hline
 $-\frac{1}{2 \sqrt{\lambda}}$&&$0$&$0$& \\ \hline 
 \end{tabular}
\end{table}
$\;$\\
from the equations for $R$ and
\begin{table}[H]
\centering
 \begin{tabular}{|c|c|c|c|c|}
\hline
  $\rho$&$S_+^a$&$S_-^a$&$\overline{S}^a_+$&$\overline{S}^a_-$ \\ \hline
 $\frac{1}{2 \sqrt{\lambda}}$&$0$&&&$0$ \\ \hline
 $-\frac{1}{2 \sqrt{\lambda}}$&&$0$&$0$& \\ \hline 
 \end{tabular}
\end{table}
$\;$ \\
from the equations for $S$. If we combine the previous solutions we find four different boundary conditions:
\begin{table}[H]
 \centering
  \begin{tabular}{|c|c|c|c|c|c|c|c|c|c|c|}
\hline
&$k$&$\rho$&$R_+^a$&$R_-^a$&$\overline{R}^a_+$&$\overline{R}^a_-$&$S_+^a$&$S_-^a$&$\overline{S}^a_+$&$\overline{S}^a_-$ \\ \hline
$I$&$\frac{1}{2 \sqrt{\lambda}}$&$\frac{1}{2 \sqrt{\lambda}}$&0&&&0&0&&&0 \\ \hline
$II$&$\frac{1}{2 \sqrt{\lambda}}$&$-\frac{1}{2 \sqrt{\lambda}}$&0&&&0&&0&0& \\ \hline
$III$&$-\frac{1}{2 \sqrt{\lambda}}$&$\frac{1}{2 \sqrt{\lambda}}$&&0&0&&0&&&0 \\ \hline
$IV$&$-\frac{1}{2 \sqrt{\lambda}}$&$-\frac{1}{2 \sqrt{\lambda}}$&&0&0&&&0&0& \\ \hline
  \end{tabular}
 \end{table}
$\;$\\
which correspond to the conditions found in the previous section when\\ $\lambda \ne 0$.\\

In regard to the Time-Reversal symmetry, we note that the BF action with a cosmological constant, which is itself T-invariant, can be rewritten as the sum of two Chern-Simons actions which are not themselves T-invariant, as we observed in the previous chapter. Nonetheless, the two terms are transposed into each over by the Time-Reversal transformation and the total action is globally T-invariant, (as it should be). Moreover, we noted in the previous section that the only acceptable boundary conditions which respect the T-invariance are those with $\lambda>0$. We can reformulate this assertion by saying that the equations of motion for $R$ in \eqref{compcsbf} are transformed into the equations of motion for $S$ and vice-versa by the Time-Reversal transformation.

\chapter{The four-dimensional BF model}
In this chapter we shall discuss the introduction of a boundary in the abelian four-dimensional BF model. This topic has been treated by some authors \cite{balachandran, cho, momen} in the literature using different approaches. In particular, the attention of these papers is focussed on the edge states of the four-dimensional BF model with a boundary. \\
In what follows, we will analyze the abelian four dimensional BF model with a boundary by using the techniques developed in the previous chapters. These techniques are very useful in order to analyze the boundary conditions and to characterize the dynamics on the three-dimensional boundary.\\
In fact, as we will see, the algebra of local boundary observables which exists due to the residual gauge invariance of the bulk theory can be interpreted as a set of canonical commutation relations from which it is possible to construct a boundary Lagrangian which describes the physics on the boundary.\\ 
The contents of this chapter are contained \cite{Amoretti:2012kb}.\\

In Section \ref{classt} the classical four-dimensional BF theory and its features are described.\\
In Section \ref{secbf42} the Symanzik's method for the introduction of the boundary is illustrated, and the boundary conditions are derived using the technique developed in Chapter 2.\\
In Section \ref{secbf43} the algebra of local boundary observables is derived and is interpreted as a set of canonical commutation relations which make it possible to describe the physics on the boundary.\\
In Section \ref{secbf44} the findings of this chapter are summarized and discussed.\\
Moreover, the complete propagators of the whole theory are given in Appendix \ref{apppropbf4}
\section{The classical theory}
\label{classt}
In the abelian case, the action of the four-dimensional BF model \cite{Birmingham, Horowitz}, which describes the interaction between the two-form $B_{\mu \nu}$ and the gauge field $A_{\mu}$, is given by:
\begin{equation}
\label{bf4}
 S_{bf}= \frac{\alpha}{2} \int d^4x \epsilon^{\mu \nu \rho \sigma} F_{\mu \nu} B_{\rho \sigma},
\end{equation}
where $F_{\mu \nu}= \partial_{\mu} A_{\nu}-\partial_{\nu} A_{\mu}$. It is well known that the BF theories, in any space-time dimensions, both in the abelian and non-abelian case, do not depend on any coupling constant. Here, 
$\alpha$ is a constant which we have introduced in
order to distinguish the boundary terms from the bulk terms; it can be eventually put equal to one at the end of the computation.\\
The action \eqref{bf4} is invariant under the symmetries:
\begin{equation}
\label{sss1}
 \begin{split}
  &\delta^{(1)}A_{\mu}=-\partial_{\mu} \theta\\
  &\delta^{(1)}B_{\mu \nu}=0
 \end{split}
\end{equation}
and
\begin{equation}
\label{sss2}
 \begin{split}
  &\delta^{(2)} A_{\mu}=0\\
  &\delta^{(2)} B_{\mu \nu}= -(\partial_{\mu} \varphi_{\nu} - \partial_{\nu} \varphi_{\mu}),\\
 \end{split}
\end{equation}
where $\theta$ and $\varphi_{\mu}$ are local parameters.\\
We remind that the action \eqref{bf4} is not the most general one
compatible with the symmetries \eqref{sss1} and \eqref{sss2}. Indeed,
a Maxwell term $\int d^4xF_{\mu\nu}F^{\mu\nu}$ could be added,
coupled to an additional parameter. In this paper, we consider the
action \eqref{bf4} alone, because we are interested in the 3D
dynamics on the edge of a TQFT. We are allowed to do that, if we
think of the action \eqref{bf4} as the abelian limit of the
non-abelian 4D BF theory \cite{maggiore1}, which, as any other TQFT,
is protected from the occurrence of non-topological terms by an
additional symmetry, called ``vector super-symmetry in
\cite{Guadagnini:1990br}. Nevertheless, the non-abelian case,
which is much richer from the fields theoretical point of view
\cite{Batalin:1981jr,Batalin:1984jr}, and the addition of a Maxwell
term are interesting extensions.

Characteristically, it is necessary to fix a gauge in order to make the theory consistent and, since in this chapter we want to study the the theory with a boundary $x_3=0$, we choose the axial gauge:
\begin{equation}
 \begin{split}
  &A_3=0\\
  &B_{i3}=0,
 \end{split}
\end{equation}
where the index $i$ denotes $0,1,2$. Continuing, in order to simplify the notation, we will denote the indices $0,1,2$ by the Latin letters and the indices $1,2,3$ by the Greek letters.\\
With these conventions, we fix the gauge by adding to the action the gauge-fixing term:
\begin{equation}
\label{gfgfgf}
 S_{gf}= \int d^4x \{b A_3 + d^i B_{i3} \},
\end{equation}
where $b$ and $d^i$ are respectively the Lagrange multipliers for the fields $A_3$ and $B_{i3}$. As usual, in the abelian case the ghost fields are decoupled from the
other fields. One of the main differences with the non-abelian case,
is the structure of the gauge fixing term, which, because of the
reducible symmetry \eqref{sss2}, involves ghosts for ghosts, and
therefore is highly non trivial \cite{Batalin:1981jr,Batalin:1984jr}.\\

As regards the discrete symmetries, once defined the action of the parity transformation on the coordinates as follows:
\begin{equation}
 \begin{split}
  &x_0 \rightarrow x_0\\
  &x_{\alpha} \rightarrow -x_{\alpha},
 \end{split}
\end{equation}
it is possible to find a transformation in the space of the fields which leaves $S_{TOT}=S_{gf}+S_{bf}$ unchanged. In what follows, we will label this transformation \textit{parity}:
\begin{equation}
 \begin{split}
  &x_0 \rightarrow x_0\\
  &x_{\alpha} \rightarrow -x_{\alpha}\\
  &A_0 \rightarrow A_0\\
  &A_{\alpha} \rightarrow -A_{\alpha}\\
  &B_{0 \alpha} \rightarrow B_{0 \alpha} \\
  &B_{\alpha \beta} \rightarrow -B_{\alpha \beta}\\
  &b \rightarrow -b\\
  &d^0 \rightarrow d^0\\
  &d^{\alpha} \rightarrow -d^{\alpha}.
\end{split}
\end{equation}
Regarding the Time-Reversal transformation, if we define the action of this symmetry on the coordinates so that $x_0$ changes sign, the action $S_{TOT}$ is invariant under the transformation:
\begin{equation}
 \begin{split}
  &x_0 \rightarrow -x_0\\
  &x_{\alpha} \rightarrow x_{\alpha}\\
  &A_0 \rightarrow -A_0\\
  &A_{\alpha} \rightarrow A_{\alpha}\\
  &B_{0 \alpha} \rightarrow B_{0 \alpha} \\
  &B_{\alpha \beta} \rightarrow -B_{\alpha \beta}\\
  &b \rightarrow b\\
  &d^0 \rightarrow d^0\\
  &d^{\alpha} \rightarrow -d^{\alpha},
 \end{split}
\end{equation}
which we will call \textit{Time-Reversal}.\\
In reality, in the abelian case, there is no unequivocal way to define the discrete symmetries. Indeed, it is possible to define two more transformations, one for the parity and the other for the Time-reversal, which leave $S_{TOT}$ unchanged. In this thesis we choose to define as parity and Time-reversal those symmetries which are exact symmetries also in the non-abelian case.\\
With the conventions adopted above, the classical action $\Gamma_c(J_{\varphi})$ is given by:
\begin{equation}
\label{zc}
\begin{split}
 \Gamma_c[J_{\varphi}]= \int d^4x \{& \alpha \epsilon^{ijk}[2\partial_i A_j B_{k3}+ (\partial_i A_3-\partial_3 A_i)B_{jk}]+
bA_3+d^iB_{i3}+\\ &+J^{ij}_{B_{ij}}B_{ij}+2J^{i3}_{B_{i3}}B_{i3}+J^i_{A_i}A_i+J_{A_3}A_3+J_bb+J_{d_i}^id_i \},
\end{split}
\end{equation}
where $J^i_{A^i},J_{B^{ij}}^{ij},J^{i3}_{B^{i3}},J^i_{d^i},J_b$ and $J_{A^3}^3$ are respectively the sources for the fields $A^i,B^{ij},B^{i3},d^i,b$ and $A^3$. The equations of motion can be derived from $\Gamma_c$, obtaining the following equations:
\begin{equation}
\label{motobf4}
 \begin{split}
  &J^i_{A_i}+\alpha \epsilon^{ijk} [2 \partial_j B_{k3}+\partial_3 B_{jk}]=0\\
  &J^{ij}_{B_{ij}}+\alpha \epsilon^{ijk}(\partial_kA_3-\partial_3A_k)=0\\
  &J^3+b-\alpha \epsilon^{ijk} \partial_i B_{jk}=0\\
  &2J^{i3}_{B_{i3}}+d^i+2 \alpha \epsilon^{ijk} \partial_j A_k=0\\
  &A_3+J_b=0\\
  &B^{i3}+J^i_d=0.
 \end{split}
\end{equation}
Moreover, we note that the gauge-fixing term \eqref{gfgfgf} does not completely fix the gauge and the action $S_{TOT}$ has a residual gauge invariance in the sub-manifold $x_3=0$, which is functionally described by the two local Ward identities, (one for each symmetries $\delta^{(1)}$ and $\delta^{(2)}$):
\begin{equation}
\label{wwf1}
 W(x)\Gamma_c[J_{\varphi}]=\partial_i J^i_{A^i}+ \partial_3 J^3_{A^3}+\partial_3 \frac{\delta Z_c}{\delta J_b}=0,
\end{equation}
\begin{equation}
\label{wwf2}
 W^i(x)\Gamma_c[J_{\varphi}]=\partial_jJ^{ij}_{B^{ij}}+\partial_3J^{i3}_{B^{i3}}+\frac{1}{2}\partial_3 \frac{\delta Z_c}{\delta J_{d^i}^i}=0.
\end{equation}
Finally, we list the mass dimensions of the quantity involved in the theory in the following table.\\
\begin{table}[H]
 \centering
   \begin{tabular}{|c|c|c|c|c|c|c|c|c|c|c|c|c|}
\hline
&$A_{i}$&$A_3$&$B_{ij}$&$B_{i3}$&$b$&$d^i$&$J_{A_{i}}^i$&$J_{A_3}^3$&$J_{B_{ij}}^{ij}$&$J_{B_{i3}}^{i3}$&$J_b$&$J_{d_i}^i$ \\ \hline
Dim&1&1&2&2&3&2&3&3&2&2&3&2 \\ \hline
 \end{tabular}
\caption{mass dimensions of the quantity involved in the theory}
\end{table}
\section{The boundary}
\label{secbf42}
In this section we will study the effects of the introduction of a boundary ($x_3=0$) in the theory. 
In order to simplify the notation, thus far we express the two-form $B^{ij}$ in terms of  its dual $\tilde{B}^i=\epsilon^{ijk}B_{jk}$. As a consequence, the sources $J^{ij}_{B^{ij}}$ and $J^i_{\tilde{B}^i}$ are related by the equation $J^{ij}_{B^{ik}}=\epsilon^{ijk}J_{\tilde{B}^k}$.\\
Having done that, the most general local boundary Lagrangian which respects the power-counting, which is covariant in the sub-manifold $x_3=0$ and which is a quadratic functional of the gauge fields is:
\begin{equation}
\label{lbx3}
 \mathcal{L}_b=\delta(x_3) \Big[ \frac{b_1}{2} \epsilon^{ijk} \partial_i A_j A_k+b_2d_iA^i+a_1 A_i \tilde{B}^i+a_2\frac{m}{2}A_iA^i +a_3b \Big],
\end{equation}
where we have chosen to denote the coefficients of the terms which violate the T-invariance  by $b_i$ and those which preserve the Time-Reversal symmetry  by $a_i$. $a_i$ and $b_i$ are constant parameters to be determined by the boundary conditions while the term proportional to $a_2$ is multiplied by a massive parameter $m$, ($[m]=1$), which exists only on the boundary $x_3=0$. (We have omitted from $\mathcal{L}_b$ the terms proportional to a negative power of the mass $m$).\\
Notice that the term $\delta'(x_3) A_i A^i$ respects the power-counting, it is a quadratic functional of the gauge fields and it is covariant in the sub-manifold $x_3=0$. Consequently, in principle we should have included this term in the boundary Lagrangian $\mathcal{L}_b$. However, we have decided not to consider this term since the equations of motion \eqref{motobf4} are first order differential equations, (they depend on the derivative of the fields with respect to $x_3$), and the term $\delta'(x_3)A^i$, which would have appeared in the equation of motion for the field $A_{\mu}$, if we had included $\delta(x_3)A_iA^i$ in $\mathcal{L}_b$, gives us information about the derivative of the field $A_i$ on the boundary and, for this reason, it is incompatible with the bulk action since we have to fix only $A_i(X,x_3=0)$ in order to solve the equations of motion. The situation would have been different if we had add the Maxwell term $\int d^4xF_{\mu \nu} F^{\mu \nu}$ to the bulk action \eqref{bf4}, 
since, in this case, the equations of motion would have been second order differential equations. Moreover, $\delta'(x_3)A_i A^i$ is incompatible with the boundary conditions as we will see in the next section.\\    

We may now derive the boundary term which must be added to the equations of motion \eqref{motobf4}  from $\mathcal{L}_b$, finding the following equations of motion with a boundary term:
\begin{equation}
\label{motozrotte}
 \begin{split}
  &J^i_{A_i}+2\alpha \epsilon^{ijk}  \partial_j B_{k3}+ \alpha \partial_3 \tilde{B}^i=-\delta(x^3)[b_1 \epsilon^{ijk}(\partial_j A_k)^++b_2d^{i+}+\\
   &+a_1\tilde{B}^{i+}+a_2mA^{i+}+a_3b^+]\\
  &\epsilon^{ijk}J_{\tilde{B}^k}+\alpha \epsilon^{ijk}(\partial_kA_3-\partial_3A_k)=-a_1\delta(x^3) \epsilon^{ijk}A_k^+\\
  &J^3+b-\alpha  \partial_i \tilde{B}^i=0\\
  &2J^{i3}_{B_{i3}}+d^i+2 \alpha \epsilon^{ijk} \partial_j A_k=0\\
  &A_3+J_b=-\delta(x^3)a_3\\
  &B^{i3}+J^i_d=-\delta(x^3)b_2A^{i+},
 \end{split}
\end{equation}
where the apex $+$ denotes the insertions of the fields of the theory on the boundary $x_3=0$ as defined in the previous chapter.\\
Notice that the approach used to derive the equations of motion \eqref{motozrotte} is different from that adopted in the previous chapters. In the three-dimensional theory we were looking for the most general boundary term to be included in the equations of motion compatible with the prescriptions illustrated by Symanzik in \cite{syma}. On the other hand, in order to derive the equations \eqref{motozrotte}, we have found the most general local boundary Lagrangian \eqref{lbx3} which respects the power-counting, which is covariant in the sub-manifold $x_3=0$ and which is quadratic in the gauge fields. Nonetheless, it is easy to see that the compatibility condition which we have used in the previous chapter is equivalent to requiring that the boundary terms are derived from an action and, in this case, is automatically satisfied, as the requirement that the breaking of the equations of motion is linear in the fields, which corresponds to set the boundary Lagrangian $\mathcal{L}_b$ to be quadratic in the gauge 
fields. Moreover, we have not imposed the equations of motion \eqref{motozrotte} to be invariant under parity since these constraints would give us relations involving the coefficients and the fields on the right side and on the left side of the boundary, and we have decided to restrict our analysis to the right side of the boundary, in accordance with the separability condition.\\

Consequently, it is possible to derive the boundary breaking of the Ward identities \eqref{wwf1} and \eqref{wwf2} from the equations of motion \eqref{motozrotte}, obtaining the following local Ward identities with a boundary term:
\begin{equation}
\label{wwr1}
\partial_i J_{A_i}^i+ \partial_3 J_{A_3}^3 + \partial_3 b =- \delta(x^3) [b_2 \partial_i d^{i+}+a_1  \partial_i \tilde{B}^{i+} + a_2m \partial_i A^{i+}],
\end{equation}
\begin{equation}
\label{wwr2}
\epsilon^{ijk}\partial_j J_{\tilde{B}^k}+ \partial_3 J_{B_{i3}}^{i3}+ \frac{1}{2} \partial_3 d^i=-\delta(x^3)a_1 \epsilon^{ijk} \partial_j A^+_k.
\end{equation}
Thus far, we postulate that $b(x_3=\pm \mathcal{1})=d^i(x_3=\pm \mathcal{1})=0$. Therefore, the Ward identities \eqref{wwr1} and \eqref{wwr2} can be rewritten in the integrated form as follows:
\begin{equation}
\label{wwrg1}
\int_{-\mathcal{1}}^{\mathcal{1}}dx^3 \partial_i J_{A_i}^i=-[b_2 \partial_i d^{i+}+a_1 \partial_i \tilde{B}^{i+} + a_2m \partial_i A^{i+}],
\end{equation}
\begin{equation}
\label{wwrg2}
\int_{-\mathcal{1}}^{\mathcal{1}}dx^3 \epsilon^{ijk} \partial_j J_{\tilde{B}^k}=-a_1 \epsilon^{ijk} \partial_j A^+_k.
\end{equation}
We stress that the breaking terms at the right hand side of \eqref{wwr1} and
\eqref{wwr2} are linear in the quantum fields, and hence a non-renormalization 
theorem ensures that they are present at the
classical level only, and do not acquire quantum corrections
\cite{brs}.
\subsection{The boundary conditions}
\label{boundarycondition}
If we apply the algebraic method illustrated in the previous chapters to the equations of motion \eqref{motozrotte}, we obtain the following algebraic system involving the insertion of the gauge fields on the boundary and the parameters $a_i$ and $b_i$:
\begin{equation}
\label{sistemacondizio}
\begin{split}
&(\alpha +a_1) \tilde{B}^{i+}=-b_1 \epsilon^{ijk}\partial_j A_k^+-b_2d^{i+}-a_2mA^{i+}-a_3b^+\\
&(\alpha-a_1)A^+_i=0\\
&a_3=0\\
&b_2A^{i+}=0.
\end{split}
\end{equation}
As illustrated in the previous chapter, when we applied the same method to the three-dimensional BF model, the solutions of the system \eqref{sistemacondizio} are acceptable if the boundary term of the corresponding Ward identities \eqref{wwrg1} and \eqref{wwrg2} does not vanish. Indeed, in Chapter \ref{capBF} we explained that the boundary term of the Ward identities, (if we suppose the condition $b(x_3=\pm \mathcal{1})=d^i(x_3=\pm \mathcal{1})=0$ to be valid), plays the role of the gauge-fixing and, if it vanished, the propagators would be non-invertible.\\
It is evident that, if we want the boundary term of the Ward identity \eqref{wwrg2} not to vanish, we must impose the condition $A^{i+} \ne 0$. Consequently we deduce two necessary constraints on the parameters:
\begin{equation}
 \begin{split}
  &b_2=0\\
  &a_1=\alpha.
 \end{split}
\end{equation}
At this point, it is possible provide a further argument which justifies the exclusion of the term $\alpha_c \delta'(x_3)A_iA^i$, ($\alpha_c$ is an arbitrary parameter), from the boundary Lagrangian $\mathcal{L}_b$. Indeed, if we had included this term in the Lagrangian, the algebraic method would have led to the following condition:
\begin{equation}
(\alpha +a_1) \tilde{B}^{i+}=-b_1 \epsilon^{ijk}\partial_j A_k^+-b_2d^{i+}-a_2mA^{i+}-a_3b^+ -\alpha (\partial_3 A^i)^++\alpha_c [\delta(x_3) A^i]_{x_3=0}
\end{equation}
In order to make the previous identity consistent, the last term of the right hand side must vanish. It is possible to reach the goal by setting $A^{i+}=0$, but this request contradicts what we have just argued. Consequently, the only solution is to set $\alpha_c=0$, in agreement with the statement made in the previous section.\\ 

This done, the system \eqref{sistemacondizio} is reduced to a single equation:
\begin{equation}
\label{sistemacondizio1}
 2 \alpha \tilde{B}^{i+}=-b_1 \epsilon^{ijk}\partial_j A_k^+-a_2mA^{i+}.
\end{equation}
The previous identity furnishes, from the algebraic point of view, four different solutions which we list in the following table.
\begin{table}[H]
\centering
\begin{tabular}{|c|c|c|c|c|}
\hline
&$b_1$&$a_2$&$A^{i+}$&$\tilde{B}^{i+}$ \\ \hline
1&0&$\neq 0$&$\neq 0$&$-\frac{a_2 m}{2 \alpha}A^{i+}$ \\ \hline
2&0&0&$\neq 0$&0 \\ \hline
3&$\neq0$&0&$\neq 0$&$-\frac{b_1}{2 \alpha}\epsilon^{ijk}\partial_jA_k^+$ \\ \hline
4&$\neq0$&$\neq0$&$\neq 0$&$-\frac{1}{2 \alpha}(b_1 \epsilon^{ijk} \partial_j A^+_k+a_2 m A^{i+})$ \\ \hline
\end{tabular}
\caption{Solutions of the equation \eqref{sistemacondizio1}}
\label{tabbu}
\end{table}
The substitution of the solutions 2 and 3 into the equations \eqref{wwrg1} and \eqref{wwrg2} leads to ill-defined Ward identities, (the boundary term of the identity \eqref{wwrg1} vanishes), and, for this reason, these solutions are not acceptable.\\

To sum up, the acceptable boundary conditions are reported in the following table.
\begin{table}[H]
\centering
\begin{tabular}{|c|c|c|c|c|}
\hline
&$b_1$&$a_2$&$A^{i+}$&$\tilde{B}^{i+}$ \\ \hline
$I$&0&$\neq0$&$\neq0$&$-\frac{a_2 m}{2\alpha}A^{i+}$ \\ \hline
$II$&$\neq0$&$\neq0$&$\neq0$&$-\frac{1}{2 \alpha}(b_1 \epsilon^{ijk} \partial_j A^+_k+a_2 m A^{i+})$ \\ \hline
\end{tabular}
\caption{acceptable solutions of the equation \eqref{sistemacondizio1}}
\label{tabbu1}
\end{table}
Notice that the presence of the massive term $m$ is necessary in order to make the theory consistent. In other words, it is necessary that the boundary Lagrangian is not scale-invariant. This fact is very important in order to study of the physics on the boundary, as we will see later.\\

To conclude the section, we make some observation about the Time-Reversal symmetry. Primarily, there are no acceptable boundary conditions which completely break T, i.e solutions with $a_i=0 \; \forall \; i$. (Remember that $a_1=\alpha \ne 0$ is a necessary condition in order for the solution to be acceptable).\\
Moreover, the solution $I$ preserves the Time-Reversal, while the solution $II$ partially breaks this symmetry and, in this case, the breaking is reflected in the fact that the field $\tilde{B}^{i+}$ does not transform coherently under T, as is evident in Table \ref{tabbu1}.\\

\section{The algebra and the physics on the boundary}
\label{secbf43}
In this section we will derive the algebra of local observables which is generated on the boundary due to the residual gauge invariance of the theory, functionally described by the two Ward identities \eqref{wwf1} and \eqref{wwf2}.\\
Next, we will argue that it is possible to describe the physics on the boundary in terms of two fields: a gauge fields $\zeta ^i$ and a scalar massless field $\Lambda$. We will derive the Lagrangian which describe the physics on the boundary by interpreting the boundary algebra as a set of canonical commutation relations for the fields $\zeta^i$ and $\Lambda$.

\subsection{The boundary algebra}
Despite the fact that the solutions listed in Table \ref{tabbu1} appear to be different, and depending on free parameters, the broken  Ward identities for
both of them are:
\begin{equation}
\label{aa}
 \begin{split}
  &\int_{-\mathcal{1}}^{\mathcal{1}}dx_3 \partial_i J^i_{A^i}=\alpha \partial_i \tilde{B}^{i+}  \\
  &\int_{-\mathcal{1}}^{\mathcal{1}}dx_3 \epsilon^{ijk} \partial_j J_{\tilde{B}^k}=-\alpha \epsilon^{ijk} \partial_j A^+_k.
 \end{split}
\end{equation}
We therefore remark that, as a matter of fact, a unique solution exists, which does not depends on any free parameter. We recall that the constant $\alpha$ was introduced in order to keep trace of the bulk dependence, but it is not a true coupling constant. We can therefore freely put $\alpha =1$ in what follows.\\
If we evaluate the previous relations at the vanishing sources, we find that the following conditions must be satisfied on the mass-shell:
\begin{equation}
\label{sornul}
 \begin{split}
  &\partial_i \tilde{B}^{i+}=0\\
  &\epsilon^{ijk} \partial_j A^+_k=0.
 \end{split}
\end{equation}
We now differentiate the first identity in \eqref{aa} with respect to $J_{A}^l(x')$, with $x'$ lying on the right side of the boundary $x_3=0$, obtaining the following equation:
\begin{equation}
\label{alg1}
 \delta^i_l \partial_i \delta^{(3)}(X'-X)= \partial_i \Big( \Delta_{A_l \tilde{B}^i}(x',x) \Big)_{x_3=x'_3=0^+}.
\end{equation}
Next, we express the propagator in \eqref{alg1} in terms of the T-ordered product as follows:
\begin{equation}
 \Big( \Delta_{A_l \tilde{B}^i}(x',x) \Big)_{x_3=x'_3=0^+}=\theta(t-t')\tilde{B}^{i+}(X)A_l^+(X')+\theta(t'-t)A_l^+(X')\tilde{B}^{i+}(X).
\end{equation}
If we substitute the previous identity in the equation \eqref{alg1}, we find that:
\begin{equation}
\begin{split}
 \delta^i_l \partial_i \delta^{(3)}(X'-X)&= \delta(t-t')[\tilde{B}^{0+}(X),A_l(X')]+\\
 & \big(\theta(t-t')\partial_i\tilde{B}^{i+}(X)A_l^+(X')+\theta(t'-t)A_l^+(X')\partial_i\tilde{B}^{i+}(X) \big).
\end{split}
\end{equation}
The second term of the right hand side of the previous equation vanishes due to the second condition in \eqref{sornul} and we obtain:
\begin{equation}
\label{llll}
  \delta(t-t')[\tilde{B}^{0+}(X),A_l(X')]=\delta^i_l \partial_i \delta^{(3)}(X'-X).
\end{equation}
If $l=1,2$, it is possible to factorize the $\delta(t-t')$, finding that:
\begin{equation}
\label{comm1}
[\tilde{B}^{0+}(X),A_{\beta}(X')]_{t=t'}= \partial_{\beta} \delta^{(2)}(X'-X),
\end{equation}
where $\beta$ denotes the indices $1,2$. By now, we will denote the spatial indices with Greek letters.\\
Next, if we differentiate the first identity in \eqref{aa} with respect to $J_{\tilde{B}}^l(x')$, with $x'$ lying on the right side of the boundary, we obtain the following relation:
\begin{equation}
\label{BB}
 \partial_i \Big( \Delta_{\tilde{B}_l \tilde{B}^i}(x',x) \Big)_{x_3=x'_3=0^+}=0.
\end{equation}
The previous arguments applied to the equation \eqref{BB} lead to the following commutation relations:
\begin{equation}
 [\tilde{B}^{0+}(X),\tilde{B}_l^+(X')]_{t=t'}=0.
\end{equation}
In particular, if $l=0$, the previous commutation relation become:
\begin{equation}
\label{comm2}
 [\tilde{B}^{0+}(X),\tilde{B}_0^+(X')]_{t=t'}=0.
\end{equation}
Let us now consider the second identity in \eqref{aa}. The differentiation of this identity with respect to $J_{A^l}(x')$, (with $x_3'=0^+$), leads to:
\begin{equation}
 \epsilon^{ijk} \Big(\Delta_{A^l A_k}(x',x) \Big)_{x_3=x'_3=0^+}=0.
\end{equation}
Taking into account the second condition in \eqref{sornul},  the previous identity furnishes the following commutation relations:
\begin{equation}
\label{comm3}
 [A^+_{\beta}(X),A^{\gamma +}(X')]_{t=t'}=0.
\end{equation}
Next, if we differentiate the second Ward identities in \eqref{aa} with respect to $J_{\tilde{B}}^l(x')$, we find the following equation:
\begin{equation}
\begin{split}
 (\partial_j \delta_k^l -\partial_k \delta_j^l)&\delta^{(3)}(X'-X)=\\
 &- \partial_j \Big( \Delta_{\tilde{B}^l A_k}(x',x) \Big)_{x_3=x'_3=0^+}+ \partial_k \Big( \Delta_{\tilde{B}^l A_j}(x',x) \Big)_{x_3=x'_3=0^+}
 \end{split}
\end{equation}
which does not provide additional commutation relations.\\
In conclusion, the commutations relation \eqref{comm1}, \eqref{comm2} and \eqref{comm3} form the following algebra of local boundary observables:
\begin{equation}
\label{algebras}
 \begin{split}
  &[\tilde{B}^{0+}(X),A_{\beta}(X')]_{t=t'}= \partial_{\beta} \delta^{(2)}(X'-X)\\
  &[\tilde{B}^{0+}(X),\tilde{B}_0^+(X')]_{t=t'}=0\\
  &[A^+_{\beta}(X),A^{\gamma +}(X')]_{t=t'}=0,
 \end{split}
\end{equation}
which will be discussed in detail in the last section of this chapter, together with the other results.

\subsection{The physics on the boundary}
The conditions \eqref{sornul} allow us to express the fields $\tilde{B}^{i+}$ and $A^{i+}$ in terms of the potentials $\Lambda$ and $\zeta^i$:
\begin{equation}
 \begin{split}
  &\partial_i \tilde{B}^{i+}=0 \; \Rightarrow \; \tilde{B}^{i+}=\epsilon^{ijk}\partial_j \zeta_k\\
  &\epsilon^{ijk} \partial_j A^+_k=0 \; \Rightarrow \; A^+_k=\partial_k \Lambda,
 \end{split}
\end{equation}
where $\Lambda(X)$ and $\zeta^i(X)$ are such that $[\Lambda]=0$ and $[\zeta^i]=1$. Notice that the symmetry,
\begin{equation}
 \begin{split}
  &\delta \Lambda=c\\
  &\delta \zeta_i=\partial_i \theta,
  \end{split}
\end{equation}
where $c$ is a constant and $\theta$ is a local parameter, leaves the fields $A_k^+$ and $\tilde{B}^{i+}$ unchanged. Consequently, $\zeta^i$ must be a gauge field.\\

Furthermore, let us consider, for the moment, the boundary condition $I$ in Table \ref{tabbu1}, which preserves the Time-reversal symmetry. We can rewrite this condition in terms of the fields $\zeta^i$ and $\Lambda$ as follows:
\begin{equation}
\label{condiaz}
 \epsilon^{ijk}\partial_j \zeta_k=-\frac{a_2 m}{2}\partial^i \Lambda.
\end{equation}
By now we set $a_2=-2$, (remember that we had not yet fixed this parameter). The massive parameter $m$ in the equation \eqref{condiaz} allows to rescale the fields $\zeta^i$ and $\Lambda$ as follows:
\begin{equation}
 \begin{split}
  &\Lambda \rightarrow \frac{\Lambda}{\sqrt{m}}\\
  &\zeta^i \rightarrow \sqrt{m} \zeta^i.
 \end{split}
\end{equation}
Consequently, the rescaled fields have the canonical dimensions of a gauge field and of a scalar field in three space-time dimensions ($[\zeta^i]=[\Lambda]=\frac{1}{2}$). With these conventions, the equation \eqref{condiaz} becomes:
\begin{equation}
\label{arat}
 \epsilon^{ijk}\partial_j \zeta_k= \partial^i \Lambda,
\end{equation}
which is exactly the duality relation between a scalar field and a gauge field which is required to construct massless fermionic fields in three dimensions via the tomographic representation \cite{aratyn}. This could be interpreted as the sign that the actual degrees of
freedom of the 3D theory obtained on the boundary are fermionic
rather than bosonic. We shall come back to this point in the conclusive Section \ref{secbf44}.\\
We now consider the solution $II$ in Table \ref{tabbu}:
\begin{equation}
\tilde{B}^{i+}=-\frac{1}{2}(b_1 \epsilon^{ijk} \partial_j A^+_k+a_2 m A^{i+}).
\end{equation}
It is evident that, if we evaluate the previous condition on the mass-shell, the term proportional to $b_1$ vanishes due to the conditions \eqref{sornul} and the previous equation become equivalent to the boundary condition $I$. In other words, the only boundary term which breaks the Time-Reversal symmetry vanishes on the mass-shell due to the conditions arising from the Ward identities evaluated at the vanishing sources and, consequently, the physics on the boundary always preserves T.\\
Moreover, via this argument we can affirm that the duality
condition \eqref{arat} always holds, and the 3D physics we are
discussing here is therefore uniquely determined. \\

We now need to find a boundary Lagrangian for the fields $\zeta^i$ and $\Lambda$ which describes the physics on the boundary and which is compatible with the condition \eqref{arat}. In what follows, we will show that it is possible to interpret the commutation relations found in the previous section as canonical commutation relations for the fields $\zeta^i$ and $\Lambda$ and to construct a Lagrangian from these relations, doing the contrary of what is commonly done, which is to  find the
canonical variables and their commutation relations from a given
Lagrangian.\\
Firstly, we consider the equation \eqref{llll} with $l=0$:
\begin{equation}
 \delta(t-t')[\tilde{B}^{0+}(X),A_0(X')]= \delta'(t-t') \delta^{(2)}(X-X').
\end{equation}
If we express the previous identity in terms of the fields $\zeta^i$ and $\Lambda$, we obtain:
\begin{equation}
 \delta(t-t')\partial'_0[\epsilon^{\beta \gamma} \partial_{\beta} \zeta_{\gamma}(X),\Lambda(X')]= \delta'(t-t') \delta^{(2)}(X-X'),
\end{equation}
where we have factorized the operator $\partial'_0$ on the right hand side since it acts only on the field $\Lambda$. It is easy to see that that $\delta(t-t')\partial'_0=-\delta'(t-t')$ and, consequently, we can factorize the $\delta'(t-t')$, finding the following commutation relation:
\begin{equation}
\label{canonical1}
 [\Lambda(X'),\epsilon^{\beta \gamma} \partial_{\beta} \zeta_{\gamma}(X)]_{t=t'}=\delta^{(2)}(X-X').
\end{equation}
Secondly, we consider the first commutation relation in \eqref{algebras}:
\begin{equation}
 [\tilde{B}^{0+}(X),A_{\beta}(X')]_{t=t'}= \partial_{\beta} \delta^{(2)}(X'-X).
\end{equation}
If we express the previous identity in terms of the fields $\Lambda$ and $\zeta^i$, we find:
\begin{equation}
 \partial_{\gamma}[\epsilon^{\gamma \beta} \zeta_{\beta}(X), \partial'_{\delta}\Lambda(X')]_{t=t'}= \delta^{\gamma}_{\delta} \partial_{\gamma} \delta^{(2)}(X'-X),
\end{equation}
and it is evident that the previous relation is compatible with the following equation:
\begin{equation}
\label{canonical2}
 [\epsilon^{\gamma \beta} \zeta_{\beta}(X), \partial'_{\delta}\Lambda(X')]_{t=t'}= \delta^{\gamma}_{\delta} \delta^{(2)}(X'-X).
\end{equation}
We are now ready to construct the Lagrangian. The commutation
relations \eqref{canonical1} and \eqref{canonical2} allow us to
interpret the fields 
$\Pi_{(\Lambda)}\equiv\epsilon^{\alpha \beta} \partial_{\alpha}
\zeta_{\beta}$ 
and 
$\Pi_{(\zeta)\alpha}\equiv\partial_{\alpha} \Lambda$ as the conjugate
momenta of the fields $\Lambda$ and
$\tilde{\zeta}^\alpha\equiv\epsilon^{\alpha \beta} \zeta_{\beta}$
respectively.\\
With these assumptions, the Lagrangian of the system is given by:
\begin{equation}
 \mathcal{L}=\sum \Pi\dot{\Phi}-H,
\end{equation}
where $H$ is the Hamiltonian of the system. If we assume that $H$
contains only the kinetic term, (i.e. $H=\sum \frac{1}{2}\Pi^2$),
$\mathcal{L}$ is given by:
\begin{equation}
 \mathcal{L}=\epsilon^{\alpha \beta} \partial_{\alpha} \zeta_{\beta}
\partial_t \Lambda+\partial_{\alpha} \Lambda \epsilon^{\alpha \beta}
\partial_t \zeta_{\beta}-\frac{1}{2}(\epsilon^{\alpha \beta}
\partial_{\alpha} \zeta_{\beta})^2-\frac{1}{2}(\partial_{\alpha}
\Lambda)^2,
\end{equation}
which is equivalent to the Lagrangian postulated in \cite{cho} for
the study of the topological insulators. Moreover, if we omit the
kinetic term, $\mathcal{L}$ is equivalent to the Lagrangian
considered in \cite{balachandran} to study the edge states of the 4D
BF theory.\\

\section{Summary and discussion}
\label{secbf44}
The main results presented in this chapter are\\
\vspace{0.7cm}
1) {\bf Ward identities in presence of a boundary}
\begin{eqnarray}
 \int_{-\infty}^{+\infty}dx_3 \partial_i J^i_{A^i} &=&
\partial_i \tilde{B}^{i+}  \label{wiconcl1}\\
 \int_{-\mathcal{1}}^{+\mathcal{1}}dx_3 \epsilon^{ijk} \partial_j
J_{\tilde{B}^k} &=&- \epsilon^{ijk} \partial_j A^+_k.
\label{wiconcl2}
\end{eqnarray}
Quite remarkably, the apparently distinct solutions I and II of Table 3
physically coincide, since they lead to the same Ward 
identities on the boundary. This is the first evidence of the 
striking electromagnetic structure which determines the physics on 
the boundary, as we shall discuss shortly. In addition, despite the fact that the solutions 
depend on free parameters, when put into the Ward identities \eqref{wiconcl1} 
and \eqref{wiconcl2}, which contain all the physical information, these disappear. The 
separability condition isolates a unique dynamics on the boundary, 
without any dependence on free parameters.\\ 
\vspace{0.7cm}
2) {\bf electromagnetism on the boundary} 
\begin{eqnarray}
\partial_i \tilde{B}^{i+}&=0 \; \Rightarrow \;&
\tilde{B}^{i+}=\epsilon^{ijk}\partial_j \zeta_k \label{max1}\\
\epsilon^{ijk} \partial_j A^+_k &=0 \; \Rightarrow \;&
A^+_k=\partial_k \Lambda\label{max2},
\end{eqnarray}
On the boundary $x_3=0$, $and$ on the mass shell 
$J_{\phi}=\left.\frac{\delta 
\Gamma_{c}}{\delta\phi}\right|_{J=0}=0$ (we stress this double constraint defining the boundary), the 4D topological BF theory 
displays Maxwell equations for an electric field and a magnetic 
field, to be identified with the boundary insertions
$\tilde{B}^{i+}$ and $A^{+}_{i}$, respectively. This is a direct 
consequence of the result 2). Consequently, two potentials can be 
introduced: an electric scalar potential $\Lambda(X)$ and a magnetic 
vector potential $\zeta^{i}(X)$, depending on the 3D coordinates on the plane $x_3=0$: $X=(x_0,x_1,x_2)$.\\
\vspace{0.7cm}
3) {\bf duality} 
\begin{equation}
\label{duality}
 \epsilon^{ijk}\partial_j \zeta_k= \partial^i \Lambda.
\end{equation}
The solutions of Table 3, $i.e.$ the possible boundary 
conditions on the fields, translates in the ``duality'' 
condition between the potentials \eqref{duality}.
This confirms the fact that the dynamics on the boundary is uniquely 
determined by the Ward identities \eqref{wiconcl1} and 
\eqref{wiconcl2}. We find here, in a well defined field theoretical 
framework, a strong motivation for a relation which is known since a 
long time \cite{aratyn}, where this duality (or ``tomographic'') relation  was introduced to 
give a Bose description of fermions in 3D. For us, this condition is nothing else 
than the unique boundary condition on the fields $A^{i+}=\tilde{B}^{i+}$, written in 
terms of electromagnetic potentials defined by the boundary Maxwell 
equations \eqref{max1} and \eqref{max2}. This strongly suggest that the actual 
degrees of freedom of the dimensionally reduced 3D theory are 
fermionic, confirming recent developments concerning the edge 
states of topological insulators, which seem to be described in terms 
of fermion fields \cite{Hasan:2010xy}.\\
\vspace{0.7cm}
4) {\bf 3D boundary algebra} 
\begin{equation}
\label{algebraconcl}
 \begin{split}
  &[\tilde{B}^{0+}(X),A_{\alpha}(X')]_{t=t'}=
\partial_{\alpha} \delta^{(2)}(X'-X)\\
  &[\tilde{B}^{0+}(X),\tilde{B}_0^+(X')]_{t=t'}=0\\
  &[A^+_{\alpha}(X),A^{\beta +}(X')]_{t=t'}=0,
 \end{split}
\end{equation}
On the boundary, the above algebra is found. It is formed by a vectorial, conserved current, whose 
3D components are the insertions of the fields on (one side of) the boundary 
($\tilde{B}^{0+}(X)$ and  $A^+_\alpha(X)\ \alpha=1,2$, related by the 
duality-boundary condition $I$ in Table \ref{tabbu1}). We stress that the conservation of the 
current is obtained
on the mass-shell, $i.e.$ at vanishing external sources $J_\phi$.  This is in perfect analogy with what happens in the 3D CS and the 3D BF theory. In all cases, the conservation of the currents comes from the Ward identities of the residual gauge invariance broken by the most general boundary term respecting Symanzik's separability condition, going on the mass-shell, and exploiting the boundary condition previously found on the quantum fields. The physical interpretation of the current conservation is different, since in the 3D CS and BF cases, it leads, thanks to the boundary conditions, to the chirality of the currents. In the 4D BF case the current conservation \eqref{max1} (again, together with the duality-boundary condition \eqref{duality}), is tightly related to the electromagnetic structure and the consequent determination of the electromagnetic potentials.\\
\vspace{0.7cm}
5) {\bf canonical commutation relations and dimensional reduction} 
\begin{eqnarray}
\left [\Lambda(X),\Pi_{(\Lambda)}(X')\right]_{t=t'} &=& \delta^{(2)}(X-X')
\\
\left[\tilde\zeta^\alpha(X),\Pi_{(\zeta)\beta}(X')\right]_{t=t'}  &=&
\delta^{\alpha}_{\beta} \delta^{(2)}(X-X'),
\end{eqnarray}
where $\Pi_{(\Lambda)}\equiv\epsilon^{\alpha \beta} \partial_{\alpha}
\zeta_{\beta}$ and $\Pi_{(\zeta)\alpha}\equiv\partial_{\alpha} \Lambda$ are the conjugate
momenta of the fields $\Lambda$ and
$\tilde{\zeta}^\alpha\equiv\epsilon^{\alpha \beta} \zeta_{\beta}$
respectively. The point to stress here, is that, written in terms of the electromagnetic potentials \eqref{wiconcl1} and \eqref{wiconcl2}, the boundary algebra \eqref{algebraconcl} can be interpreted as a set of canonical commutation relations, for the canonically conjugate variables. Once realized this, it is almost immediate to write down the corresponding 3D Lagrangian, which is uniquely determined by our procedure. Indeed this analysis can be viewed as a systematic way to find (D$-1$)-dimensional Lagrangian out of D-dimensional bulk theories. 
It is a surprising and welcome result, that this new way of dimensionally reducing D-dimensional theories originates from the algebraic structure found on the boundary, interpreted as a set of canonical commutation relations, and which comes from the Ward identities describing the residual gauge invariance on the boundary and broken (by this latter) in the most general (and, as it turns out, unique !) way compatible with the Symanzik's simple criterion of separability. \\
\vspace{0.7cm}
6) {\bf 3D Lagrangian} 
\begin{eqnarray}
\mathcal{L} &=& \sum \Pi\dot{\Phi}-H \nonumber \\
&=& \epsilon^{\alpha \beta} \partial_{\alpha} \zeta_{\beta}
\partial_t \Lambda+\partial_{\alpha} \Lambda \epsilon^{\alpha \beta}
\partial_t \zeta_{\beta}-\frac{1}{2}(\epsilon^{\alpha \beta}
\partial_{\alpha} \zeta_{\beta})^2-\frac{1}{2}(\partial_{\alpha}
\Lambda)^2\label{3dlagrangian}
\end{eqnarray}
This is the 3D Lagrangian obtained on the mass-shell boundary of the 4D topological BF theory. It is the unique solution compatible with the QFT request of locality, power counting and with the Symanzik's request of separability. It is left invariant by gauge and translational transformations. It is non-covariant, and its dynamical variables (scalar and vector potentials) are coupled in a non-trivial way. Quite remarkably, this action, uniquely derived here by very general QFT principles, coincides with the one studied in \cite{balachandran} for the edge states of the 4D BF theory, where, the same algebraic origin is stressed. In a completely different 
theoretical framework, the action \eqref{3dlagrangian} is employed to study the surface of 4D (3+1) topological insulators \cite{cho}. The duality relation \eqref{duality} is there exploited to extract the desired fermionic degrees of freedom.

\appendix{

\chapter{The propagators of the three-dimensional BF model}
\label{appA}
In this appendix we furnish some details on the calculations of the propagators for the three-dimensional BF model.\\
The free equations of motion for the theory are:
\begin{equation}
\label{motolib}
 \begin{split}
   & \overline{\partial}B_u^a-\partial_u \overline{B}^a+J_A^a=\delta(u) [ \alpha_1 (\overline{A}^a_++\overline{A}^a_-)+\alpha_2 \overline{B}^a_++\alpha_3 \overline{B}^a_- ]\\
&\partial_u B^a-\partial B_u^a+J^a_{\overline{A}}=\delta(u) [\alpha_1(A^a_++A^a_-)+\alpha_3B^a_++\alpha_2B^a_-]\\
&\partial \overline{B}^a-\overline{\partial}B^a + b^a+J^a_{A_u}=0\\
&A_u^a+J_b^a=0\\
&\overline{\partial}A_u^a-\partial_u \overline{A}^a+J^a_B=\delta(u) [\alpha_3 \overline{A}^a_++ \alpha_2 \overline{A}^a_-+ \alpha_4 (\overline{B}^a_++\overline{B}^a_-)]\\
&\partial_u A^a-\partial A_u^a +J_{\overline{B}}^a=\delta(u)[\alpha_2A^a_++\alpha_3A^a_-+\alpha_4(B^a_++B^a_-)]\\
&\partial \overline{A}^a-\overline{\partial} A^a+d^a+J^a_{B_u}=0\\
&B_u^a+J^a_d=0.
 \end{split}
\end{equation}
If we differentiate the previous equations with respect to the sources $J_{\varphi}^b(x')$, we obtain a system of equations involving the propagators of the theory $\Delta^{ba}_{AB}(x',x)$, the solution of which is reported at the end of this appendix.\\
In what follows, we will derive one of the algebraic compatibility conditions between the propagators and the equations of motion \eqref{condizionieq} and one of the compatibility conditions between the propagators and the Ward identities \eqref{condizioniward}, in order to make more transparent the steps done in Chapter \ref{capBF}.\\
For example, let us derive the eighth relations in \eqref{condizionieq}. In order to do this, we differentiate the second equation in \eqref{motolib} with respect to $J_B^b(x')$, finding that:
\begin{equation}
\begin{split}
\partial_u \Delta_{BB}^{ba}(x',x)-\partial \Delta^{ba}_{B B_u}(x',x)=&\delta(u)[\alpha_1(\Delta^{ba}_{BA}(x',u=0^+)+\Delta^{ba}_{BA}(x',u=0^-))\\
                                                                     &+\alpha_3 \Delta^{ba}_{BB}(x',u=0^+)+\alpha_2 \Delta^{ba}_{BB}(x',u=0^-)].
\end{split}
\end{equation}
Then, we substitute the general solution for the propagators \eqref{solsols} in the previous equation and we expand the derivative with respect to $u$, finding the following relation:
\begin{equation}
 \begin{split}
  &\delta(u)(\theta(u')-\theta(-u')) \frac{a_8}{2 \pi i (z-z')^2}=\\
  &\delta(u) \Big[ \alpha_1 ( \theta_++\theta_-) \frac{a_4}{2 \pi i (z-z')^2} +(\alpha_3 \theta_+ +\alpha_2 \theta_-) \frac{a_8}{2 \pi i (z-z')^2} \Big].
 \end{split}
\end{equation}
If we require that $u'>0$, the previous relation become:
\begin{equation}
\label{rr}
 (\alpha_1 a_4+a_8(\alpha_3-1))\delta(u) \frac{1}{2 \pi i (z-z')^2}=0.
\end{equation}
Consequently, if we impose the equation \eqref{rr} to be valid for every pair of point $x$ and $x'$, we obtain:
\begin{equation}
 a_8(1-\alpha_3)-\alpha_1 a_4=0,
\end{equation}
which is the relation we are searching for.\\
The remaining equations in \eqref{condizioniward} can be derived in the same way, by differentiating the equations of motion with respect to the appropriate source and by substituting the general solution for the propagators in the equation found.\\
As regards the compatibility conditions between the propagators and the Ward identities, we derive the second equation in \eqref{condizioniward}.\\
Let us also consider the following integrated Ward identity with a boundary term:
\begin{equation}
\begin{split}
 \int_{-\mathcal{1}}^{\mathcal{1}} du (\partial J^a_A(x)+\overline{\partial}J^a_{\overline{A}}(x)-\sum_{\varphi} f^{abc} J^b_{\varphi} \varphi^c)=&-\alpha_1(\partial \overline{A}^a_++\partial \overline{A}^a_-+\overline{\partial}A^a_++\overline{\partial}A^a_-)\\
&-\alpha_2(\partial \overline{B}^a_++\partial \overline{B}^a_-+\overline{\partial}B^a_++\overline{\partial}B^a_-).\\
\end{split}
\end{equation}
If we differentiate the previous identity with respect to $J^b_{\overline{A}}(x')$ we obtain, keeping into account the relation $\delta^{ab}f^{abc}=0$:
\begin{equation}
\begin{split}
 &\delta(z-z')\delta'(\overline{z}-\overline{z}')\delta^{ab}=\delta^{ab} \Big[\\
&-\alpha_1\theta_+ (\partial \Delta^{ba}_{\overline{A}\overline{A}}(x',u=0^+)+\overline{\partial} \Delta^{ba}_{\overline{A}A}(x',u=0^+))+\\
&-\alpha_1\theta_-(\partial \Delta^{ba}_{\overline{A}\overline{A}}(x',u=0^-)+\overline{\partial}\Delta^{ba}_{\overline{A}A}(x',u=0^-))+\\
&-\alpha_2\theta_+(\partial \Delta^{ba}_{\overline{A}\overline{B}}(x',u=0^+)+\overline{\partial}\Delta^{ba}_{\overline{A}B}(x',u=0^+))+\\
&-\alpha_2\theta_-(\partial \Delta^{ba}_{\overline{A}\overline{B}}(x',u=0^-)+\overline{\partial} \Delta^{ba}_{\overline{A}B}(x',u=0^-)) \Big].
\end{split}
\end{equation}
This accomplished, we substitute the general solutions for the propagators in the previous equation and we set $u'>0$. Keeping in mind the identity $\frac{1}{(\overline{z}-\overline{z}')^2}=-\overline{\partial}\frac{1}{\overline{z}-\overline{z}'}$, we find:
\begin{equation}
\begin{split}
  &\delta(z-z')\delta'(\overline{z}-\overline{z}')\delta^{ab}=\delta^{ab} \Big[\\
 &-(\alpha_1 a_5+\alpha_2(a_6-a_7))\overline{\partial}\partial\frac{1}{2 \pi i (\overline{z}-\overline{z}')}+(a_2 \alpha_1+\alpha_2(1+a_7))\delta(z-z')\delta'(\overline{z}-\overline{z}') \Big].
\end{split}
\end{equation}
Then, if we impose the previous relation to be valid for every pairs of point $x$ and $x'$, we find:
\begin{equation}
\label{fine}
\alpha_1 a_2+(1+a_7)\alpha_2-a_5 \alpha_1-(a_6-a_7) \alpha_2=1,
\end{equation}
where we have used the relation $\partial\frac{1}{2 \pi i (\overline{z}-\overline{z}')}=\delta^{(2)}(Z-Z')$. The equation \eqref{fine} is the compatibility condition we are searching for.\\
The remaining equations in \eqref{condizioniward} can now be derived from the integrated Ward identities \eqref{wardbf1} and \eqref{wardbf2} in the same way.\\

\begin{landscape}
\begin{changemargin}{-2.5cm}{-1cm}
\begin{equation}
\label{solsols}
\begin{split}
&\Delta_{+AB}=\\
&\begin{pmatrix}
 \frac{a_1}{2 \pi i (z-z')^2} & a_2 \delta^{(2)} & 0 & \partial T_{a_3}(x',x) & \frac{a_4}{2 \pi i (z-z')^2} & T_{a_3-a_4}(x',x) & 0 & (a_1+a_2) \partial \delta^{(2)}\\
 a_2 \delta^{(2)}& \frac{a_5}{2 \pi i (\overline{z}-\overline{z}')^2}&0&-\overline{\partial}T_{a_6}(x,x')&T_{a_7}(x,x')& \frac{a_6-a_7}{2 \pi i (\overline{z}-\overline{z}')^2}&0& -(a_2+a_5)\overline{\partial}\delta^{(2)}\\
0&0&0&-\delta^{(3)}&0&0&0&0\\
-\partial T_{a_3}(x,x') & \overline{\partial} T_{a_6}(x',x) & -\delta^{(3)}(x-x') &(2a_9+a_8+a_{10}=\partial \overline{\partial} \delta^{(2)}& -(a_8+a_9)\partial \delta^{(2)} & (a_9+a_{10}) \overline{\partial} \delta^{(2)} & 0 & (1+a_3+a_6) \partial \overline{\partial} \delta^{(2)}\\
\frac{a_4}{2 \pi i (z-z')^2} & T_{a_7}(x',x)& 0 & (a_8+a_9) \partial \delta^{(2)} & \frac{a_8}{2 \pi i (z-z')^2}& a_9 \delta^{(2)} & 0 & \partial T_{a_4+a_7}(x',x)\\
T_{a_3-a_4}(x,x') & \frac{a_6-a_7}{2 \pi i (\overline{z}-\overline{z}')^2}&0& -(a_9+a_10 \overline{\partial} \delta^{(2)} & a_9 \delta^{(2)} & \frac{a_{10}}{2 \pi i (\overline{z}-\overline{z}')^2}&0& -\overline{\partial} T_{a_3-a_4+a_6-a_7}(x,x')\\
0 & 0 & 0 & 0 & 0 & 0 & 0 & - \delta^{(3)}\\
-(a_1+a_2) \partial \delta^{(2)} & (a_2+a_5) \overline{\partial} \delta^{(2)} & 0 & (1+a_3+a_6) \partial \overline{\partial} \delta^{(2)} & - \partial T_{a_4+a_7}(x,x') & \overline{\partial}T_{a_3-a_4-a_6-a_7}(x',x) & -\delta^{(3)}& (2a_2+a_1+a_5 \partial \overline{\partial} \delta^{(2)}
\end{pmatrix}
\end{split}
\end{equation}
\\
The indices $A$ and $B$ denote the ordered set of fields $\{A, \overline{A}, A_u, b, B, \overline{B}, B_u,d \}$. As was previously done, $\Delta_{-AB}$ can be obtained from $\Delta_{+AB}$ by a parity transformation. $T_{\rho}$ and $\frac{1}{2 \pi i (z-z')^2}$ are the tempered distributions defined in \eqref{Trho} and \eqref{unosuzeta}.
\end{changemargin}
\end{landscape}

\chapter{The propagators of the four-dimensional BF model}
\label{apppropbf4}
In this appendix we will illustrate how to derive the propagators of the four-dimensional BF model, keeping into account the boundary conditions obtained Section \ref{boundarycondition}.\\

As usual, due to the separability condition, the propagators of the theory assume the following form:
\begin{equation}
 \Delta_{\varphi_1 \varphi_2}(x,x')=\theta(x_3)\theta(x_3')\Delta^+_{\varphi_1 \varphi_2}(x,x')+\theta(-x_3)\theta(-x_3')\Delta^-_{\varphi_1 \varphi_2}(x,x'),
\end{equation}
where $\Delta^+_{\varphi_1 \varphi_2}$ and $\Delta^-_{\varphi_1 \varphi_2}$ are solutions of the system of equations for the propagators obtained by differentiating the equations of motion \eqref{motobf4} with respect to the sources of the fields. They must be compatible with the equations of motion \eqref{motozrotte} and with the Ward identities \eqref{wwr1} and \eqref{wwr2}. Since $\Delta^+_{\varphi_1 \varphi_2}$ and $\Delta^-_{\varphi_1 \varphi_2}$ are transformed into each other by a parity transformation, in this Appendix, we will derive a solution for $\Delta^+_{\varphi_1 \varphi_2}$, where $x_3,x_3' \ge 0$. (In what follows we omit the apex $+$).\\
If we differentiate the equations of motion \eqref{motobf4} with respect to the sources of the fields, and we evaluate the expressions obtained at the vanishing sources, we find a system of equations involving the propagators of the theory:
\begin{equation}
\label{propbf4}
 \begin{split}
  &\Delta_{A_3 \psi}(x,x')=0 \qquad \qquad \forall \; \psi(x') \neq b(x')\\
 &\Delta_{A_3 b}(x,x')=-\delta^{(4)}(x-x')\\
 &\Delta_{B^{i3} \psi} (x,x')=0 \qquad \qquad \forall \; \psi(x') \neq d^i(x')\\
 &\Delta_{B^{i3} d_l} (x,x')=- \delta^i_{\;l} \delta^{(4)}(x-x')\\
 &\partial_3 \Delta_{A_l \tilde{B}^i}(x',x)=-\frac{\delta^i_{\;l}}{\alpha}\delta^{(4)}(x-x')\\
 &\partial_3 \Delta_{\tilde{B}^l \tilde{B}_i}(x',x)=0\\
 &\partial_3 \Delta_{b \tilde{B}^i}(x',x)=0\\
 &\partial_3 \Delta_{d_l \tilde{B}^i}(x',x)=2 \epsilon^{ij}_{\;\;l} \partial_j \delta^{(4)}(x-x')\\
 &\partial_3 \Delta_{A_l A^i}(x',x)=0\\
 &\partial_3 \Delta_{\tilde{B}_l A^i}(x',x)=\frac{\delta^i_{\;l}}{\alpha}\delta^{(4)}(x-x')\\
 &\partial_3 \Delta_{b A^i}(x',x)=-\partial_i \delta^{(4)}(x'-x)\\
 &\partial_3 \Delta_{d_l A^i}(x',x)=0\\
 &\Delta_{A_l b}(x',x)=\alpha \partial_i \Delta_{A_l \tilde{B}^i}(x',x)\\ 
 &\Delta_{\tilde{B}_l b}(x',x)=\alpha \partial_i \Delta_{\tilde{B}_l \tilde{B}^i}(x',x)\\
 &\Delta_{b b}(x',x)=\alpha \partial_i \Delta_{b \tilde{B}^i}(x',x)\\
 &\Delta_{d_l b}(x',x)=\alpha \partial_i \Delta_{d_l \tilde{B}^i}(x',x)\\
 &\Delta_{A_l d^i}=-2 \alpha \epsilon^{ijk}\partial_j \Delta_{A_l A_k}(x',x)\\
 &\Delta_{\tilde{B}_ld^i}(x',x)=-2 \alpha \epsilon^{ijk}\partial_j \Delta_{\tilde{B}_l A_k}(x',x)\\
 &\Delta_{d_l d^i}(x',x)=-2 \alpha \epsilon^{ijk} \partial_j \Delta_{d_l A_k}(x',x)\\
 &\Delta_{b d^i}(x',x)=-2\alpha \epsilon^{ijk} \partial_j \Delta_{b A_k}(x',x).
 \end{split}
\end{equation}
Notice that it follows directly from the gauge conditions, i.e., from the last two equations in \eqref{motobf4}, that the Green functions containing $A_3$ and $B_{i3}$ are all zero except $\Delta_{A_3 b}(x,x')=-\delta^{(4)}(x-x') $ and $\Delta_{B^{i3} d_l} (x,x')=- \delta^i_{\;l} \delta^{(4)}(x-x')$ and, for this reason, we do not list these propagators in the following developments.\\
Consequently, the most general solution of the previous system is:
\begin{equation}
\label{desolp}
\begin{split}
&\Delta_{AB}(x',x)=\\
&\begin{pmatrix} 
 \Xi_l^{\;i}(X,X') & -\frac{\delta^i_l}{\alpha}T_c(x,x') & -2 \alpha \epsilon^{ij}_{\; \; k} \partial_j \Xi_l^{\;k}(X',X) & - \partial_l T_c(x,x')\\ 
 \; & \; & \; & \;\\
 -\frac{\delta^i_l}{\alpha}T_{c_1}(x',x) & \Omega_l^{\;i}(X,X') & 2\epsilon^{ij}_{\; \; l} \partial_j T_{c_1}(x',x) & \alpha \partial_i \Omega_l^{\;i}(X,X')\\
 \; & \; & \; & \;\\
 -2 \alpha \epsilon^{ij}_{\; \; k} \partial_j \Xi_l^{\;k}(X',X) & 2 \epsilon^{ij}_{\; \; l} \partial_j T_{\delta}(x,x') & 4 \alpha^2 \epsilon^{ij}_{\; \; k} \partial_j \epsilon^{kp}_{\; \; q} \partial_p \Xi_c^{\;p}(X,X') & 0 \\
 \; & \; & \; & \;\\
 \partial_i T_{\gamma}(x',x) & \alpha \partial^l \Omega_l^{\;i}(X',X) & 0 & \alpha \partial^l \partial_i \Omega_l^{\;i}(X,X')
\end{pmatrix}
\end{split}
\end{equation}

As above, we have used the matrix notation and the indices $A$ and $B$ denote the ordered set of fields $\{ A^i, \tilde{B}^i, d^i,b \}$. $T_{\psi}(x,x')$ is the tempered distribution $(\theta(x_3-x_3')+\psi)\delta^3(X-X')$, $\Xi_l^{\;i}(X,X')$ and $\Omega_l^{\;i}(X,X')$ are arbitrary function of the transverse coordinates $X$ and $c,c_1, \gamma$, and $\delta$ are arbitrary constant parameters.\\

Let us consider the boundary conditions $I$ and $II$ in Table \eqref{tabbu1}. For both these solutions the Ward identities \eqref{wwrg1} and \eqref{wwrg2} take the following form:
\begin{equation}
\label{wibfsol}
 \begin{split}
  &\int_{-\mathcal{1}}^{\mathcal{1}}dx_3 \partial_i J^i_{A^i}=\alpha \partial_i \tilde{B}^{i+}  \\
  &\int_{-\mathcal{1}}^{\mathcal{1}}dx_3 \epsilon^{ijk} \partial_j J_{\tilde{B}^k}=-\alpha \epsilon^{ijk} \partial_j A^+_k.
 \end{split}
\end{equation}
If we differentiate the equations \eqref{wibfsol} with respect to the sources $J_{A^l}(x'),$ $J_{\tilde{B}^l}(x'),J_{d^l}(x')$ and $J_b(x')$, we obtain eight relations involving the propagators of the theory:
\begin{equation}
\label{wiconditionbf4}
 \begin{split}
 &\partial_i \delta^{i}_{\;l} \delta^{(3)}(X'-X)=\alpha \partial_i \Big(\Delta_{A_l \tilde{B}^i}(x',x) \Big)_{x_3=0}\\
 &\partial_i\Big(\Delta_{\tilde{B}_l \tilde{B}^i}(x',x) \Big)_{x_3=0}=0\\
 &\partial_i \Big(\Delta_{d_l \tilde{B}^i}(x',x) \Big)_{x_3=0} =0\\
 &\partial_i \Big(\Delta_{b \tilde{B}^i}(x',x) \Big)_{x_3=0}=0\\
 &\epsilon^{ijk} \partial_j \Big(\Delta_{A_l A^k}(x',x) \Big)_{x_3=0}=0\\
 &\epsilon^{ijk} \partial_j \delta_{kl} \delta^{(3)}(X'-X)=-\alpha \epsilon^{ijk} \partial_j \Big(\Delta_{\tilde{B}_l A_k}(x',x) \Big)_{x_3=0}\\ 
 &\epsilon^{ijk}\partial_j \Big(\Delta_{d_l A_k}(x',x) \Big)_{x_3=0}=0\\
 &\epsilon^{ijk} \partial_j \Big(\Delta_{b A_k}(x',x) \Big)_{x_3=0}=0.
 \end{split}
\end{equation}
If we substitute the propagators \eqref{desolp} in the previous system of differential equations we obtain the following constraints on the parameters $c$ and $c_1$:
\begin{equation}
\label{psymmetric}
 \begin{split}
  &c=-1\\
  &c_1=0.
 \end{split}
\end{equation}
Moreover, it is necessary that the Green functions involving the Lagrange multipliers $b$ and $d^i$ are compatible with our choice on the behavior of these fields in the limit $x_3 \rightarrow \mathcal{1}$, (remember that we  postulated the condition $b(x_3= \pm \mathcal{1})=d^i(x_3=\pm \mathcal{1})=0$). In other words, we have to impose the following constraints:
\begin{equation}
 \begin{split}
  &\lim_{x_3 \rightarrow + \mathcal{1}} 2 \epsilon^{ij}_{\; \; l} \partial_j T_{\delta}(x,x')=0\\
  &\lim_{x_3 \rightarrow + \mathcal{1}} \partial_i T_{\gamma}(x',x)=0,
 \end{split}
\end{equation}
which yield the following conditions on the parameters $\gamma$ and $\delta$:
\begin{equation}
 \begin{split}
  &\delta=0\\
  &\gamma=-1.
 \end{split}
\end{equation}
Regarding the propagators $\Delta_{A_l A^i}(x',x)$ and $\Delta_{\tilde{B}_l \tilde{B}^i}(x',x)$, we have decided to require that these two Green functions are symmetric in the exchange $\{x,i\} \leftrightarrow \{x',l\}$ since, in this case, the two fields involved are equal (in fact they are diagonal terms of the matrix of the propagators). As a consequence, keeping into account the second and seventh equation in \eqref{wiconditionbf4}, we obtain that the functions $\Xi_l^{\;i}(X,X')$ and $\Omega_l^{\;i}(X,X')$ take the following form:
\begin{equation}
 \begin{split}
  &\Xi_l^{\;i}(X,X')=\partial_i \partial_l \eta (X-X')\\
  &\Omega_l^{\;i}(X,X')=\epsilon^{ijk} \partial_j \epsilon_l^{\;rs} \partial_r \varphi_{ks} (X-X'),
 \end{split}
\end{equation}
where $\eta$ is a dimensionless scalar function of $X-X'$ and $\varphi_{ks}$ is a function with two vector indices such that $[\varphi]=2$.\\
Consequently, the propagators \eqref{desolp} become:
\begin{equation}
\label{desolp1}
\begin{split}
&\Delta_{AB}(x',x)=\\
&\begin{pmatrix} 
 \partial_i \partial^l \eta(X-X') & -\frac{\delta^i_l}{\alpha}T_{-1}(x,x') & 0 & - \partial_l T_{-1}(x,x')\\ 
 \; & \; & \; & \;\\
 -\frac{\delta^i_l}{\alpha}T_{0}(x',x) & \epsilon^{ijk} \partial_j \epsilon_l^{\;rs} \partial_r \varphi_{ks} (X-X') & 2\epsilon^{ij}_{\; \; l} \partial_j T_{0}(x',x) & 0\\
 \; & \; & \; & \;\\
 0 & 2 \epsilon^{ij}_{\; \; l} \partial_j T_{0}(x,x') & 0 & 0 \\
 \; & \; & \; & \;\\
 \partial_i T_{-1}(x',x) & 0 & 0 & 0
\end{pmatrix}
\end{split}
\end{equation}

Finally, we make a comment on the method used to derive the propagators in this Appendix. In Chapter 2 we found the propagators by solving the system of equations for the propagators arising from the free equations of motion and by fixing the arbitrary parameters of the solution found by means of the compatibility relations resulting from the Ward identities and from the equations of motion with a  boundary term. Conversely, here we have used the equations of motion with a boundary term in an indirect manner, i.e., by substituting the boundary conditions found with the algebraic method into the Ward identities and by using the relations obtained to fix the arbitrary parameters of the propagators. It is apparent that the two approaches coincide. In fact, while in Chapter 2 we read the boundary conditions from the propagators, in this Appendix we use the condition found with the algebraic method to obtain propagators which are compatible with these boundary 
conditions.}

\chapter*{Conclusions}
\addcontentsline{toc}{chapter}{Conclusions}
\chaptermark{Conclusions}
In this thesis we studied the Symanzik's method for the introduction of the boundary in a field theory and, specifically, we applied this method to three Topological Field Theories of the Shwartz type: the non-abelian Chern-Simons model, the non-abelian three-dimensional BF theory and its abelian four-dimensional version.\\
In particular, the original findings of this thesis comes from the analysis of the BF model in three and four space-time dimensions.\\
Regarding the three-dimensional model we compared two methods known in the literature for the computation of the boundary conditions, \cite{maggiore,magnoli}. In the method used in \cite{maggiore} the direct calculation of the propagators is necessary while the technique used in \cite{magnoli} avoids the computation of the Green functions. However, the second method have some complications since it provides a certain number of additional boundary conditions which are non-acceptable. As illustrated in Chapter 2, we found that the boundary term in the action plays the role of the gauge fixing of the residual gauge invariance of the theory on the boundary. This original interpretation of the boundary term of the action allow us to establish a set of criteria in order to discard the additional boundary conditions furnished by this method without the need to compare the result with those obtained by calculating the propagators.\\
These criteria were extremely useful in the analysis of the four-dimensional BF model, as illustrated in Chapter 3, (for a detailed discussion of the results of Chapter 3 see Chapter 3, Section \ref{secbf44}). Our main findings here is the characterization of the dynamics on the boundary of the model. 
In fact we found that, due to the residual gauge invariance of the model, on the boundary and on the mass shell, the 4D BF theory displays Maxwell equations for an electric field and a magnetic field. Consequently two potentials can be introduced: an electric scalar potential and a magnetic vector potential, depending on the 3D coordinates on the boundary.\\
Furthermore, the only acceptable boundary condition which we found using the criteria established in Chapter 2, translates in a ``duality'' condition between these two potentials. This duality condition is exactly the relation which is necessary to give a Bose description of fermions in 3D. This strongly suggest that the actual degrees of freedom of the dimensionality reduced 3D theory are fermionic, confirming recent developments concerning the edge states of topological insulators, which seem to be described in terms of fermion fields.\\ 
Moreover the algebra of local boundary observables, which exists due to the residual gauge invariance of the bulk theory, can be interpreted as a set of canonical commutation relations which allowed us to construct a boundary Lagrangian and to complete the description of the dynamics on the boundary.\\
Both the boundary algebra and the boundary Lagrangian which we found were already studied in literature \cite{balachandran,cho}, but our main finding is to derive these quantities solely on the basis of principles of field theory without the need to introduce any additional assumption, as was done in \cite{balachandran,cho}.\\

Regarding the future developments, it will be possible to study the non-abelian four-dimensional BF model with a boundary in order to verify whether this model has some differences compared to the abelian case, or to study the five-dimensional BF model in order to analyze the dynamics on its four-dimensional boundary.

\cleardoublepage

\chapter*{Acknowledgements}
\addcontentsline{toc}{chapter}{Acknowledgements}
\chaptermark{Acknowledgements}
I would like to thank everyone who contributed to the writing of this thesis, especially my thesis advisors Nicola Maggiore and Nicodemo Magnoli for the time they dedicated to me and for the beautiful experience of this past year.


\addcontentsline{toc}{chapter}{Bibliography}
\begin{thebibliography}{999}
\bibitem{schwarz1978}
  A.~S.~Schwarz,
  ``The Partition Function of Degenerate Quadratic Functional and Ray-Singer Invariants,''
  Lett.\ Math.\ Phys.\  {\bf 2} (1978) 247.
\bibitem{Witten:1982im}
  E.~Witten,
  ``Supersymmetry and Morse theory,''
  J.\ Diff.\ Geom.\  {\bf 17} (1982) 661.
\bibitem{donaldson}
  S.~K.~Donaldson,
  ``An Application of gauge theory to four-dimensional topology,''
  J.\ Diff.\ Geom.\  {\bf 18} (1983) 279.
\bibitem{Schwarz:1979ae}
  A.~S.~Schwarz,
  ``The Partition Function of a Degenerate Functional,''
  Commun.\ Math.\ Phys.\  {\bf 67} (1979) 1.
\bibitem{Birmingham}
  D.~Birmingham, M.~Blau, M.~Rakowski and G.~Thompson,
  ``Topological field theory,''
  Phys.\ Rept.\  {\bf 209} (1991) 129.
\bibitem{Horowitz}
  G.~T.~Horowitz and M.~Srednicki,
  ``A Quantum Field Theoretic Description Of Linking Numbers And Their Generalization,''
  Commun.\ Math.\ Phys.\  {\bf 130} (1990) 83.
\bibitem{Moore}
  G.~W.~Moore and N.~Seiberg,
  ``Taming the Conformal Zoo,''
  Phys.\ Lett.\ B {\bf 220} (1989) 422.
\bibitem{Witten:1988hc}
  E.~Witten,
  ``(2+1)-Dimensional Gravity as an Exactly Soluble System,''
  Nucl.\ Phys.\ B {\bf 311} (1988) 46.
\bibitem{Aneziris:1990gm}
  C.~Aneziris, A.~P.~Balachandran, L.~Kauffman and A.~M.~Srivastava,
  ``Novel Statistic For Strings And String 'chern-simons' Terms,''
  Int.\ J.\ Mod.\ Phys.\ A {\bf 6} (1991) 2519.
\bibitem{Diamantini:1995yb}
  M.~C.~Diamantini, P.~Sodano and C.~A.~Trugenberger,
  Nucl.\ Phys.\ B {\bf 474} (1996) 641
  [hep-th/9511168].
\bibitem{Cattaneo:1995tw}
  A.~S.~Cattaneo, P.~Cotta-Ramusino, J.~Frohlich and M.~Martellini,
  ``Topological BF theories in three-dimensions and four-dimensions,''
  J.\ Math.\ Phys.\  {\bf 36} (1995) 6137
  [hep-th/9505027].
\bibitem{Witten:1988hf}
  E.~Witten,
  ``Quantum Field Theory and the Jones Polynomial,''
  Commun.\ Math.\ Phys.\  {\bf 121} (1989) 351.
\bibitem{kac}
  V.~G.~Kac,
  ``Simple graded algebras of finite growth,''
  Funct.\ Anal.\ Appl.\  {\bf 1} (1967) 328.
\bibitem{Witten:1983ar}
  E.~Witten,
  ``Nonabelian Bosonization in Two-Dimensions,''
  Commun.\ Math.\ Phys.\  {\bf 92} (1984) 455.
\bibitem{Deser:1983nh}
  S.~Deser and R.~Jackiw,
  ``Three-Dimensional Cosmological Gravity: Dynamics of Constant Curvature,''
  Annals Phys.\  {\bf 153} (1984) 405.
\bibitem{Banados:1992wn}
  M.~Banados, C.~Teitelboim and J.~Zanelli,
  ``The Black hole in three-dimensional space-time,''
  Phys.\ Rev.\ Lett.\  {\bf 69} (1992) 1849
  [hep-th/9204099].
\bibitem{Banados:1998ta}
  M.~Banados, T.~Brotz and M.~E.~Ortiz,
  ``Boundary dynamics and the statistical mechanics of the (2+1)-dimensional black hole,''
  Nucl.\ Phys.\ B {\bf 545} (1999) 340
  [hep-th/9802076].
\bibitem{Park:1998qk}
  M.~-I.~Park,
  ``Statistical entropy of three-dimensional Kerr-de Sitter space,''
  Phys.\ Lett.\ B {\bf 440} (1998) 275
  [hep-th/9806119].
\bibitem{maggiore}
  N.~Maggiore and P.~Provero,
  ``Chiral current algebras in three-dimensional BF theory with boundary,''
  Helv.\ Phys.\ Acta {\bf 65} (1992) 993
  [hep-th/9203015].
\bibitem{Balachandran:1992qg}
  A.~P.~Balachandran and P.~Teotonio-Sobrinho,
  ``The Edge states of the BF system and the London equations,''
  Int.\ J.\ Mod.\ Phys.\ A {\bf 8} (1993) 723
  [hep-th/9205116].
\bibitem{balachandran}
  A.~P.~Balachandran, G.~Bimonte and P.~Teotonio-Sobrinho,
  ``Edge states in 4-d and their 3-d groups and fields,''
  Mod.\ Phys.\ Lett.\ A {\bf 8} (1993) 1305
  [hep-th/9301120].
\bibitem{momen}
  A.~Momen,
  ``Edge dynamics for BF theories and gravity,''
  Phys.\ Lett.\ B {\bf 394} (1997) 269
  [hep-th/9609226].
\bibitem{Tsui:1999zz}
  D.~C.~Tsui,
  ``Nobel Lecture: Interplay of disorder and interaction in two-dimensional electron gas in intense magnetic fields,''
  Rev.\ Mod.\ Phys.\  {\bf 71} (1999) 891.
\bibitem{Kane:2005zz}
  C.~L.~Kane and E.~J.~Mele,
  ``Z-2 Topological Order and the Quantum Spin Hall Effect,''
  Phys.\ Rev.\ Lett.\  {\bf 95} (2005) 146802.
\bibitem{Bernevig:2006zz}
  B.~A.~Bernevig and S.~-C.~Zhang,
  ``Quantum Spin Hall Effect,''
  Phys.\ Rev.\ Lett.\  {\bf 96} (2006) 106802.
\bibitem{konig:2007}
  M. ~Konig et al.,
  ``Quantum Spin Hall Insulator State in HgTe Quantum Wells,''
  Science\ (2007).
\bibitem{Hasan:2010xy}
  M.~Z.~Hasan and C.~L.~Kane,
  ``Topological Insulators,''
  Rev.\ Mod.\ Phys.\  {\bf 82} (2010) 3045
  [arXiv:1002.3895 [cond-mat.mes-hall]].
\bibitem{Zhang:1992eu} 
  S.~-C.~Zhang,
  ``The Chern-Simons-Landau-Ginzburg theory of the fractional quantum Hall effect,''
  Int.\ J.\ Mod.\ Phys.\ B {\bf 6}, 25 (1992).
\bibitem{Wen:1995qn}
  X.~-G.~Wen,
  ``Topological orders and edge excitations in FQH states,''
  PRINT-95-148 (MIT).
\bibitem{cho}
  G.~Y.~Cho and J.~E.~Moore,
  ``Topological BF field theory description of topological insulators,''
  Annals Phys.\  {\bf 326} (2011) 1515
  [arXiv:1011.3485 [cond-mat.str-el]].
\bibitem{Santos:2011bf}
  L.~Santos, T.~Neupert, S.~Ryu, C.~Chamon and C.~Mudry,
  ``Time-reversal symmetric hierarchy of fractional incompressible liquids,''
  Phys.\ Rev.\ B {\bf 84} (2011) 165138
  [arXiv:1108.2440 [cond-mat.str-el]].
\bibitem{magnoli}
  A.~Blasi, A.~Braggio, M.~Carrega, D.~Ferraro, N.~Maggiore and N.~Magnoli,
  ``Non-Abelian BF theory for 2+1 dimensional topological states of matter,''
  New J.\ Phys.\  {\bf 14} (2012) 013060
  [arXiv:1106.4641 [cond-mat.mes-hall]].
\bibitem{syma}
  K.~Symanzik,
  ``Schrodinger Representation and Casimir Effect in Renormalizable Quantum Field Theory,''
  Nucl.\ Phys.\ B {\bf 190} (1981) 1.
\bibitem{emery}
  S.~Emery and O.~Piguet,
  ``Chern-Simons theory in the axial gauge: Manifold with boundary,''
  Helv.\ Phys.\ Acta {\bf 64} (1991) 1256.
\bibitem{collina}
  A.~Blasi and R.~Collina,
  ``The Chern-Simons model with boundary: A Cohomological approach,''
  Int.\ J.\ Mod.\ Phys.\ A {\bf 7} (1992) 3083.
\bibitem{maxw}
  A.~Blasi, N.~Maggiore, N.~Magnoli and S.~Storace,
  ``Maxwell-Chern-Simons Theory With Boundary,''
  Class.\ Quant.\ Grav.\  {\bf 27} (2010) 165018
  [arXiv:1002.3227 [hep-th]].
\bibitem{ferraro}
  A.~Blasi, D.~Ferraro, N.~Maggiore, N.~Magnoli and M.~Sassetti,
  ``Symanzik's Method Applied To The Fractional Quantum Hall Edge States,''
  Annalen Phys.\  {\bf 17} (2008) 885
  [arXiv:0804.0164 [hep-th]].
\bibitem{bassetto}
  A.~Bassetto, G.~Nardelli and R.~Soldati,
  ``Yang-Mills theories in algebraic noncovariant gauges: Canonical quantization and renormalization,''
  Singapore, Singapore: World Scientific (1991) 227 p
\bibitem{swartz}
 A.~Schwarz,
 ``New topological invariants arising in the theory of quantized fields,''
 Abstract of the International Topological Conference (1978).
\bibitem{witten}
  E.~Witten,
  ``Topological Quantum Field Theory,''
  Commun.\ Math.\ Phys.\  {\bf 117} (1988) 353.
\bibitem{weinberg}
  S.~Weinberg,
  ``The Quantum theory of fields. Vol. 1: Foundations,''
  Cambridge, UK: Univ. Pr. (1995) 609 p
\bibitem{sassarini}
  A.~Blasi, R.~Collina and J.~Sassarini,
  ``Finite Casimir effect in quantum field theory,''
  Int.\ J.\ Mod.\ Phys.\ A {\bf 9} (1994) 1677.
\bibitem{dirac}
  P.~A.~M.~Dirac,
  ``Forms of Relativistic Dynamics,''
  Rev.\ Mod.\ Phys.\  {\bf 21} (1949) 392.
\bibitem{brandhuber}
  A.~Brandhuber, S.~Emery, K.~Landsteiner and M.~Schweda,
  ``The Three-dimensional BF model with cosmological term in the axial gauge,''
  Helv.\ Phys.\ Acta {\bf 68} (1995) 126
  [hep-th/9502147].
\bibitem{maggiore2}
  N.~Maggiore and S.~P.~Sorella,
  ``Finiteness of the topological models in the Landau gauge,''
  Nucl.\ Phys.\ B {\bf 377} (1992) 236.
\bibitem{sorella}
  E.~Guadagnini, N.~Maggiore and S.~P.~Sorella,
  ``Supersymmetry of the three-dimensional Einstein-Hilbert gravity in the Landau gauge,''
  Phys.\ Lett.\ B {\bf 247} (1990) 543.
\bibitem{Amoretti:2012kb}
  A.~Amoretti, A.~Blasi, N.~Maggiore and N.~Magnoli,
  ``3D Dynamics of 4D Topological BF Theory With Boundary,''
  arXiv:1205.6156 [hep-th].
\bibitem{maggiore1}
  N.~Maggiore and S.~P.~Sorella,
  ``Perturbation theory for antisymmetric tensor fields in four-dimensions,''
  Int.\ J.\ Mod.\ Phys.\ A {\bf 8} (1993) 929
  [hep-th/9204044].
\bibitem{Guadagnini:1990br}
  E.~Guadagnini, N.~Maggiore and S.~P.~Sorella,
  ``Supersymmetric structure of four-dimensional antisymmetric tensor fields,''
  Phys.\ Lett.\ B {\bf 255} (1991) 65.
\bibitem{Batalin:1981jr}
  I.~A.~Batalin and G.~A.~Vilkovisky,
  ``Gauge Algebra and Quantization,''
  Phys.\ Lett.\ B {\bf 102} (1981) 27.
\bibitem{Batalin:1984jr}
  I.~A.~Batalin and G.~A.~Vilkovisky,
  ``Quantization of Gauge Theories with Linearly Dependent Generators,''
  Phys.\ Rev.\ D {\bf 28} (1983) 2567
   [Erratum-ibid.\ D {\bf 30} (1984) 508].
\bibitem{brs}
C.~Becchi, A.~Rouet and R.~Stora,
``1975 Lectures given at Ettore Majorana Summer School (Erice,
Sicily) (Erice
Math. Phys. Conf. 1975)'' (Erice: Ettore Majorana Summer School) p
299.
\bibitem{aratyn}
  H.~Aratyn,
  Phys.\ Rev.\ D {\bf 28} (1983) 2016.
\end{thebibliography}
\end{document}